\documentclass{aa}  
\usepackage{epstopdf}
\usepackage{graphicx}
\usepackage{lscape}
\usepackage{natbib} 
\usepackage{siunitx}
\usepackage{url}
\usepackage[varg]{txfonts}
\usepackage{longtable}
\usepackage{supertabular}
\usepackage{pdfpages}
\usepackage{multirow}
\usepackage{pgfplotstable,tikz,filecontents}
\usepackage{pdfpages}

\bibpunct{(}{)}{;}{a}{}{,} 
\def\changed{}

\begin{document}

\title{Stellar population of the superbubble N\,206 in the LMC}

   \subtitle{II. Parameters of the OB and WR stars, and the total massive star feedback}

   \author{Varsha Ramachandran\inst{1}
          \and W.-R. Hamann\inst{1}
          \and R. Hainich \inst{1}
          \and L. M. Oskinova\inst{1}
          \and T. Shenar\inst{1}
          \and A. A. C. Sander\inst{1}
          \and H. Todt\inst{1}
          \and J. S. Gallagher\inst{2}      
          }

   \institute{Institut f\"ur Physik und Astronomie,
              Universit\"at Potsdam,
              Karl-Liebknecht-Str. 24/25, D-14476 Potsdam, Germany \\
              \email{varsha@astro.physik.uni-potsdam.de}
              \and Department of Astronomy, University of Wisconsin - Madison, WI, USA }       
   \date{Received <date> / Accepted <date>}


\abstract
%
%
{Clusters or associations of early-type stars are often associated with a `superbubble' of hot gas. The formation of such superbubbles is caused by the feedback from massive stars. The complex N\,206 in the Large Magellanic Cloud exhibits a superbubble and a rich massive star population.
} 
%
%
{Our goal is to perform quantitative spectral analyses of all massive stars associated with the N\,206 superbubble in order to determine their stellar and wind parameters. We compare the superbubble energy budget to the stellar energy input and discuss the star formation history of the region.
}
%
%
{
We observed the massive stars in the N\,206 complex using the multi-object spectrograph FLAMES  at ESO's Very Large Telescope (VLT).  Available ultra-violet (UV) spectra from archives are also used. The spectral analysis is performed with Potsdam Wolf-Rayet (PoWR) model atmospheres by reproducing the observations with the synthetic spectra.
} 
%
%
{
We present the stellar and wind parameters of the OB stars and the two Wolf-Rayet (WR) binaries in the N\,206 complex. Twelve percent of the sample show Oe/Be type emission lines, although most of them appear to rotate far below critical. We found eight runaway stars based on their radial velocity. The wind-momentum luminosity relation of our OB sample is consistent with  the expectations. The Hertzsprung-Russell diagram (HRD) of the OB stars reveals a large age spread ($1-30$\,Myr), suggesting different episodes of star formation in the complex. The youngest stars are concentrated in the inner part of the complex, while the older OB stars are scattered over outer regions. We derived the present day mass function for the entire N\,206 complex as well as for the cluster NGC\,2018. The total ionizing photon flux produced by all massive stars in the N\,206 complex is $Q_{0} \approx 5 \times 10^{50}\, \rm{s^{-1}}$, and the mechanical luminosity of their stellar winds amounts to $ L_{\rm{mec}} = 1.7 \times 10^{38}\, \rm{erg \,s^{-1}}$. Three very massive Of stars are found to dominate  the feedback among 164 OB stars in the sample. The two WR winds alone release about as much mechanical luminosity as the whole OB star sample.
 The cumulative mechanical feedback from all massive stellar winds is comparable to the combined mechanical energy of the supernova explosions that likely occurred in the complex. Accounting also for the WR wind and supernovae, the mechanical input over the last five Myr is $\approx 2.3 \times 10^{52}\,\rm{erg}$.
}
%
%
{The N206 complex in the LMC has undergone star formation episodes since more than 30\,Myr ago. From the spectral analyses of its massive star population, we derive a current star formation rate of $2.2 \times 10^{-3} M_\odot\,\mathrm{yr}^{-1}$.  From the combined input of mechanical
energy from all stellar winds, only a minor fraction is emitted in the form of X-rays. The corresponding input accumulated over a long time also exceeds the current energy content of the complex by more than a factor of five. The morphology of the complex suggests a leakage of hot gas from the superbubble. 
}


\maketitle

\section{Introduction}
\label{sect:intro}

Understanding the feedback of stars on their environment is one of the key problems in star and galaxy formation. Massive stars are the main feedback agents, altering the surrounding environment on local, global, and cosmic scales. They dynamically shape the interstellar medium (ISM) around them on timescales of a few Myr. With their winds, ionizing radiation, and supernova explosions, ensembles of massive stars cause the largest structures of the ISM such as giant or multiple \ion{H}{ii} regions, superbubbles, and supergiant shells.

The interaction via winds and ionizing  radiation  of a single massive star with its surroundings is usually referred to as an interstellar bubble \citep{Weaver1977}. A massive star first forms an \ion{H}{ii} region around itself through its strong ionizing radiation field, and its stellar wind interacts with this ionized gas. The supersonic stellar wind (typically 2000 km/s) drives a shock into the ambient medium, while the reverse shock decelerates the wind material, and the kinetic energy of the shocked stellar wind becomes thermal energy, leading to  a hot ($T > 10^{6} $\,K), very low density bubble that emits soft X-rays. In a cluster, these hot bubbles around stars can interact with each other and form a superbubble \citep{Tenorio-Tagle1988,Oey2001,Chu2008}. Superbubbles are observed around many young massive star clusters with sizes  of  hundreds of parsecs \citep{Tenorio-Tagle1988,Sasaki2011}. 

 \begin{figure*}
 \centering
 \includegraphics[scale=0.5,trim={0cm 0 0 0}]{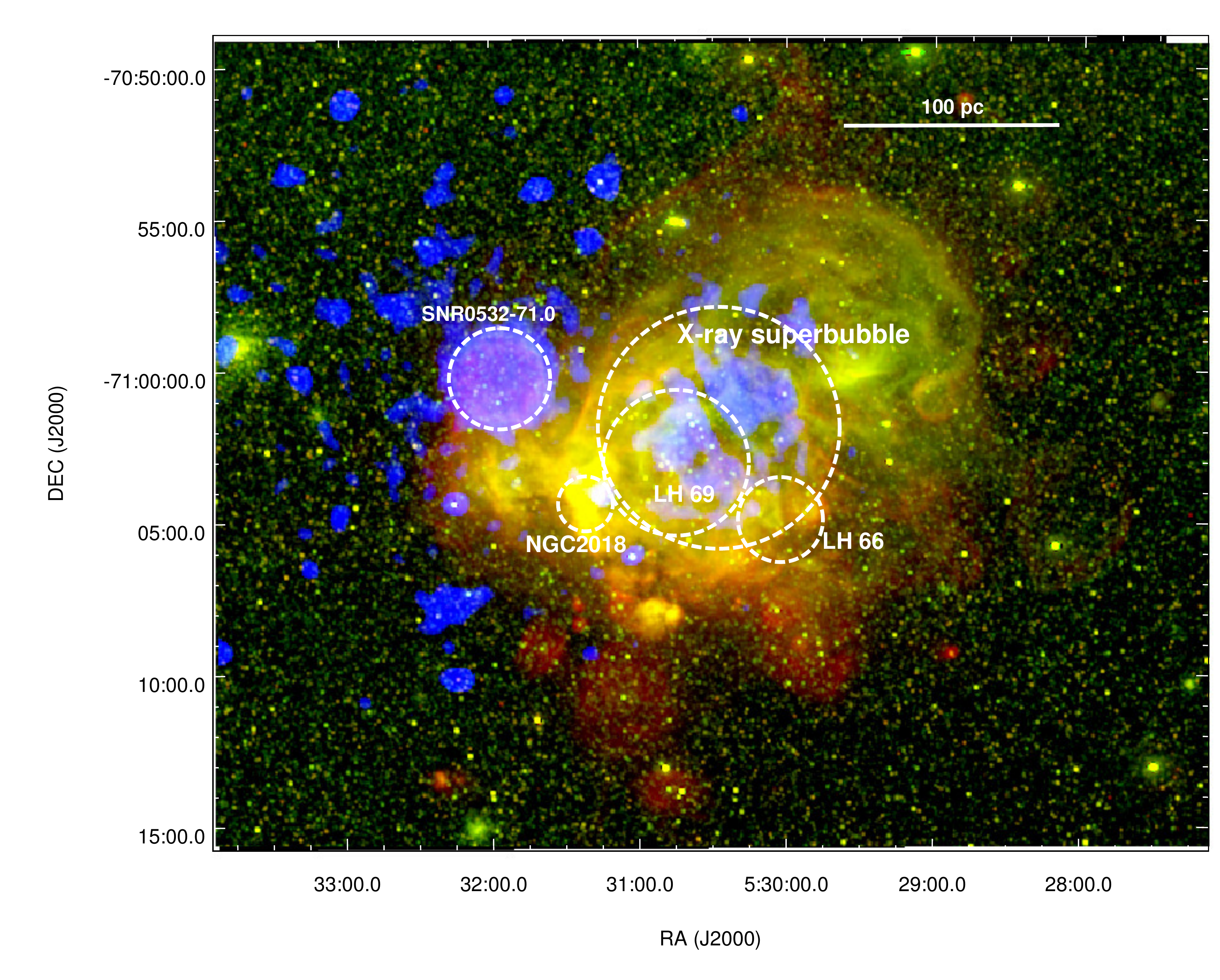}
 \caption{Location of the X-ray superbubble, SNR\,B0532-71.0, cluster NGC\,2018, and OB associations LH\,66 and LH\,69 in the N\,206 complex. The three-color composite image (H$\alpha$ (red) + [\ion{O}{iii}] (green) + X-ray (blue)) shown in the background is from the Magellanic Cloud Emission-Line Survey \citep[MCELS, ][]{Smith2005} and \textit{XMM-Newton} ($0.3-1$\,keV).
 } 
 \label{fig:bubble}
 \end{figure*}
In a series of papers, we study the superbubble associated with the giant \ion{H}{ii} region N\,206 in the Large Magellanic Cloud (LMC).
The LMC is one of the closest galaxies to the Milky Way with a distance modulus of only DM = 18.5\,mag \citep{Madore1998}, allowing for detailed spectroscopy of its bright stars. Another advantage of the LMC is the marginal foreground reddening along the line of sight \citep{Subramaniam2005,Haschke2011}.
The LMC is chemically less evolved than  the Milky Way with a metallicity around $Z=0.5\,Z_{\odot}$ \citep{Rolleston2002}. This allows us to study how stellar feedback works at sub-solar metallicity.

Our target of study is the high-mass star-forming complex N\,206 (alias: Henize\,206, LHA\,120-N\,206, and DEM L 221) in the outskirts of the LMC that surrounds the star forming cluster NGC\,2018 (LHA\,120-N\,206A). This \ion{H}{ii} region was first cataloged in an H$ \alpha $ objective prism survey by Karl Henize in 1956 \citep{Henize1956}.  The N\,206 complex also harbors a supernova remnant (SNR) B0532-71.0 \citep{Mathewson1973,Williams2005} in the eastern part. The N\,206 complex has been studied in multiwavelengths by various authors. \citet{Dunne2001} reported the expansion velocity of the H$ \alpha $ shell to be $ \sim $30\,km/s. Since the superbubble has expanded to a diameter of $\sim$112\,pc, this suggests an age of $ \sim $2\,Myr. More than one hundred  young stellar object (YSO) candidates were identified in this region using infrared data from \textit{Spitzer}, and  a star formation rate (SFR) of $0.25\times 10 ^{-2}\,M_{\odot}$\,yr$ ^{-1}$ has been estimated \citep{Carlson2012,Romita2010,Gorjian2004}.

 \begin{figure*}
 \centering
 \includegraphics[scale=0.52,trim={0cm 0cm 0 0cm}]{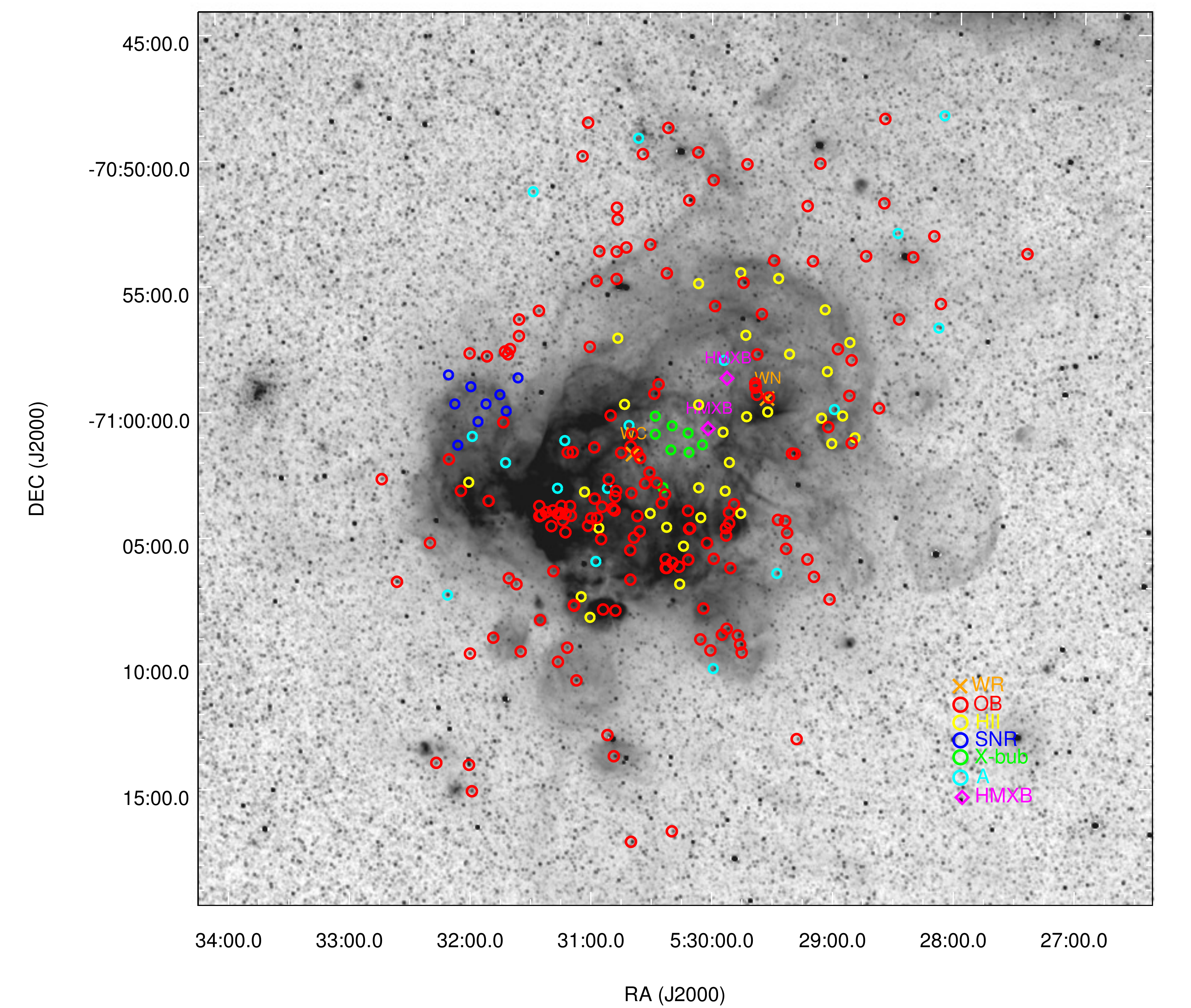}
 \caption{Position of all VLT-FLAMES MEDUSA fibers on the N\,206 complex, including the OB-type stars (red circle), WR (orange cross), \ion{H}{ii} region (yellow circle), SNR (blue circle), X-ray superbubble (green circle), A-type stars (cyan circle), and candidate HMXBs (magenta diamond). The background image shown is H$\alpha$ emission from the Magellanic Cloud Emission-Line Survey \citep[MCELS, ][]{Smith2005}.  
}
 \label{fig:n206_ob}
 \end{figure*}

The central part of the region is filled with hot gas and has been detected in X-rays. This X-ray superbubble is excited by the winds of the massive stars in the young cluster NGC\,2018  as well as the LH\,66 and LH\,69 OB associations \citep{Lucke1970} (see Sect.\,\ref{sect:structure} for more details of the structure of the region).
\citet{Dunne2001} concluded that current star formation is taking place around the X-ray superbubble.
 A detailed study of the complex in X-rays using the \textit{XMM-Newton} telescope was published by \citet{Kavanagh2012}. They compared the thermal energy stored in the X-ray emitting gas of the superbubble and the mechanical input supplied by the stellar population. Moreover, they found that the pressure of the hot gas drives the expansion of the shell into the surrounding \ion{H}{i} cloud. They estimated the overall mass and thermal energy content of the superbubble and blowout region as 741$ \pm $85\,M$ _{\odot} $ and (3.5$ \pm1.3) \times 10^{51} $\,erg from soft diffuse X-ray emission and H$ \alpha $ emission. However, this study was lacking information about the energetics of the complete massive star population in this region.

For a better understanding of a star-forming complex, we need to probe its stellar content in detail. In terms of brightness and youth, OB stars are excellent tracers of star formation. Their feedback effects can also lead to sequential star formation in these regions \citep{Gouliermis2008,Smith2010,Chen2007}. Therefore, the quantitative spectroscopy of OB star populations associated with superbubbles is essential to provide constraints on their fundamental parameters. By analyzing the massive star content of N\,206 using sophisticated model atmospheres, we can derive the stellar and wind parameters of the individual stars as well as their ionizing fluxes. This, in turn, will enable us to compare the energy budget of the cluster with the energy stored in the X-ray superbubble.


This is the second paper in the series that presents the spectroscopic analysis of the entire massive star population in the N\,206 superbubble. In the first paper of the series, we focused on the Of-type stars \citep[][hereafter Paper\,I]{Ramachandran2018}. Here we present the spectroscopic analysis of all remaining  massive stars in this region, and finally discuss the total stellar feedback  and the energy budget of the superbubble.

Section\,\ref{sect:structure} briefly describes the spatial structure of the complex.
The spectroscopic observations and spectral classifications are presented in 
Sections\,\ref{sect:spec} and \ref{sect:class}. In Sect.\,\ref{sect:analysis} and Sect.\,\ref{sec:wranalysis}, we quantitatively analyze the OB star spectra and the Wolf-Rayet (WR) binary spectra using Potsdam Wolf-Rayet (PoWR) atmosphere models. In Sect.\,\ref{sect:results}, the results are presented and discussed. The final Sect.\,\ref{sect:summary} provides a summary and general conclusions. The Appendices encompass tables with all stellar parameters and figures with the spectral fits of the analyzed stars.

\section{Spatial structure of  the N\,206 complex }
\label{sect:structure}
A color composite image of the N\,206 superbubble in H$\alpha$ (red), \ion{O}{iii} (green), and in $0.3-1$\,keV X-ray (blue) is shown in Fig.\,\ref{fig:bubble}. The H$\alpha$ emission basically traces the \ion{H}{ii} region in the complex. The entire structure is distributed over an approximately circular region with a diameter of $\approx$32$\arcmin$, corresponding to 465\,pc. The complex encompasses an X-ray superbubble, a supernova remnant (SNR), the young cluster NGC\,2018, and the OB associations LH\,66 and LH\,69.

The superbubble has been studied by many authors using Chandra and \textit{XMM-Newton} data. It has a  circular shell of diameter $\approx 8\arcmin$ (120\,pc). The brightest H$\alpha $ emission originates in the eastern and southern sides of the bubble \citep{Dunne2001}. According to \citet{Dunne2001}, the brightest X-ray emitting region of the superbubble is confined by an H$\alpha$ structure. A soft diffuse X-ray emission is detected across both the remnant and the superbubble. \cite{Kavanagh2012} noticed that no X-ray emission is detected from the major part of the superbubble region, and the hot X-ray emitting gas is distributed in a non-uniform manner.

The X-ray superbubble is surrounded by and overlaps with the young cluster NGC\,2018 and the OB associations LH\,66 and LH\,69. This region contains 41 OB stars including four Of-type stars (N206-FS\,111, 131, 162, and 178 analyzed in Paper\,I). Two WR binaries are present in the N\,206 complex. The WC binary is located nearly at the center and the WN binary near the edge of the X-ray superbubble. The OB associations LH\,66 and LH\,69 occupy an approximately circular region of 3.3$\arcmin$ and 4.8$\arcmin$ in diameter. They harbor 12 and 29 OB stars, respectively. The entire LH\,69 OB association is located within the boundaries of the X-ray superbubble region. The young cluster NGC\,2018 is very bright in H$\alpha$. The emission spans over a region of $\approx2\arcmin$ in diameter. This cluster encompasses 14 OB stars, mostly young O stars including three Of-type stars (N206-FS\,180, 187, and 193).

The 30\,pc$\times$30\,pc SNR\,B0532-71.0 is located on the eastern side of the nebular complex and has a faint circular H$\alpha$ shell \citep{Dunne2001}. \citet{Williams2005} have done a detailed study of this SNR in X-ray and radio. Their estimate of the thermal energy would yield an initial explosion energy of $8 \times 10^{50}$\,erg. They estimated the age of the SNR to be in the range of 17\,000-40\,000 years.

\section{Spectroscopy}
\label{sect:spec}

The presented study is largely based on spectroscopic data obtained with the Fiber Large Array Multi-Element Spectrograph (FLAMES) at ESO-VLT.
Accounting for a typical color excess of $E_{\rm B-V} \approx 0.1-0.2$\,mag, implying an extinction of $ A_{V} \approx 0.3-0.6 $\,mag, we targeted the blue hot stars (i.e.,\ with spectral subtypes earlier than B2V) by selecting all sources with $ B-V <0.20$\,mag and $V<16 $\,mag. Therefore, for spectral types later than B2V  our sample is incomplete. 
Apart from this, a few blue stars in the dense parts of the region were missed because the allocation of the Medusa fibers is constrained by the physical size of the fiber buttons. More details of the observations and the data reduction are given in Paper\,I. In total, 2918 spectra (including multiple exposures) were normalized, co-added, and cleaned for cosmic rays. The final spectra refer to 234 individual fiber positions as indicated in Fig.\,\ref{fig:n206_ob}. 
The sample of spectra consists of: 
\begin{itemize}
\item Nine Of stars (analyzed in Paper\,I)
\item 155 other OB-type stars (see Sect.\,\ref{sect:analysis})
\item 17 A-type stars (not considered in this work)
\item Two WR binaries WN+O and WC+O, see Sect.\,\ref{sec:wranalysis}
\item Two candidate high mass X-ray binary (HMXB) positions (see Sect.\,\ref{Subsubsec:SNfb})
\item 32 fiber positions in the \ion{H}{ii} region  (not considered in this work)
\item Eight fiber positions in the X-ray superbubble (not considered in this work)
\item Nine fiber positions in the SNR  (not considered in this work).
\end{itemize}

From the fibers that have been placed at the diffuse background, one can extract and analyze the nebular emission lines. This will be subject to a separate study.

All the objects observed in this survey are labeled by a prefix N206-FS (N\,206 FLAMES Survey) along with a number corresponding to the ascending order of their right ascension (1--234).
 We identified and analyzed all the OB-type stars among them. Their coordinates and spectral types are listed in Table 1. More details on the spectral classification schemes are given Sect.\,\ref{sect:class}.

Apart from the VLT optical spectra, eight of the OB stars in our sample have available UV spectra in the Mikulski Archive for Space Telescopes (MAST\footnote{ http://archive.stsci.edu/.}). The stars N206-FS\,58, 60, 65, 93, and 134 have available Hubble Space Telescope (HST) /STIS (Space Telescope Imaging Spectrograph) spectra. These were taken with the far ultra-violet Multi-Anode Microchannel Array (FUV-MAMA) detector (G140L grating), with 25$\times$25 arcsecond square field of view (FOV), operating in the near ultraviolet from 1140 to 1730\,\AA. 
 We retrieved an IUE (International Ultraviolet Explorer) short-wavelength spectrum in the wavelength range 1150--2000\,\AA\ from the archive for the star N206-FS\,147, which is taken with a large aperture ($21\arcsec \times 9\arcsec$). We also used available  far-UV FUSE (Far Ultraviolet Spectroscopic Explorer) spectra of the stars N206-FS\,90 and 119 in the wavelength range 905--1187\,\AA, taken with a medium aperture ($ 4\arcsec \times 20\arcsec$). It should be noted that the UV spectra from the archive are flux-calibrated, while our VLT-FLAMES spectra are not.

The WR binary N206-FS\,128 has available FUSE, IUE short, and IUE long (2000--3200\,\AA) spectra. Additionally, a FOcal Reducer/low dispersion Spectrograph 2 (FORS2) spectrum in the 4590--9290\,\AA\ wavelength range is available from the ESO archive.

In addition to the spectra, we used various photometric data from the VizieR 
archive to construct the spectral energy distribution (SED). UV and optical ($U, 
B, V, $ and $ I$) photometry was taken from \citet{Massey2002}, \citet{Zaritsky2004}, \citet{Girard2011}, and \citet{Zacharias2012}. The infrared 
magnitudes ($JHK_{s}$ and \textit{Spitzer}-$IRAC$) of the sources are taken from the 
catalogs \citet{Kato2007}, \citet{Bonanos2009}, and  \citet{Cutri2012}.


\begin{center}
\begin{table}
\caption{Coordinates and spectral types of OB stars in our sample} 
\label{table:obstars}     
\begin{tabular}{cccl}

\hline
\hline
\noalign{\vspace{1mm}}
 N206-FS & RA (J2000)   & DEC (J2000)  & Spectral type \\
 \#  & (\degr) & (\degr) &  \\
\noalign{\vspace{1mm}}
\hline 
\noalign{\vspace{1mm}}
1 	&  81.861792 	&  -70.896000    &	B1.5 V           \\
3 	&  82.036292 	&  -70.929639    &	B2.5  IV         \\
5 	&  82.050750 	&  -70.884806    &	B1 V             \\
6 	&  82.093167 	&  -70.898833    &	B2.5  IV         \\
7 	&  82.120750 	&  -70.940083    &	B0.5 V           \\
9 	&  82.151542 	&  -70.807222    &	B1.5  (III)e     \\
10 	&  82.152417 	&  -70.863194    &	B0.5 V           \\
11 	&  82.160250 	&  -70.999139    &	B2  IV           \\
12 	&  82.188458 	&  -70.898389    &	B2.5  IV         \\
14 	&  82.215750 	&  -71.022361    &	B2  (IV)e        \\
15 	&  82.216792 	&  -70.967389    &	B2.5  IV         \\
17 	&  82.221000 	&  -70.990972    &	B2  IV           \\
19 	&  82.244750 	&  -70.960083    &	B2.5  IV         \\
22 	&  82.259167 	&  -71.126028    &	B2.5 V           \\
23 	&  82.263500 	&  -71.011778    &	O8 V             \\
27 	&  82.282167 	&  -70.837083    &	O9.5 IV          \\
28 	&  82.291167 	&  -71.111111    &	B7 IV            \\
29 	&  82.296208 	&  -70.901861    &	B5 IV            \\
30 	&  82.304833 	&  -71.099583    &	B0.5 V           \\

\hline
\end{tabular}
\tablefoot{Continues in Table\,\ref{table:App_coord}.} 
\end{table}
\end{center}

\begin{figure*}
\centering
\includegraphics[scale=0.85,trim={0 0cm 3cm 11cm}]{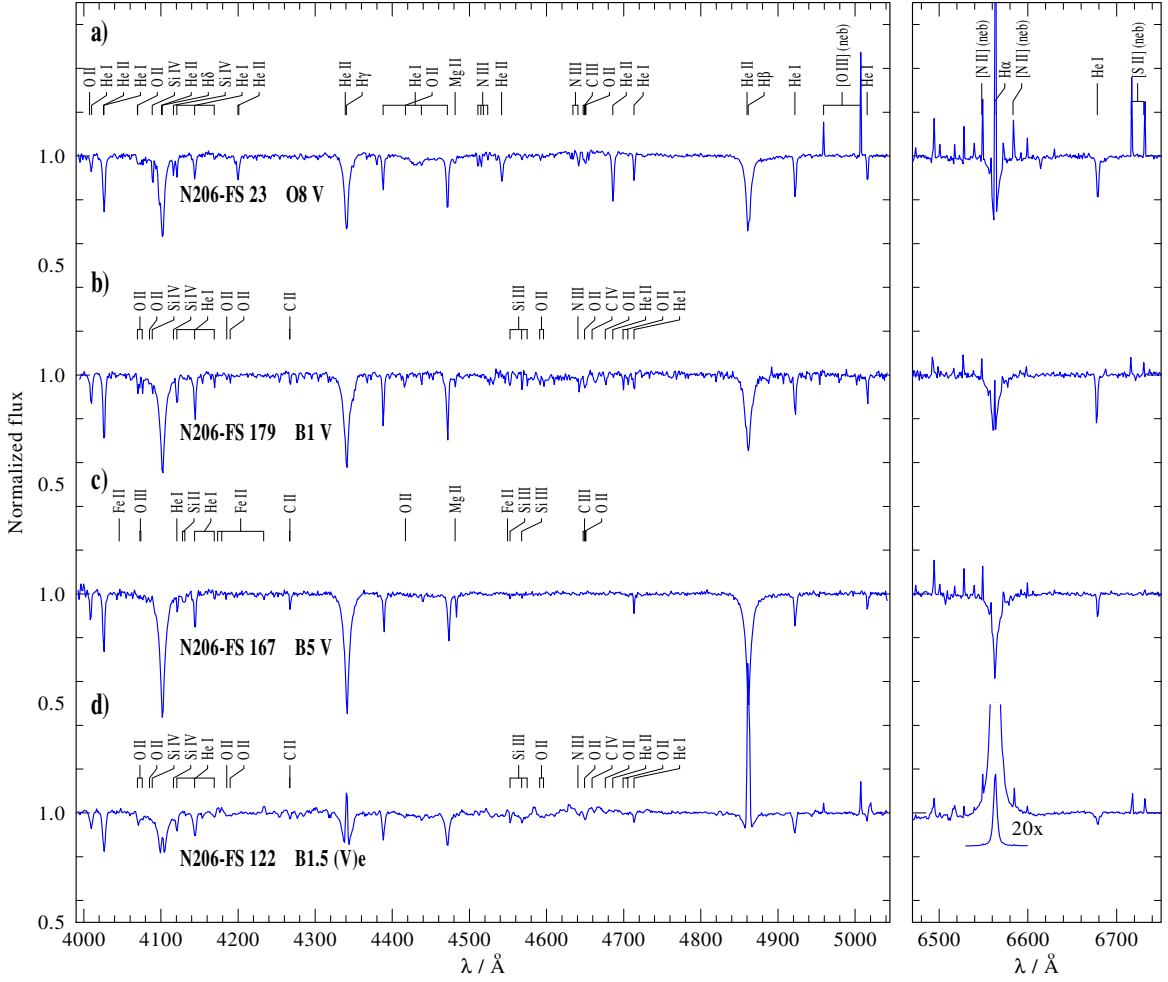}
\caption{Example spectra of O, early B, late B, and Be stars in the N\,206 superbubble. The [\ion{O}{iii}], [\ion{N}{ii}], and [\ion{S}{ii}] lines are from nebular emission 
(neb). }
\label{fig:OBspec}
\end{figure*}


\section{Spectral classification}
\label{sect:class}
The spectral classification of OB stars is primarily based on the blue optical wavelength range. We mainly follow the classification schemes proposed in \citet{Sota2011}, \citet{Sota2014}, and \citet{Walborn2014}. 


\subsection{O stars}
We identified 31 O stars in the whole sample (excluding the nine Of stars described in Paper\,I). 
The main diagnostic line ratios used for their subtype classification are \ion{He}{i\,$\lambda4388$}/\,\ion{He}{ii\,$\lambda4542$},  \ion{He}{i\,$\lambda4144$}/\,\ion{He}{ii\,$\lambda4200$} and \ion{He}{i\,$\lambda4713$}/\,\ion{He}{ii\,$\lambda4686$}.  An example of an O star spectrum is shown in Fig.\,\ref{fig:OBspec} (a).

The luminosity-class criteria of O stars are mainly based on the strength of \ion{He}{ii}\,4686. Two O supergiants are identified in the sample, N206-FS\,90 and N206-FS\,134. Absent or very weak \ion{He}{ii\,$\lambda4686$} absorption together with H$\alpha$ emission reveals their supergiant nature. The \ion{Si}{iv} absorption lines are very strong in these spectra. Compared to these objects, N206-FS\,62 and N206-FS\,192 have stronger \ion{He}{ii\,$\lambda4686$} absorption, and therefore are classified as giants. Additionally, these spectra are contaminated with disk emission, suggesting their Oe nature. Furthermore, two main sequence O stars are also identified with Oe features.

We found six stars that exhibit special characteristics typical for the Vz 
class, namely, N206-FS\,64, 107, 145, 154, 184, and 198.
According to \citet{Walborn2006}, these objects may be near or on the zero-age 
main sequence (ZAMS). As described in \citet{Walborn2014}, the main 
characteristic of the Vz class is the prominent \ion{He}{ii\,$\lambda4686$} absorption 
feature that is stronger than any other \ion{He}{} line in the blue-violet region. 

\subsection{Early B stars}
Stars with spectral subtype B0 to B2 are considered as early B stars in this paper. We identified 102 such stars in the whole sample. The main classification schemes for B stars are taken from \citet{Evans2015} and \citet{McEvoy2015}. The classification is based on the  ionization equilibrium of helium and silicon. The spectral subtypes are mainly determined from \ion{Si}{iii\,$\lambda$4553}/\,\ion{Si}{iv\,$\lambda$4089} line ratio and the strength of \ion{He}{ii\,$\lambda4686$}, \ion{He}{ii\,$\lambda4542$}, and \ion{Mg}{ii\,$\lambda4481$}. An example of an early B-type spectrum is shown in Fig.\,\ref{fig:OBspec} (b). Here the  \ion{Si}{iii}  lines are stronger and the \ion{He}{ii} lines are weaker than in the O star spectrum in Fig.\,\ref{fig:OBspec} (a).

For B stars, luminosity classes were mainly determined from the width of the Balmer lines and from the intensity of the silicon absorption lines (\ion{Si}{iv} and \ion{Si}{iii}). The B supergiants and bright giants show rich absorption-line spectra. For more details on the luminosity and subtype classification, we refer the reader to Table\,1 and Table\,2 of \citet{Evans2015}.

\subsection{Late B stars}
Stars with spectral subtypes ranging from B2.5 to B9 are considered as late B stars in this paper. We identified a total of 22  such objects. It should be noted that this sample is not complete because of its color and magnitude cut-offs (see Sect.\,\ref{sect:spec}). The main diagnostic lines for late B stars are  \ion{Si}{ii\,$\lambda\lambda$4128-4132}. We used the line ratios of \ion{Si}{ii\,$\lambda\lambda$4128-4132} to \ion{He}{i\,$\lambda4144$} and of \ion{Mg}{ii\,$\lambda4481$} to  \ion{He}{i\,$\lambda4471$} for classifying their spectral subtypes. An example of a late B-type star (B5\,V) is shown in Fig.\,\ref{fig:OBspec} (c).

\begin{figure}
\centering
\includegraphics[scale=0.48,trim={1cm 0 0 0cm}]{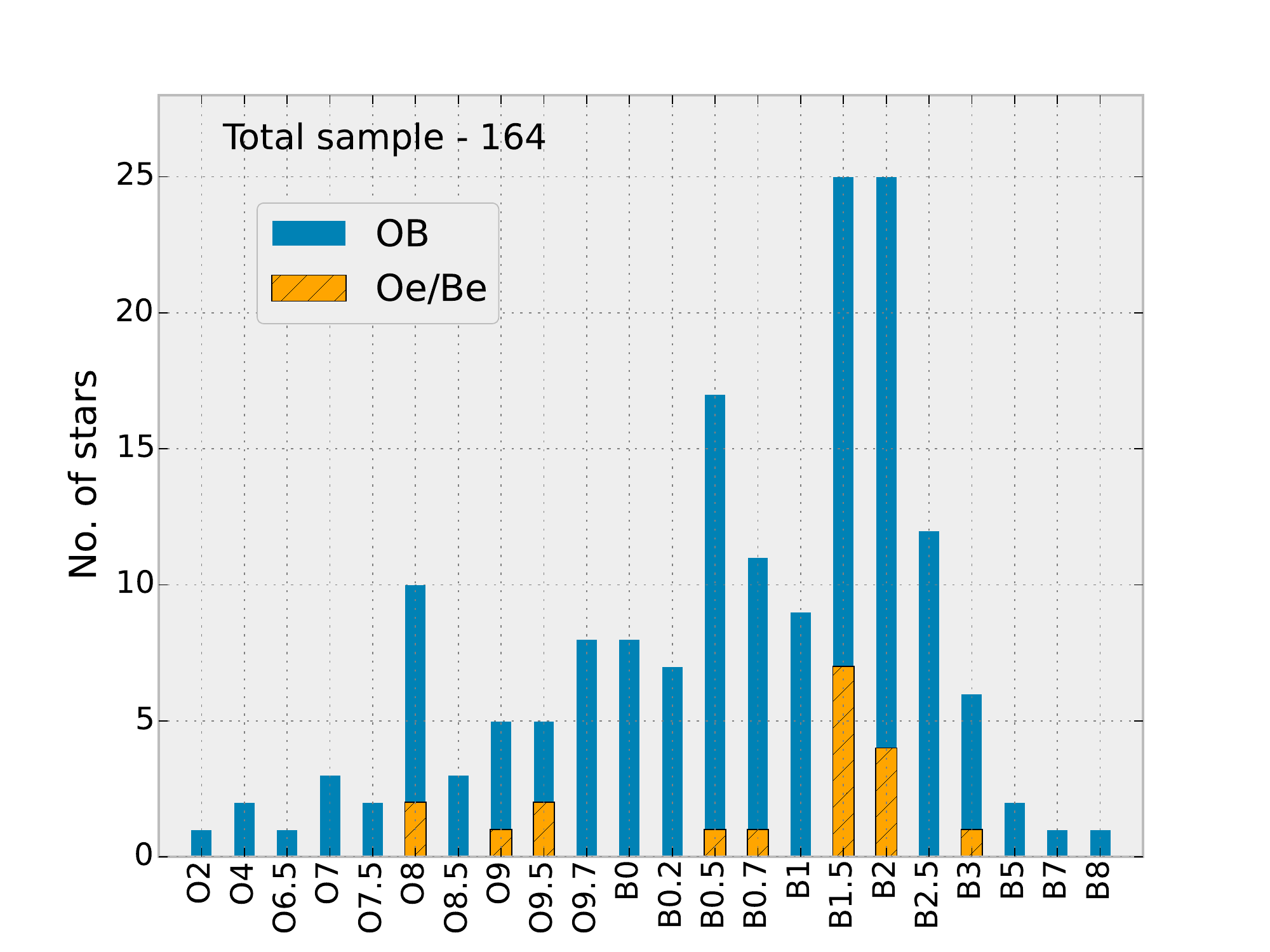}
\caption{Spectral subtype distribution of all OB stars in our sample. 
For each spectral subtype, the number of included Oe/Be star is represented in a different color.}
\label{fig:spec}
\end{figure}


\subsection{Oe/Be stars}
\label{subsec:bestars}
 The characteristic features of Oe/Be stars are the strong and broad emission of H$\alpha$ and other Balmer lines (see Fig.\,\ref{fig:OBspec}(d)), which are attributed to a circumstellar decretion disk \citep{Struve1931} fed by the ejected matter from the central star. We identified 19 such disk-originating  emission-line stars in the whole sample, where five of them are classified as Oe stars (including the Oef star in Paper\,I). Among this Oe/Be sample, three stars are giants or bright giants, while the rest are dwarfs. Most of the Be stars have spectral subtype B1.5--B2.

 Thus, the fraction of Oe/Be stars in this region is $\sim$12\%. Since the Be stars have a transient nature and the emission line profiles are usually variable, this fraction is just a lower limit. According to \citet{McSwain2008}, approximately 25 to 50\% of the Be stars may go undetected in a single epoch spectroscopic observation.

 The H$\alpha$ and H$\beta$ lines of these Be stars also show different profiles depending on the disk viewing angle \citep[see][for more details]{Rivinius2013}. 
 Seven stars (N206-FS\,9, 119, 121, 122, 181, 192, 233) show H$\alpha$ and H$\beta$ line profiles close to pole-on view.
 All the other twelve Oe/Be stars show line profiles that indicate higher inclinations because the H$\alpha$ and H$\beta$ emission lines are double-peaked. Four of these stars (N206-FS\,62, 113, 186, 192) additionally show characteristics of B[e] stars, defined by Balmer lines in emission plus forbidden emission lines \citep{Lamers1998}. 
 
\bigskip
 A histogram of the spectral types of all investigated stars in the superbubble N\,206 (including the Of stars from Paper\,I) is shown in Fig.\,\ref{fig:spec}. The number of stars gradually increases with spectral subtypes starting from O2, until B2. One exception is a local maximum at spectral subtype O8. The most populated spectral subtypes are B1.5 and B2. The sudden decrease in the number of stars towards later B subtypes is due to the incompleteness of the sample.
\section{Analysis of OB stars}
\label{sect:analysis}
We performed spectral analyses of all OB stars in our sample, determining their physical parameters. This was achieved by reproducing the observed spectra with synthetic spectra, calculated with the Potsdam Wolf-Rayet (PoWR) model atmosphere code.

 \begin{figure*}
\centering
\vspace{8mm}
\includegraphics[scale=0.8]{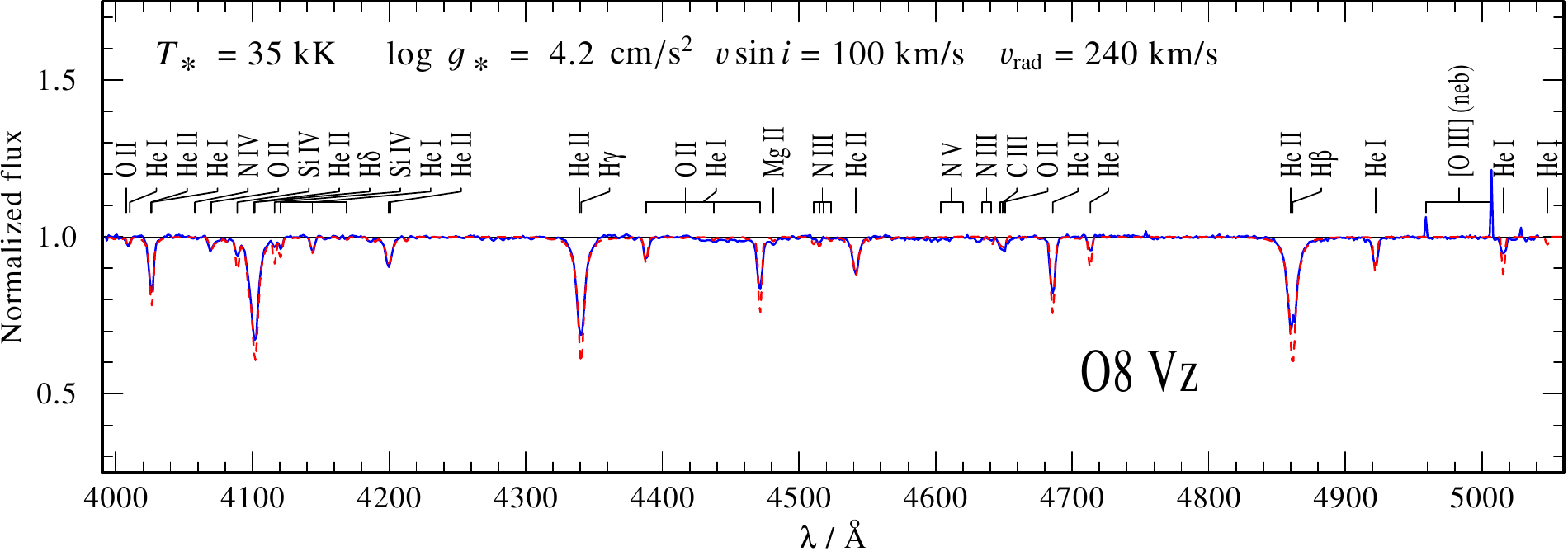}
\includegraphics[scale=0.8]{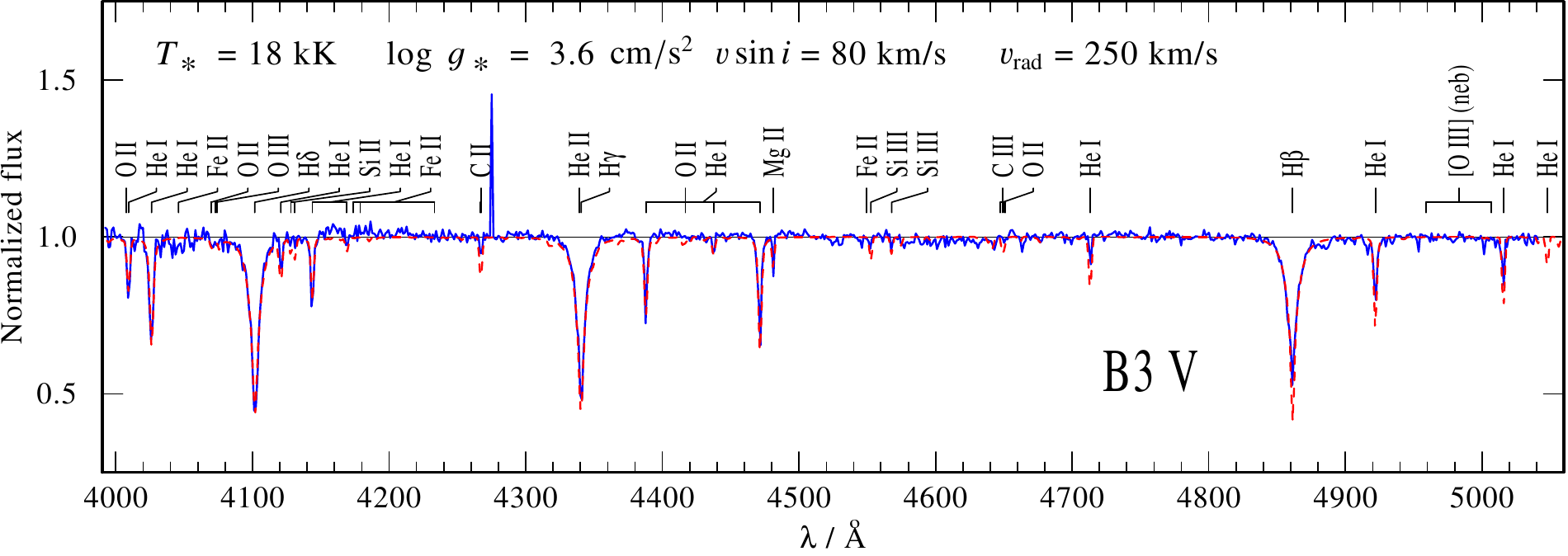}
\caption{Normalized VLT-FLAMES blue optical spectra (blue solid) for the O star N206-FS\,64 (upper panel) and the B star N206-FS\,221 (lower panel), respectively. Over-plotted are the synthetic PoWR model spectra (red dashed).  The stellar parameters of these best fitting models are given in each panel. The observed spectra are contaminated with nebular emission lines as indicated by ``(neb)''.}
\label{fig:spec_opt}
\end{figure*}
 
\subsection{The models}
\label{subsect:models}
The PoWR code for expanding stellar atmospheres is an advanced non-local thermodynamic equilibrium (non-LTE) code  that accounts for mass loss, line blanketing, and wind clumping. It can be employed for a wide range of hot stars at arbitrary metallicities \citep[e.g.][]{Hainich2014,Hainich2015,Oskinova2011,Shenar2015}, since the hydrostatic and wind regimes of the atmosphere are treated consistently \citep{Sander2015}. The non-LTE radiative transfer is calculated in the co-moving frame. Any model can be specified by its luminosity $L$, stellar temperature $T_\ast$, surface gravity $g_\ast$, and mass-loss rate $ \dot{M} $ as main parameters. 

In the subsonic region, the velocity field is defined such that a hydrostatic density stratification is approached \citep{Sander2015}. In the supersonic region, the wind velocity field $\varv(r)$ is pre-specified assuming the so-called $\beta$-law \citep{CAK1975} with $\beta=0.8$ \citep{Puls2008}. 
For establishing the non-LTE population numbers, the Doppler broadening $\varv_{\mathrm{Dop}}$ is set to 30 km s$ ^{-1} $ throughout the atmosphere. In the formal integration that yields the synthetic spectra, the Doppler broadening varies with depth and consists of the thermal motion and a ``microturbulence'' $\xi(r)$. We adopt  $\xi(r) = \rm max(\xi_{min},\, 0.1\,\varv(r))$ for OB star models with $\rm \xi_{min}= 20 $ km s$ ^{-1} $ \citep{Shenar2016}. For all the OB stars in our study, we account for wind clumping assuming that clumping starts at the sonic point, increases outwards, and reaches a density contrast of $D=10$ at a radius of $R_{\mathrm{D}} = \, 10\,R_\ast$. The models account for complex atomic data of H, He, C, N, O, Mg, Si, P, S, and Fe group elements. More details of the PoWR models used for the analysis of our program stars are described in Paper\,I.

\subsection{Spectral fitting}
\label{subsec:specfit} 
We performed the spectral analysis by iteratively fitting the observed spectra with PoWR models. In order to do that, we constructed OB-star grids with the stellar temperature $T_\ast$ and the surface gravity $\log\,g_\ast$ as the main parameters.
For the metallicity of the LMC, we adopted the values from \cite{Trundle2007}.
Since the model spectra of OB stars are mostly sensitive to  $T_\ast$, $\log\,g_\ast$, and $L$, these parameters are varied to find the best-fit model systematically. The other parameters, namely the chemical composition and the mass-loss rate, are kept constant within the model grids. The stellar mass $M$ and luminosity $L$ in the grid models are chosen according to the evolutionary tracks calculated by \citet{Brott2011}. The LMC OB star grid\footnote{www.astro.physik.uni-potsdam.de/PoWR.html}, which is also available online, spans from $T_\ast$\,=\,13\,kK to 54\,kK with a spacing of 1\,kK, and $\log\,g_\ast$\,=\,2.2 to 4.4 with a spacing of 0.2 dex. For individual stars, we also calculated models with adjusted C, N, O, and Si abundances when necessary. More details about deriving the stellar and wind parameters are described in the following subsections. 
 
 \subsubsection{Effective temperature }
 
We constrain the stellar temperature mainly from the silicon and helium ionization balance. 
In the temperature range from 20 to 30\,kK, the line ratio \ion{Si}{iii\,$\lambda$4553}/\,\ion{Si}{iv\,$\lambda$4089} decreases with an increase in temperature. In the case of hotter stars ($>$30\,kK), the temperature determination is mainly based on the helium line ratios \ion{He}{i\,$\lambda4388$}/\,\ion{He}{ii\,$\lambda4542$},  \ion{He}{i\,$\lambda4144$}/\,\ion{He}{ii\,$\lambda4200$} and \ion{He}{i\,$\lambda4713$}/\,\ion{He}{ii\,$\lambda4686$}. 

For the hottest O stars, where the \ion{He}{i} lines are absent or very weak,  the temperature diagnostic must rely on nitrogen line ratios (see Paper\,I). For late B subtypes (13-20\,kK), \ion{He}{ii} lines are absent. The main diagnostic line, in this case, is the multiplet \ion{Si}{ii\,$\lambda\lambda$4128-4132}, which increases as the stellar temperature goes down from 20\,kK to 10\,kK. Furthermore, the line ratio \ion{Mg}{ii\,$\lambda4481$} to \ion{He}{i\,$\lambda4471$} increases with decreasing temperatures and is employed for an accurate determination of the stellar temperature in this range.

As examples, the FLAMES spectra of an O star and a B star are shown in Fig.\,\ref{fig:spec_opt} together with the corresponding PoWR model. The upper panel shows the spectrum of a typical O star with strong \ion{He}{ii} lines. The \ion{Si}{iv} lines are weak, \ion{Si}{iii} lines are absent, and the \ion{He}{i}/\,\ion{He}{ii} ratio is small indicating a high stellar temperature. The observation is reproduced best by a model with $T_\ast$ = 35\,kK. The lower panel depicts a typical late B star spectrum fitted by a model with $T_\ast$= 18\,kK. The main indicators are the presence of  \ion{Si}{ii} absorption lines, strong \ion{Mg}{ii} lines, and weak \ion{Si}{iii} lines. Here \ion{He}{ii} is completely absent and the \ion{He}{i} lines are prominent (see Fig.\,\ref{fig:spec_opt}).

The fit procedure usually gives a clear preference for one specific grid model. Hence the uncertainty in the temperature determination is limited by  the grid resolution of $ \pm $1\,kK. However, the uncertainty in $\log\,g_\ast$ (see below) also propagates to the temperature and leads to a total uncertainty of about $\pm $2\,kK.

In the case of Oe/Be stars, the stellar temperature is not of the same precision, because \ion{He}{i}, \ion{Si}{iii}, and \ion{Si}{ii} lines can be partially filled with emission. Furthermore, the Balmer wings are affected by the disk emission, so that a larger uncertainty in $\log\,g_\ast$ additionally affects the temperature estimates. 
 
 \begin{figure*}
\centering
\vspace{8mm}
\includegraphics[scale=0.85]{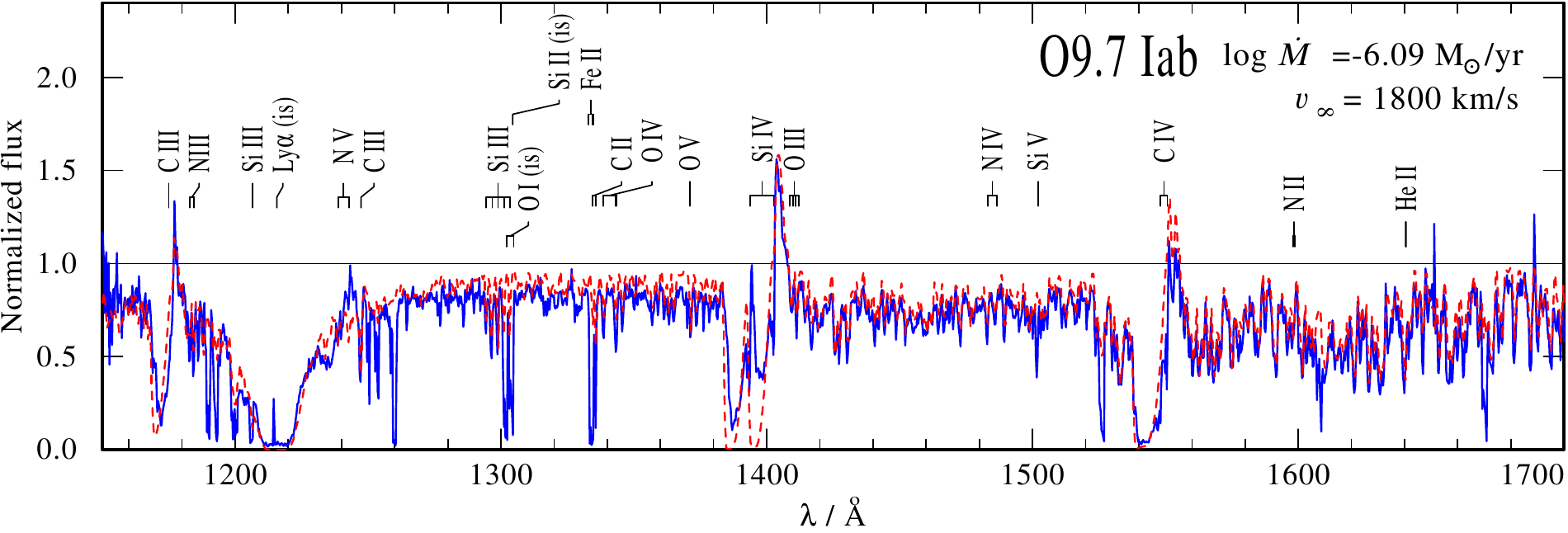}
\caption{Normalized UV (HST/STIS) spectrum (blue solid line) of an O supergiant (N206-FS\,134) is over-plotted with the PoWR model (red dashed line). The wind parameters of this best-fit model are given in the panel. The observed spectrum also contains interstellar absorption lines (is).}
\label{fig:spec_uv}
\end{figure*}
 
\begin{figure}[htpb]
\centering
\includegraphics[scale=0.8]{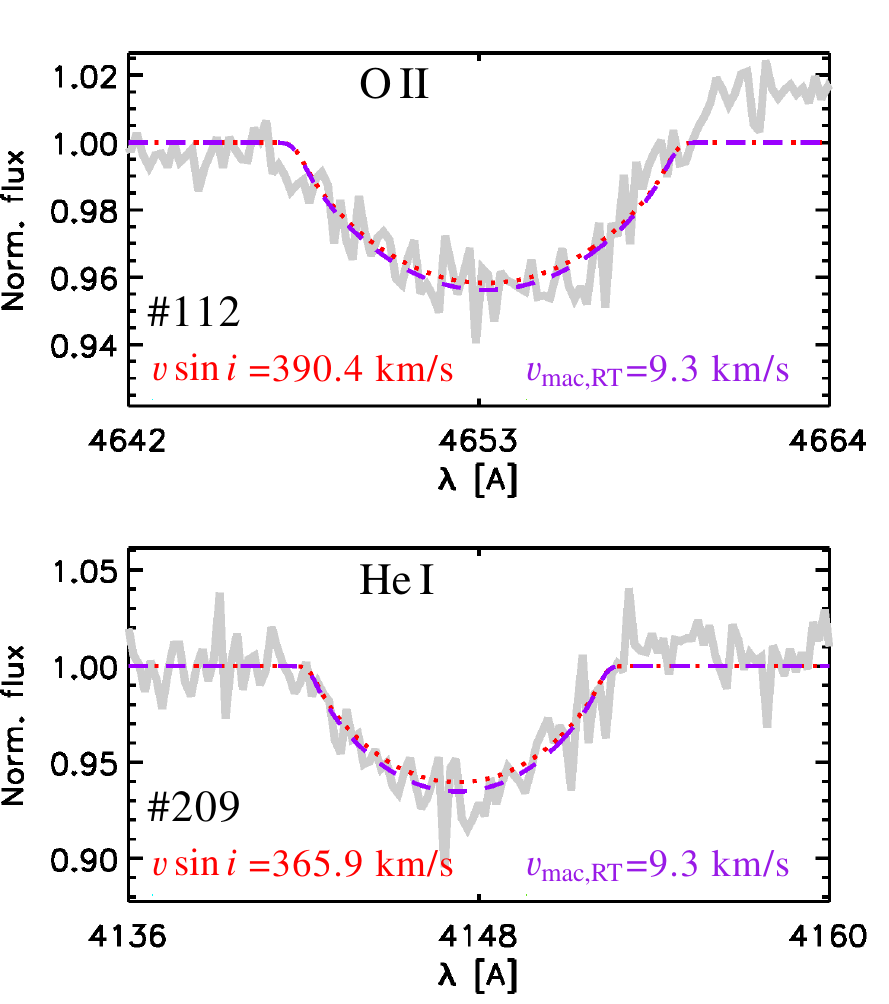}
\caption{Rotation velocity $\varv$\,sin\,$i$ from the line profile fitting 
of two fast rotating stars (N206-FS\,112 and 209) in the sample using the \texttt{iacob-broad} tool. The gray curves show the observed line profile. The $\varv$\,sin\,$i$ is calculated from line profile 
fitting based on the Fourier transform method (red dotted line) and goodness-of-fit 
analysis (violet dashed line). See text for more details.}
\label{fig:rotaion}
\end{figure}


  \subsubsection{Surface gravity}
  The Balmer lines are broadened by the Stark effect; we mainly use their wings to measure the surface gravity $\log\,g_\ast$. Since the H$\alpha$ line is often affected by wind emission, H$\gamma$ and H$\delta$ are better suited  for this purpose. The typical uncertainty for $\log\,g_\ast$ is $\pm$0.2\,dex. For example, the Balmer lines of the star shown in the upper panel of Fig.\,\ref{fig:spec_opt} are broad, and the best model fit gives $\log\,g_\ast$ = 4.2. The Balmer absorption lines in the observation are less deep than in the model because they are partially filled with nebular emission. Similarly, broad Balmer lines of the B-type star in the lower panel are fitted with a model for $\log\,g_\ast$= 3.6, since the stellar temperature is much lower for this object compared to an O star.
  
  For the Oe/Be stars in our sample, the  methods described above to measure stellar temperature and $\log\,g_\ast$ are not always successful because the equivalent widths of the Balmer lines may be reduced by disk emission.
  Therefore, the uncertainty in the surface gravity of these stars is relatively high.
  
\begin{figure*}
\centering
\includegraphics[scale=0.8]{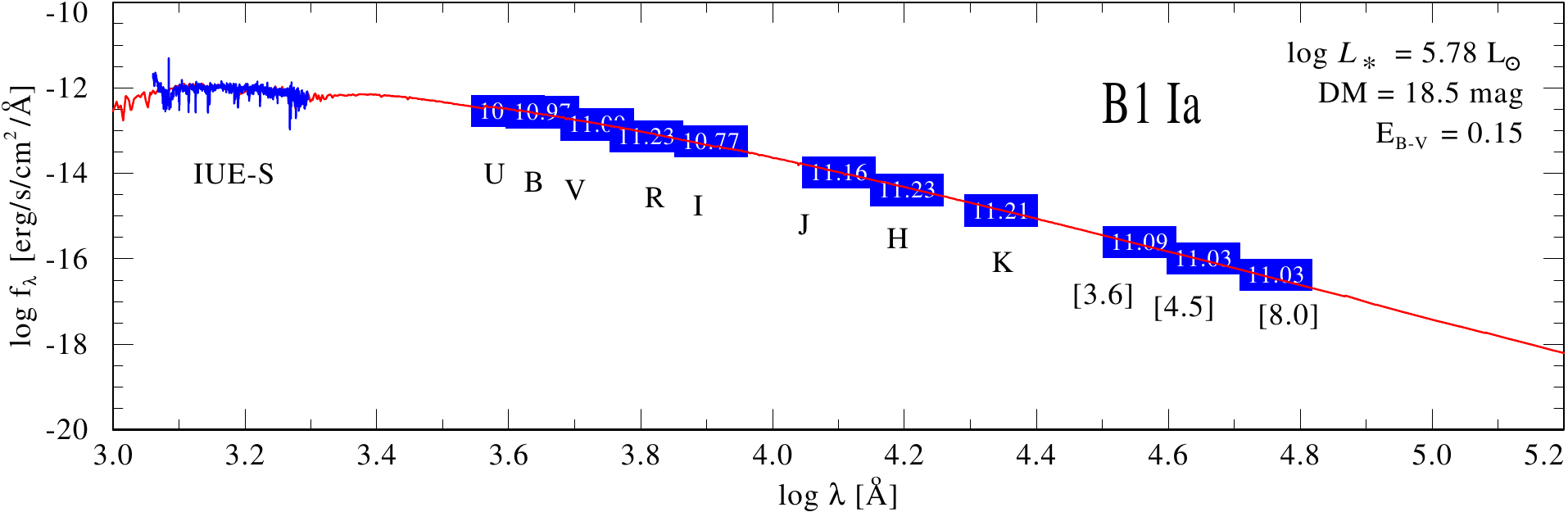}
\caption{Spectral energy distribution fit of a sample B star (N206-FS\,147). The model SED (red) is fitted to the available photometry from optical ($UBVRI$), near-infrared ($JHK_{s}$), and mid-infrared ($IRAC$ 3.6, 4.5, and 8.0 $\rm \mu m$) bands (blue boxes) as well as the flux calibrated UV spectra from the HST. The luminosity $L$ and the color excess $E_{\rm B-V} $ of the best-fitting model are given in the figure.}
\label{fig:sed}
\end{figure*}

   \subsubsection{Mass-loss rate}

  We calculated two model grids: one with a ``high'' mass-loss rate of $ 10^{-7} M_{\odot}\,\mathrm{yr}^{-1}$ and another one with a ``low'' value of $ 10^{-8} $ $ M_{\odot}\, \mathrm{yr}^{-1}$. 
The mass-loss rate can be estimated from the P-Cygni profiles of the UV resonance lines. For eight of our sample stars, UV spectra are available. The main diagnostic lines are the resonance doublets \ion{C}{iv\,$\lambda\lambda$1548--1551} and \ion{Si}{iv\,$\lambda\lambda$1393--1403} in the HST\,/ IUE range.
Useful wind lines in the FUSE range are \ion{P}{v\,$\lambda\lambda$1118--1128}, \ion{C}{iv\,$\lambda$1169}, and \ion{C}{iii\,$\lambda$1176}.

As an example, the UV spectrum of the O supergiant N206-FS\,134 is shown in Fig.\,\ref{fig:spec_uv}. The normalized high-resolution HST/STIS data are fitted with a PoWR model. This spectrum was consistently normalized  with the reddened model continuum. The mass-loss rate is estimated to be $\log \dot{M}$ = -6.1 [$ M_{\odot}\, \mathrm{yr}^{-1}$], on the basis of the shown UV fit and in accordance with H$\alpha$.
  
Without UV spectra, the mass-loss rate estimates can only rely on wind emission in H$\alpha$ and \ion{He}{ii\,$\lambda4686$}. 
In case we have UV observations or, at least, the H$\alpha$ line is in emission, we calculate additional models with adjusted mass-loss rates for the individual stars and precisely determine $\dot{M}$. The error in $\log \dot{M}$ is approximately $\pm$0.1\,dex if obtained with the help of UV spectra, and  $\pm$0.2\,dex if only based on H$\alpha$ and \ion{He}{ii$\lambda4686$}.

Most of the stars in our sample have neither H$\alpha$ in emission nor available UV spectra. In these cases, we adopt the ``high'' mass-loss rate of $ 10^{-7} M_{\odot}\,\mathrm{yr}^{-1}$ for the O stars and the``low'' $\dot{M}=10^{-8}  M_{\odot}\, \mathrm{yr}^{-1}$ for B stars and consider these values as upper limits.

  \subsubsection{Terminal velocity}
 
For stars with available UV spectra, we measured the terminal velocities ($\varv_\infty$) from the blue edge of the absorption trough of the P-Cygni profiles and recalculated the models accordingly.  The main diagnostic P-Cygni profiles are from the doublets \ion{C}{iv\,$\lambda\lambda$1548--1551}, \ion{P}{v\,$\lambda\lambda$1118--1128}, and \ion{S}{v\,$\lambda\lambda$1122--1134}. The typical uncertainties in $\varv_\infty$ from these measurements are $\pm$\,100 km\,s$^{-1}$. Figure\,\ref{fig:spec_uv} shows an example UV spectrum, where $\varv_\infty$ is measured from the best fit of the blue edge of \ion{C}{iv\,$\lambda\lambda$1548--1551} as $1800 \pm 50$ km\,s$^{-1}$. 
Since this line is saturated, it is not sensitive to clumping.

 For those stars for which we have only optical spectra, the terminal velocities are estimated theoretically from the escape velocity $\varv_{\mathrm{esc}}$. The ratio between terminal and escape velocity has been studied for Galactic massive stars both theoretically and observationally and found to be $\varv_\infty$/$\varv_{\mathrm{esc}}\simeq$\,2.6 for stars with $T_\ast\geq$\,21\,kK, while for stars with $T_\ast <$\,21\,kK the ratio is $\approx$1.3 \citep{Lamers1995,Kudritzki2000}. The terminal velocity is expected to depend slightly on metallicity, $\varv_\infty \, \propto (Z/Z_{\odot})^{q}$, where $q=0.13$ \citep{Leitherer1992}, and we adopted this scaling to account for the LMC metallicity.  

\subsubsection{Radial velocity}

We measured the radial velocity of all sample stars based on the absorption lines in the VLT-FLAMES spectra. These measurements are performed manually by fitting a number of line centers of the synthetic spectra to the observation. 
Generally, hydrogen lines are poor indicators for radial velocity measurements, because they are broad, sensitive to wind effects, and possibly affected by nebular emission \citep{Sana2013}. Metal lines, on the other hand, are very weak in O-star spectra. So, we can  use Si and Mg lines to measure radial velocity only in B- star spectra . Mainly, we use the \ion{He}{i} lines at 4026, 4387, 4471, and 4922 \AA\ and the \ion{He}{ii} lines at 4200 and 4541 \AA\ for measuring the radial velocities. Examples can be seen in Fig.\,\ref{fig:spec_opt}, where the radial velocity for the O star (upper panel) is estimated to be 240\,km\,s$^{-1}$ and for the B star to be 250\,km\,s$^{-1}$ (lower panel).

The typical uncertainty of $\varv_{\rm rad}$ is about $\pm$10 km\,s$^{-1}$. However, in a few cases, where stars have large rotational velocities or the spectra are noisy, the precise determination of the line centers is more difficult and the uncertainty can increase up to $\pm$20 km\,s$^{-1}$.

\subsubsection{Projected rotational velocity}
We constrain the projected rotation velocity $\varv\,\sin i$ of all OB stars from the line profile shapes, using the \texttt{iacob-broad} tool written in IDL \citep{Simon-diaz2014}. The $\varv\,\sin i$ measurements are mainly based on absorption lines in the blue optical wavelengths. We used both implemented methods, the combined Fourier transform (FT) and the goodness-of-fit (GOF) analysis, as described in Paper\,I.

 The primary lines selected for applying these methods are \ion{He}{i}, \ion{Si }{iv}  and \ion{Si }{iii}. The lines used for measuring $\varv\,\sin i$  of O-stars are \ion{He}{i} lines at 4026, 4387, 4471, and 4922\,\AA, and \ion{Si }{iv} lines at 4089 and 4116\,\AA. For B stars, we used \ion{He}{i} lines at 4144, 4387, 4471 and 4922\,\AA, the \ion{C}{ii} line at 4267\,\AA, the \ion{Mg}{ii} line at 4481\,\AA, \ion{Si}{iii} lines at 4552 to 4575\,\AA, the \ion{O}{ii} line at 4649\,\AA, and the \ion{N}{iii} line at 4641\,\AA.
 
 Figure\,\ref{fig:rotaion} shows rotationally broadened lines from two stars of our sample with the fastest rotation, N206-FS\,112 and N206-FS\,209 respectively. These stars are rotating close to their critical velocity (approximately $0.7\,\varv_{\rm critical}$). The typical uncertainty in $\varv\,\sin i$ is $ \sim$\,10\%. Finally, we adopt these measured velocities and convolve our model spectra to account for rotational and macroturbulent broadening. This leads to line profile fits consistent with the observations (e.g.,\ Fig\,\ref{fig:spec_opt}) in all cases.

 \subsubsection{Luminosity and reddening }
 
The luminosity $L$ and the color excess $E _{\rm B-V} $ of the individual OB stars were 
determined by fitting the model SED to the photometry. The model flux is diluted 
with the LMC distance modulus of 18.5\,mag \citep{Pietrzynski2013}. The uncertainty in the luminosity depends on the uncertainty in the color excess (which is typically small for LMC stars), the temperature, and the observed photometry. The total uncertainty is about 0.2 in $\log\,L$. If flux-calibrated UV spectra (HST, IUE, or FUSE) are available, this gives additional information on the SED, and the uncertainty in the luminosity is only $\approx$0.1 dex in these cases.

An example SED fit of a B supergiant is given in Fig.\,\ref{fig:sed}. The figure shows the theoretical SED fitted to multi-band photometry and flux calibrated UV spectra. We appropriately adjusted the reddening and the luminosity of the model's SED to the observed data. Reddening includes the contribution from the Galactic foreground 
($E_{\rm B-V} $ = 0.04\,mag) adopting the reddening law from \citet{Seaton1979}, and from the LMC using the reddening law described in \citet{Howarth1983} with $R_{V}=3.2$. The total $E_{\rm B-V} $ is a fitting parameter. 

\bigskip
Table\,\ref{table:lines} summarizes the main diagnostic lines used for the different aspects of the fitting process. This iterative procedure yields the stellar and wind parameters of the O, early B and late B stars in our sample. In the course of this procedure, individual models with refined stellar parameters and abundances are calculated for each of these stars. The fitting process is iterated until no further improvement of the fit is possible.

\begin{large}

\begin{table*}
\caption{Main diagnostics used in our spectral fitting process.} 
\label{table:lines}      
\centering
\setlength{\tabcolsep}{5pt}
\begin{tabular}{cccc}
\hline
\hline
\noalign{\vspace{1mm}}
Parameter  & O stars  & Early B stars & Late B stars  \\
\noalign{\vspace{1mm}}
\hline 
\noalign{\vspace{1mm}}
$T  _\ast$      &\ion{He}{i}/\ion{He}{ii} line ratios& \ion{Si}{iii}/\ion{Si}{iv} line ratios&\ion{Si}{ii}/\ion{Si}{iii}, \ion{Mg}{ii}/\ion{He}{i} line ratios\\
\noalign{\vspace{1mm}}
\hline 
\noalign{\vspace{1mm}}
$\log\,g_\ast$       &H$\gamma$ and H$\delta$ line wings  &H$\gamma$ and H$\delta$ line wings&H$\gamma$ and H$\delta$ line wings \\
\noalign{\vspace{1mm}}
\hline 
\noalign{\vspace{1mm}}
$\dot{M}$   &  H$\alpha$, \ion{He}{ii}\,4686, \ion{Si}{iv\,1393--1403}   

 &H$\alpha$, \ion{He}{ii}\,4686, \ion{Si}{iv\,1393--1403}
 &H$\alpha$, \ion{He}{ii}\,4686, \ion{Si}{iv\,1393--1403}
\\
 
 & \ion{P}{v\,1118--1128}, \ion{C}{iv} 1169, \ion{C}{iii} 1176 &\ion{C}{iv\,1548--1551}\\
 & \ion{C}{iv\,1548--1551}\\
 \noalign{\vspace{1mm}}
\hline 
\noalign{\vspace{1mm}}
$\varv_\infty$ &  H$\alpha$, \ion{N}{v\,1238--1242}, \ion{C}{iv\,1548--1551}& H$\alpha$, \ion{C}{iv\,1548--1551}&H$\alpha$, \ion{Si}{iv\,1393--1403}\\
 &  \ion{Si}{iv\,1393--1403}, \ion{P}{v\,1118--1128}& \ion{Si}{iv\,1393--1403}\\
 
 \noalign{\vspace{1mm}}
\hline 

\noalign{\vspace{1mm}}
$\varv\,\sin i$& \ion{He}{i} lines, \ion{N}{iii\,4510--4525}& \ion{He}{i} lines, \ion{N}{iii\,4641}   &\ion{He}{i} lines, \ion{C}{ii\,4267}\\
& \ion{Si}{iv\,4089--4116} & \ion{Si}{iii\,4552--4575}, \ion{O}{ii\,4649} &\ion{Mg}{ii\,4481} \\
\noalign{\vspace{1mm}}
\hline 
\noalign{\vspace{1mm}}
$\varv_{\rm rad}$& \ion{He}{i} and \ion{He}{ii} lines & \ion{He}{i} lines, \ion{Si}{iii\,4552--4575}&  \ion{He}{i} lines, \ion{C}{ii\,4267}, \ion{Mg}{ii\,4481}
\\
\noalign{\vspace{1mm}}
\hline 

\end{tabular}

\end{table*}
\end{large}


\begin{table*}
\caption{Stellar parameters of OB stars in N\,206} 
\label{table:stellarparameters}
\centering
\setlength{\tabcolsep}{3pt}
\begin{tabular}{llccccccccccccr}
\hline 
\hline
\noalign{\vspace{1mm}}
N206-FS & Spectral type &	$T _\ast$ 	& $\log\,L$ &	$\log\,g_\ast$ &	$\log\,\dot{M}$ &	$E_{\rm B-V} $&	$M_{\mathrm{V}}$&$R _\ast$  &	
 $\varv_\infty$ 	& $\varv$\,sin\,$i$& $\varv_{\rm rad}$&$M  _\ast$ &	$\log\,Q_{0}$ &$L_{\mathrm{mec}}$\\ 

\#& & [kK] & [$L _{\odot}$]&[cm s$ ^{-2} $]& [$M _{\odot}\,\mathrm{yr}^{-1} $] & 
[mag] &[mag]& [$R _{\odot}$] & [km\,s$^{-1}$]&[km\,s$^{-1}$]&[km\,s$^{-1}$]&[$M _{\odot} $]&[s$ ^{-1} $] 
&[$L _{\odot}$]\\

\noalign{\vspace{1mm}}
\hline 
\noalign{\vspace{1mm}}
1 	  &	B1.5 V              & 20.0 	& 4.23 	& 3.6 	& -7.78 	& 0.18 	& -3.96 	& 10.9  	& 800 	& 100 	& 280 	& 17 	& 45.7 	& 0.9 \\ 
3 	  &	B2.5  IV            & 18.0 	& 3.93 	& 3.6 	& -8.00 	& 0.12 	& -3.29 	& 9.5  	& 800 	& 130 	& 250 	& 13 	& 45.4 	& 0.5 \\ 
5 	  &	B1 V                & 25.0 	& 4.06 	& 4.2 	& -7.78 	& 0.12 	& -2.67 	& 5.7  	& 2500 	& 100 	& 270 	& 19 	& 46.3 	& 8.7 \\ 
6 	  &	B2.5  IV            & 20.0 	& 4.13 	& 3.6 	& -7.85 	& 0.12 	& -3.50 	& 9.7  	& 800 	& 180 	& 280 	& 14 	& 45.6 	& 0.7 \\ 
7 	  &	B0.5 V              & 29.0 	& 4.44 	& 4.2 	& -6.78 	& 0.08 	& -3.64 	& 6.6  	& 2500 	& 229 	& 240 	& 25 	& 47.2 	& 86.6 \\ 
9 	  &	B1.5  (III)e        & 20.0 	& 4.48 	& 3.2 	& -7.00 	& 0.25 	& -4.20 	& 14.5  	& 600 	& 130 	& 290 	& 12 	& 46.5 	& 3.0 \\ 
10 	  &	B0.5 V              & 27.0 	& 4.22 	& 4.2 	& -6.80 	& 0.05 	& -3.12 	& 5.9  	& 2400 	& 90 	& 250 	& 20 	& 46.6 	& 74.5 \\ 
11 	  &	B2  IV              & 21.0 	& 4.20 	& 3.8 	& -7.70 	& 0.10 	& -3.61 	& 9.5  	& 2000 	& 120 	& 280 	& 21 	& 45.8 	& 6.6 \\ 
12 	  &	B2.5  IV            & 20.0 	& 4.38 	& 3.6 	& -7.66 	& 0.24 	& -3.98 	& 12.9  	& 900 	& 80 	& 260 	& 24 	& 45.9 	& 1.5 \\ 
14 	  &	B2  (IV)e           & 20.0 	& 4.43 	& 3.6 	& -7.62 	& 0.18 	& -4.03 	& 13.7  	& 1000 	& 130 	& 260 	& 27 	& 46.0 	& 2.0 \\ 
15 	  &	B2.5  IV            & 19.0 	& 3.93 	& 3.8 	& -8.00 	& 0.13 	& -3.24 	& 8.5  	& 1000 	& 90 	& 250 	& 17 	& 45.3 	& 0.8 \\ 
17 	  &	B2  IV              & 20.0 	& 4.33 	& 3.6 	& -7.70 	& 0.23 	& -3.71 	& 12.2  	& 900 	& 160 	& 260 	& 22 	& 45.8 	& 1.3 \\ 
19 	  &	B2.5  IV            & 18.0 	& 4.03 	& 3.4 	& -7.93 	& 0.12 	& -3.42 	& 10.7  	& 700 	& 160 	& 280 	& 10 	& 45.4 	& 0.5 \\ 
22 	  &	B2.5 V              & 20.0 	& 4.13 	& 3.6 	& -7.85 	& 0.08 	& -3.29 	& 9.7  	& 800 	& 80 	& 240 	& 14 	& 45.6 	& 0.7 \\ 
23 	  &	O8 V                & 35.0 	& 4.86 	& 4.2 	& -6.65 	& 0.24 	& -3.79 	& 7.3  	& 2500 	& 60 	& 230 	& 31 	& 48.3 	& 115.5 \\ 
27 	  &	O9.5 IV             & 32.0 	& 4.76 	& 4.0 	& -6.93 	& 0.20 	& -4.11 	& 7.8  	& 1900 	& 21 	& 280 	& 22 	& 47.9 	& 35.4 \\ 
28 	  &	B7 IV               & 14.0 	& 3.86 	& 3.4 	& -8.11 	& 0.22 	& -3.47 	& 14.5  	& 800 	& 90 	& 240 	& 19 	& 44.9 	& 0.4 \\ 
29 	  &	B5 IV               & 17.0 	& 3.60 	& 3.6 	& -8.15 	& 0.12 	& -2.78 	& 7.3  	& 700 	& 120 	& 280 	& 8 	& 45.0 	& 0.3 \\ 
30 	  &	B0.5 V              & 27.0 	& 4.16 	& 4.2 	& -6.85 	& 0.08 	& -2.86 	& 5.5  	& 2300 	& 75 	& 250 	& 18 	& 46.6 	& 61.7 \\

\hline
\noalign{\vspace{1mm}}
\end{tabular}
\tablefoot{Continues in Table\,\ref{table:App_stellarparameters}.} 
\end{table*}

\section{Analysis of WR binaries}
\label{sec:wranalysis}
The two WR binaries present in this region are analyzed using PoWR models. The detailed analysis of the binary system N206-FS\,45 (BAT99\,49) with spectral type WN4:b+O  will be given in Shenar et al.\ (in prep.), and we are adopting the derived parameters in this paper. The second binary, N206-FS\,128 (BAT99\,53), is classified as WC4+O \citep{Bartzakos2001,Kavanagh2012}. We perform the analysis of this WC star using our LMC-WC grid models\footnote{http://www.astro.physik.uni-potsdam.de/$\sim$wrh/PoWR/LMC-WC/.}. Most of the descriptions of PoWR models given in Sect.\,\ref{subsect:models} are also applicable in the case of WR stars, with few special aspects as given below.

\subsection{The model}
In the case of WR models, the parameter $\log\,g_\ast$ is not as important as for OB star models, since the spectral lines originate primarily in the wind. The outer boundary is taken to be $R_{\rm max}$=1000 $R_{*}$ for WR models. Accounting for the very strong microturbulence in WR winds, the Doppler velocity is set to 100\,km/s \citep[e.g.][]{Hamann2006}. For the velocity law exponent (see Equation\,2 in Paper\,I), we adopt $\beta$ = 1 as usual for WR winds.

 For the chemical composition of the WC star, we assume mass fractions of 55\% helium, 40\% carbon, 5\% oxygen, 0.1\% neon, and 0.07\% iron-group elements. For the published WC grid, the terminal wind velocity was set to 2000 km/s. For the WC star in N206-FS\,128, we calculated models with higher terminal velocities to fit the spectrum.


\begin{table*}
\caption{Stellar parameters of WR binaries in N\,206} 
\label{table:stellarparametersWR}
\centering
\setlength{\tabcolsep}{3pt}
\begin{tabular}{llccccccccccccr}
\hline 
\hline
\noalign{\vspace{1mm}}
N206-FS  & Spectral type &	$T _\ast$ 	& $\log\,L$ &	$\log\,g_\ast$ &	$\log\,\dot{M}$ &	$E_{\rm B-V} $&	$D$&$R _\ast$  &	
$\varv_\infty$ 	& $\varv$\,sin\,$i$& $\varv_{\rm rad}$&$M  _\ast$ &	$\log\,Q_{0}$ &$L_{\mathrm{mec}}$\\ 

\#& & [kK] & [$L _{\odot}$]&[cm s$ ^{-2} $]& [$M _{\odot}\,\mathrm{yr}^{-1} $] & 
[mag] & & [$R _{\odot}$] & [km\,s$^{-1}$]&[km\,s$^{-1}$]&[km\,s$^{-1}$]&[$M _{\odot} $]&[s$ ^{-1} $] 
&[$L _{\odot}$]\\

\noalign{\vspace{1mm}}
\hline 
\noalign{\vspace{1mm}}
\multirow{ 2}{=10mm}{45\tablefootmark{(1)}}& WN4:b &100 &5.4 &5 &-5.4& 0.1&10 & 1.6& 1700& - &250  & 18 &49.3 &955\\ 
                    &O8 III& 33& 5.2& 3.6& -7.0&0.1&10 &12.2& 2100&280  & 250& 27&48.8 &37\\
\\

\multirow{ 2}{=10mm}{128}& WC4 &158 & 5.35 &6.0&-4.67&0.1& 40 & 0.6& 3400 & -&220 & 13&  49.1&20400\\
                     &O9 V& 33 &5.22 &3.8 &-7.0 &0.1& 10&11.3 & 2400&200  &220 & 29&48.5&27\\

\hline
\noalign{\vspace{1mm}}
\end{tabular}
\tablefoot{
\tablefoottext{1}{Parameters are taken from Shenar et al.\ (in prep.)}}
\end{table*}


\begin{figure*}
\centering
\includegraphics[scale=0.66]{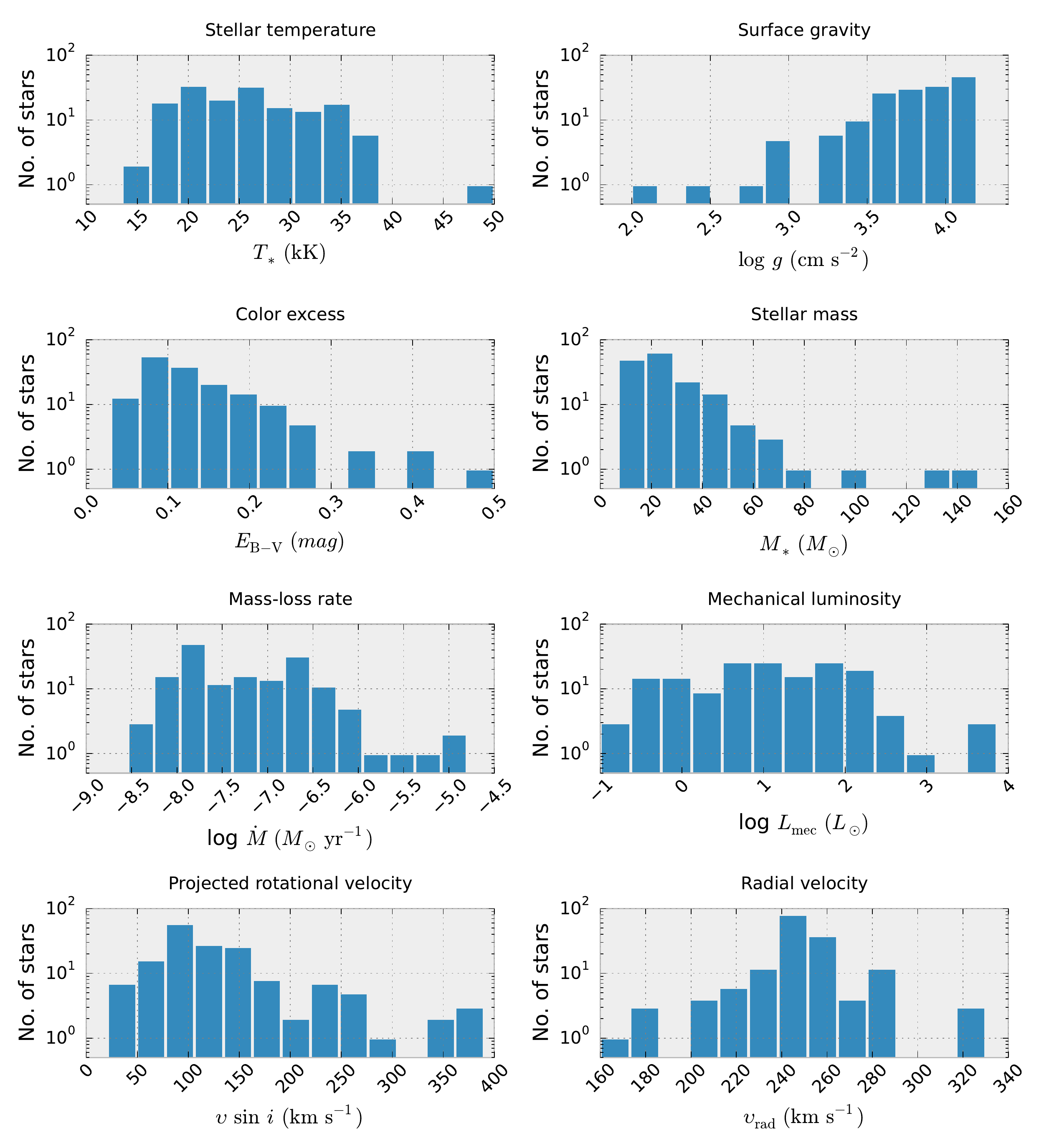}
\caption{Histograms of stellar temperature, surface gravity, color excess, projected rotational velocity, radial velocity, stellar mass, mass-loss rate, and mechanical luminosity of OB stars in N\,206 superbubble. The number of stars in the y-axis are shown in logarithmic scale. We used the square root (of data size) estimator method to calculate the optimal bin width and consequently the number of bins. }
\label{fig:hist}
\end{figure*}


\subsection{Spectral fitting}

The observed spectrum is a sum of a WC spectrum and an O star spectrum (see Fig.\,\ref{fig:N206FS128_1} and \ref{fig:N206FS128_2}). In order to reproduce this, we add the fluxes of a WC and an O star model. The light ratio is mainly constrained by the strength of the \ion{He}{i} absorption features which are clearly associated with the O star component (secondary), while the broad emission features are from the WC star (primary). For fitting the light ratio, we separately adjust the luminosities of both stars, under the constraint that the sum of the fluxes, after reddening, fits the observed SED (see Fig.\,\ref{fig:N206FS128_1}). 

The main diagnostic of the temperature of the WC star is the \ion{C}{iii} to \ion{C}{iv} line ratio. The \ion{C}{iv\,$\lambda$5808} emission is stronger than the \ion{C}{iii\,$\lambda$4648} emission. The strong \ion{O}{v} line emission also indicates a very high stellar temperature ($>$100\,kK). The best-fit model has a temperature of 158\,kK.
The terminal velocity is determined to be $\sim 3400$\,km\,$\rm s^{-1}$, based on the width of the prominent \ion{C}{iii} and \ion{C}{iv} features in the UV and optical spectra. The clumping contrast $D$ must be chosen as high as 40. With the standard value of $D=10$ \citep{Sander2012}, the model predicts much stronger electron scattering wings of the blend complex at $\lambda \approx$4686\,\AA\ and the \ion{C}{iv} line at 7721\,\AA.
Most of the  emission lines in the spectra are reproduced by a model with a mass-loss rate of $2 \times 10^{-5}\,M _{\odot}\,\mathrm{yr}^{-1} $. 
The effective temperature of the O star, $T_\ast=33$\,kK, is derived from the ratio between the \ion{He}{i} and \ion{He}{ii} absorption lines.

For the WC component, our final model has a luminosity of $\log\,L=5.35\,L_{\odot}$, which is relatively low compared to that of single WC stars in the LMC \citep{Crowther2002}. However, stars of the same spectral type in the Milky Way show such low luminosities \citep{Sander2012}. The O star component has a luminosity of $\log\,L=5.22\,L_{\odot}$, which corresponds to luminosity class V.

\section{Results and discussions}
\label{sect:results}

\subsection{Stellar parameters}
\label{sect:sparameters}

 The fundamental parameters for the individual stars are given in Table\,\ref{table:stellarparameters}. The rate of hydrogen ionizing photons ($\log\,Q_{0}$) and the rate at which the kinetic energy is carried away by the stellar winds (mechanical luminosity $L_{\mathrm{mec}}\,=\,0.5\,\dot{M}\,\varv_\infty^{2}$) are also tabulated. The derived parameters of the WR binaries are compiled in Table\,\ref{table:stellarparametersWR}.

\begin{figure}
\centering
\includegraphics[scale=0.7,trim={8cm 16cm 4cm 1cm}]{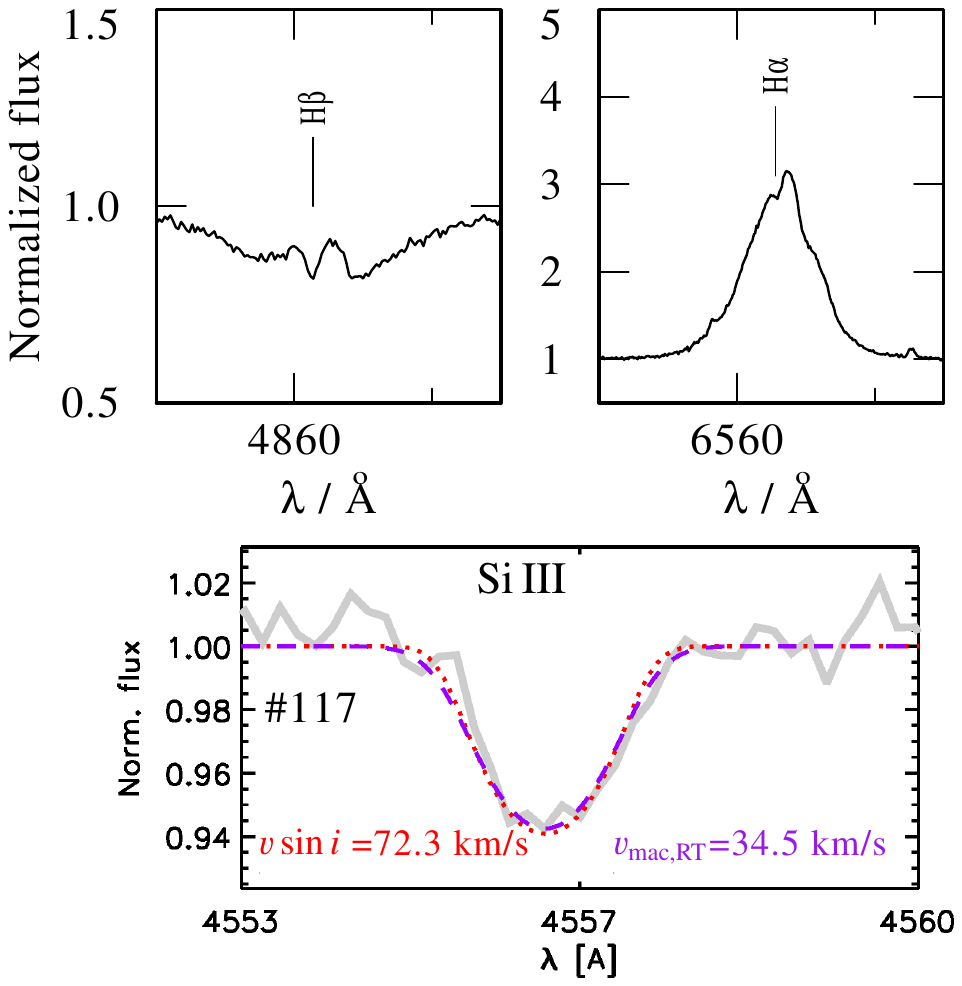}
\caption{Upper panel shows the H$\beta$ and H$\alpha$ line profiles of one of the slow rotating Be stars, N206-FS\,117. The lower panel shows $\varv$\,sin\,$i$ from the profile fitting of the \ion{Si}{iii} absorption line using the \texttt{iacob-broad} tool. The gray curves show the observed line profile. The $\varv$\,sin\,$i$ is calculated from line profile fitting based on the Fourier transform method (red dotted line) and goodness-of-fit analysis (violet dashed line).}
\label{fig:beslow}
\end{figure}

For the whole sample, we plot histograms for the distributions of stellar temperature, surface gravity, color excess, projected rotational velocity, radial velocity, stellar mass, mass-loss rate, and mechanical luminosity (Fig.\,\ref{fig:hist}). The stellar temperature of the OB stars (including the Of stars) ranges from 14 to 50\,kK. 
 In the surface gravity histogram (Fig.\,\ref{fig:hist}), most of the stars are found at surface gravities $\log\,g_\ast$ between 4.0 and 4.2. Only 15\% of the stars have log\,g $\leq$ 3.4\,cm\,s$^{-2}$, indicating giants or supergiants. The color excess histogram reveals that most of the stars in the N\,206 superbubble have a very low color excess of $E_{\rm B-V} \approx 0.1$. Stars that belong to the young cluster NGC\,2018 are found to have comparatively higher extinction, with five cluster members showing $E_{\rm B-V} > 0.25$. These young stars in the cluster are  still surrounded by the reminder of their parental cloud. This is consistent with the far infrared Herschel images and CO intensity maps that reveal a distribution of cold dense gas around this cluster (see Fig.\,\ref{fig:CO}).

The histogram of the stellar masses (Fig.\,\ref{fig:hist}) refers to spectroscopic masses, calculated from $\log\,g_\ast$ and $R_\ast$ ($g_\ast=G\,M_\ast\,R_\ast^{-2}$). These masses vary in the range of $7-150\,M _{\odot} $. The number of objects decreases with increasing mass. We note that the lowest mass bin is not complete.

From the mass-loss rate histogram, we can see that the OB stars in the sample have log\,$\dot{M}$ [$M_{\odot} \rm{yr^{-1}}$] in the range of $-8.5$ to $-4.8$.
Stars with the highest mass-loss rates are either super-giants or bright giants. The statistics of the mechanical luminosity,  $L_{\rm mec} = 0.5 \dot{M} \varv_\infty^{2}$, are also plotted as a histogram in Fig.\,\ref{fig:hist}. This distribution suggests that most of the OB stars in our sample  release $L_{\rm mec} <20\,L_{\odot}$ to the surrounding ISM. This is $10-100$ times lower than the mechanical luminosities of any of the Of stars analyzed in Paper\,I. The mechanical luminosities of the WR stars are even larger (see below).

\begin{figure}
\centering
\includegraphics[scale=0.52]{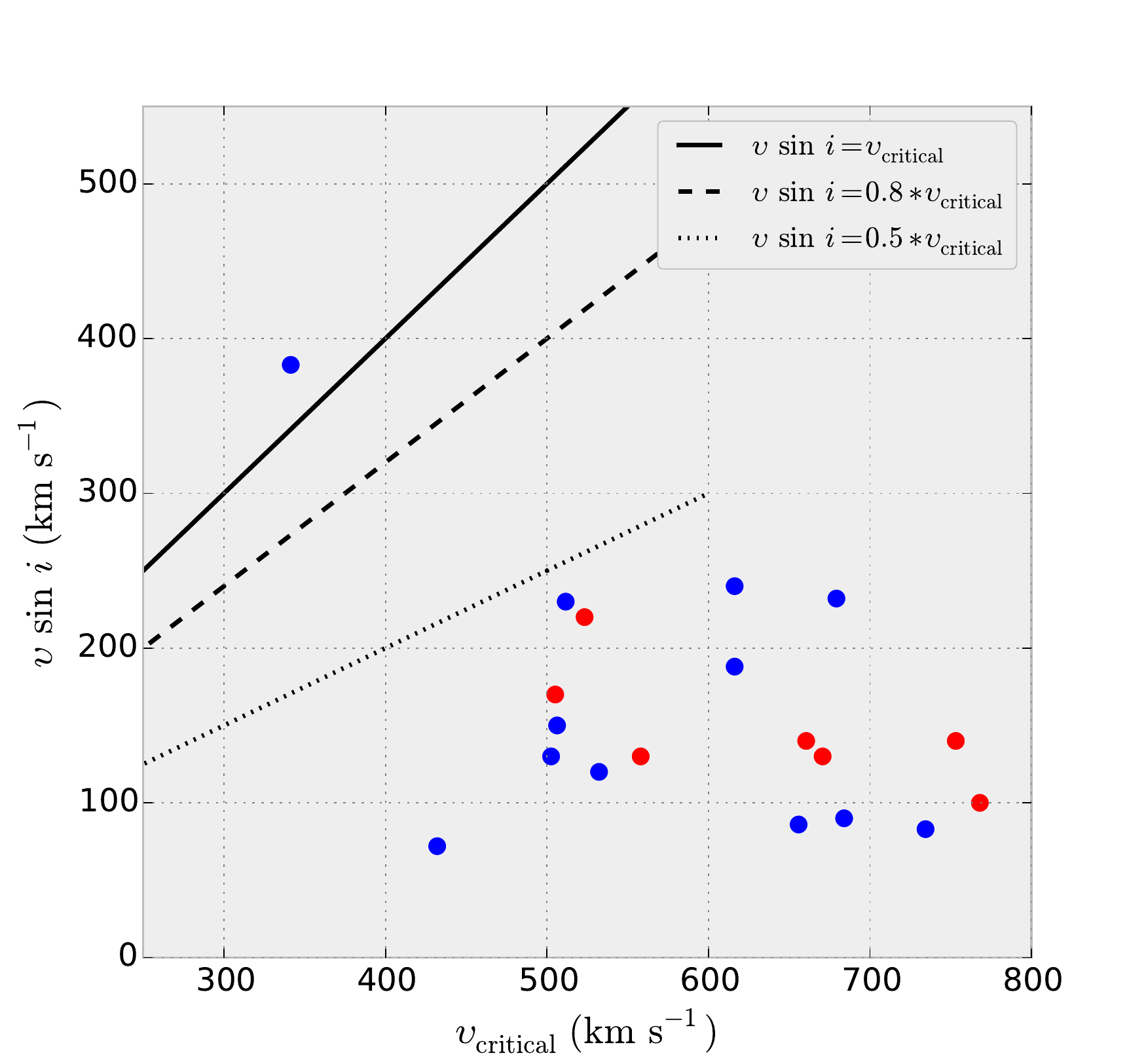}
\caption{Projected rotational velocity ($\varv\sin i$) versus critical velocity ($\varv_{\rm critical}$) of Oe/Be stars. The red dots represent stars with a pole-on view as deduced from the H$\beta$ and H$\alpha$ line profiles, while the remaining stars are represented by blue dots. The solid, dashed, and dotted lines represent 100\%, 80\%, and 50\% of $\varv_{\rm critical}$, respectively.}
\label{fig:be_rot}
\end{figure}


\subsubsection{Stellar rotation}
The distribution of projected rotational velocities ($\varv\sin i$) covers 30 to 400\,km\,s$^{-1}$ (Fig.\,\ref{fig:hist}, bottom). Most of the stars have a $\varv\sin i$ of around $\sim$100\,km\,s$^{-1}$. The presence of a low-velocity peak and a high-velocity tail is consistent with studies of other massive star forming regions \citep{Ramirez-Agudelo2013,Penny1996}.

Fifteen OB stars of the sample  are rotating faster than 200\,km\,s$^{-1}$, and five of them (3\%) exhibit very fast rotation with $\varv\sin i$ in the range 340-400\,km\,s$^{-1}$. Interestingly, only the Oe star, N206-FS\,62, is rotating very fast. All other 18 Oe/Be stars in our sample show only a moderately enhanced rotation with an average $\varv\sin i$ of about 160\,km\,$\rm s^{-1}$, while the average for the other B stars is $\approx 120$\,km\,$\rm s^{-1}$. As an example, the line profiles and projected rotational velocity measurements of one of the sample Be stars with slow rotation are shown in Fig.\,\ref{fig:beslow}.

For the Oe/Be stars in our sample,  we compared $\varv\sin i$ with the critical rotational velocity 
\(
\label{eq:crit}
\varv_{\rm critical} = (2GM_\ast/(3R_\ast))^{0.5}
\) (see Fig.\,\ref{fig:be_rot}). 
  Interestingly, most of the Oe/Be stars rotate significantly below their critical velocity with $\varv\sin i$ $\sim (0.1 \ldots 0.5)\, \varv_{\rm critical}$. Here $\varv_{\rm critical} $ is calculated from the spectroscopic masses.  For a comparison, we also calculated the  critical velocities using evolutionary masses, which are generally lower (see Sect.\,\ref{sect:hrd}). Even in this case, 75\% of the Oe/Be stars are rotating with  $\varv\sin i\,< 0.5\,\varv_{\rm critical}$. 
   
\begin{figure}
\centering
\includegraphics[scale=0.34,trim={9cm 1cm 10.8cm 1cm}]{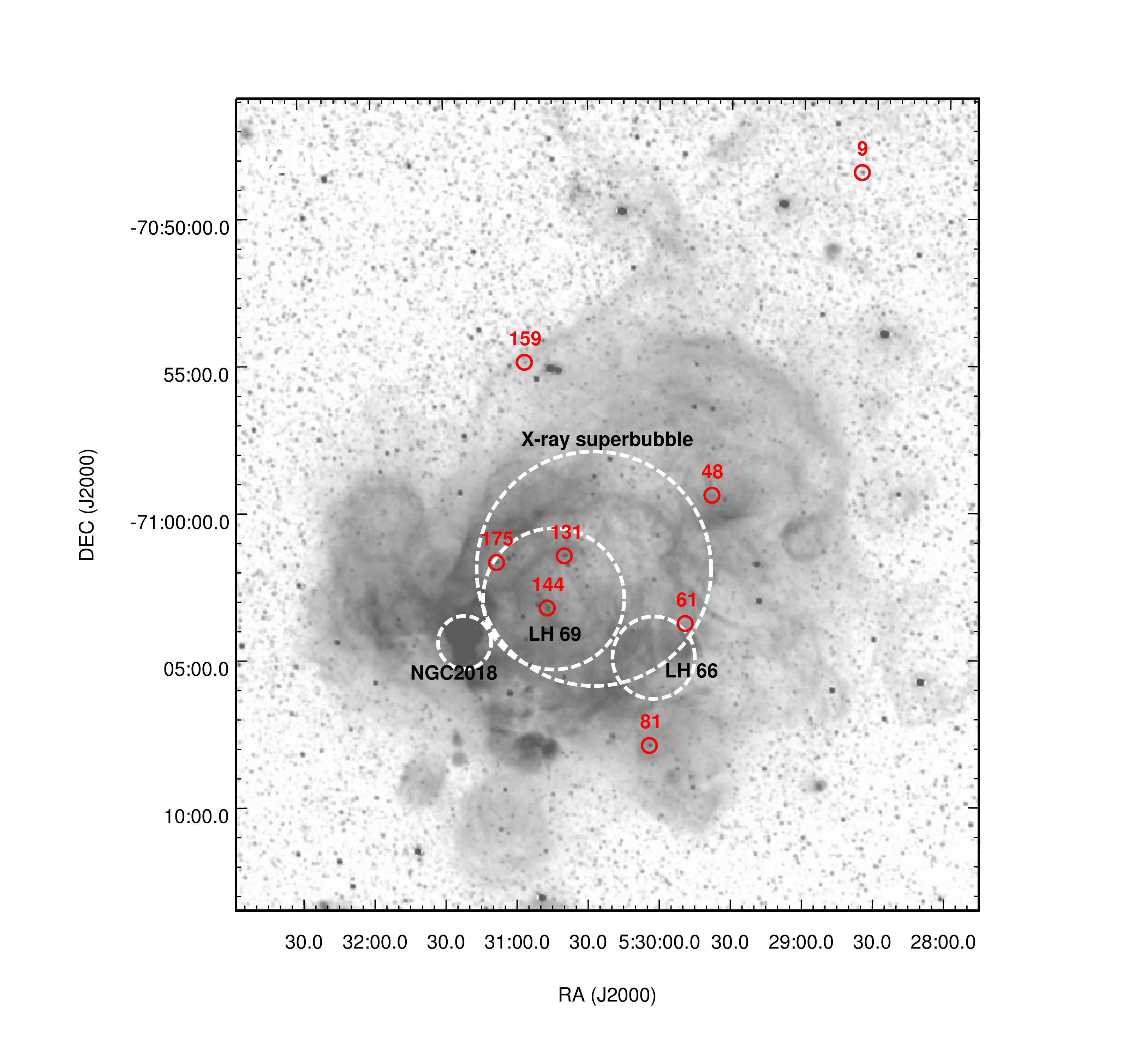}
\caption{Eight candidate runaway OB stars in the sample. The underlying H$\alpha$ image is from the Magellanic Cloud Emission-Line Survey \citep[MCELS, ][]{Smith2005}. }
\label{fig:run}
\end{figure}


\begin{table}
\caption{Candidate runaway OB stars in the N\,206 superbubble} 
\label{table:runstars}     
\centering
\begin{tabular}{cccl}
\hline
\hline
\noalign{\vspace{1mm}}
 N206-FS & $\varv_{\rm rad}$  &  $\varv_{\rm rad}$ -  $\bar{\varv}_{\rm rad}$& spectral type\\
 \#  & (km\,s$^{-1}$) & (km\,s$^{-1}$) &\\
\noalign{\vspace{1mm}}
\hline 
\noalign{\vspace{1mm}}
9   & 300 & 52 &B1.5  (IV)e \\
48  & 180&$-68$& B0.7 V \\
61  & 180 &$-68$&B0.7 V \\
81  & 160&$-88$& B1 II  \\
131\tablefootmark{1}$\!\!\!$ &185&$-63$&  O6.5 II (f) \\
144 &320&72& B0 II Nwk\\
159 &320&72& B2 V\\
175 &330&82& B0 III\\

\hline
\end{tabular}
\tablefoot{\tablefoottext{1}{One of the Of stars analyzed in Paper\,I}}
\end{table}

The Oe and Be stars in our sample in general rotate  sub-critically. Our result is in line with the statistical study by \citet{Cranmer2005} who concluded that this strongly constrains the physical models of angular momentum deposition in Be star disks. However, many observations of Be stars in the Galaxy and Magellanic Clouds suggest that most of them are rotating at almost their critical velocity \citep{Rivinius2013,Martayan2006,Martayan2010}.

One possible reason for finding a low $\varv\sin i$ could be an accidental pole-on view ($\sin i \approx 0$). From inspecting the Balmer emission line profiles, we distinguish between double-peak disk emissions and single peaks. For the latter, one may suggest a low inclination angle. Typically, the disk emission of H$\alpha$ shows up by double peaks that are separated by up to  4\,\AA\ in the case of large inclination. Given the limited spectral resolution of our data, we cannot resolve double peaks with separations below 0.8\,\AA. Therefore, our category of pole-on stars (red dots in Fig.\,\ref{fig:be_rot}) comprises stars with $\sin i \leq 0.2 $ corresponding to $ i \leq 10\degr$.

However, the statistically probability of observing an inclination below the angle $i$ scales with $1 - \cos i$. This implies that the chance of catching a star with an inclination  $i \leq 10\degr$ is only about 2\%. The likely reason that we find seven out of 19 Oe/Be stars with apparently low inclination is the contamination by nebular emission, so that the double peaks of the Balmer profiles cannot be recognized. Nevertheless, it is puzzling that nearly all   Oe/Be stars of the sample with higher inclination also show $\varv\sin i < 0.5\,\varv_{\rm critical}$. The only exception is N206-FS\,62, which rotates close to its critical velocity.

\subsubsection{Radial velocity and candidate runaway stars}
The radial velocity of OB stars in our N\,206 sample ranges from 160 to 330\,km\,s$^{-1}$ (see Fig.\,\ref{fig:hist}), with a peak of the distribution at about $\sim$240-270\,km\,s$^{-1}$. 
Runaway stars are usually defined by peculiar velocities in excess of
40 km/s \citep{Blaauw1961} compared to the systemic velocity.
Their velocity is either a result of dynamical ejection
from a young cluster or of ejection from a binary system due to a supernova.

\begin{figure}
\centering
\includegraphics[scale=0.5]{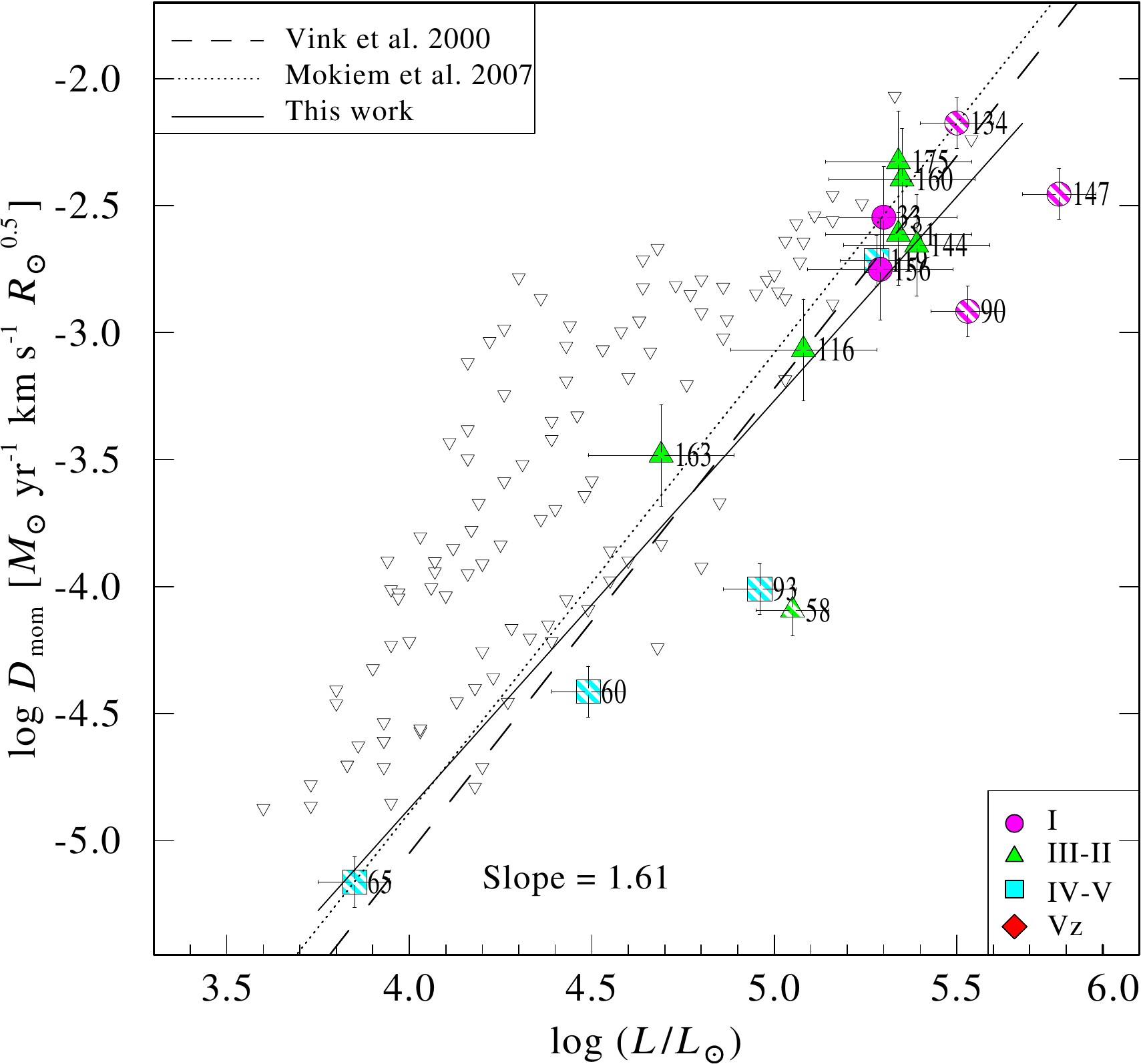}
\caption{Modified wind momentum ($D_{\mathrm{mom}}$) in units of $M _{\odot} $/yr km s$^{-1} \, R_{\odot}^{0.5} $ as a function of stellar luminosity for  analyzed LMC OB stars. The filled and hatched symbols are wind momenta determined from H$\alpha$ or nitrogen emission lines and from UV P-Cygni profiles with individual error bars, while open triangles (up-side down) indicate upper limits. A power-law (solid line) fit to the observations yields a slope of 1.61. The theoretical WLR from \citet{Vink2000} (dashed line) and an empirical WLR from \citet{Mokiem2007A} for LMC stars are also plotted.
Luminosity classes are distinguished by different symbols as given in the legend.}
\label{fig:wlr}
\end{figure}


\begin{figure}
\centering
\includegraphics[scale=0.5]{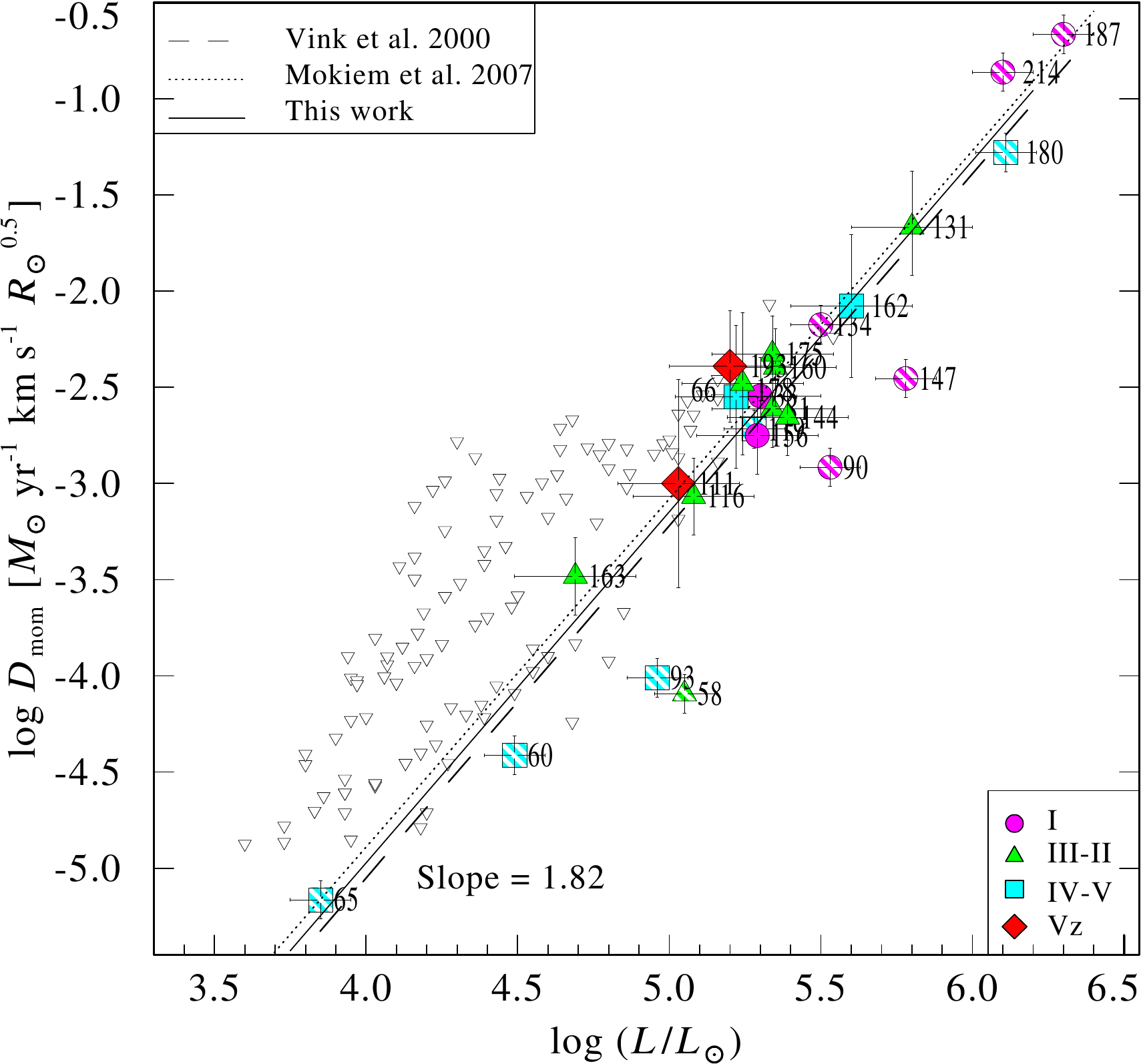}
\caption{Same as Fig.\,\ref{fig:wlr}, but including the Of stars from Paper\,I.}
\label{fig:wlrfull}
\end{figure}


\begin{figure*}[htpb]
\centering
\includegraphics[width=0.65\textwidth]{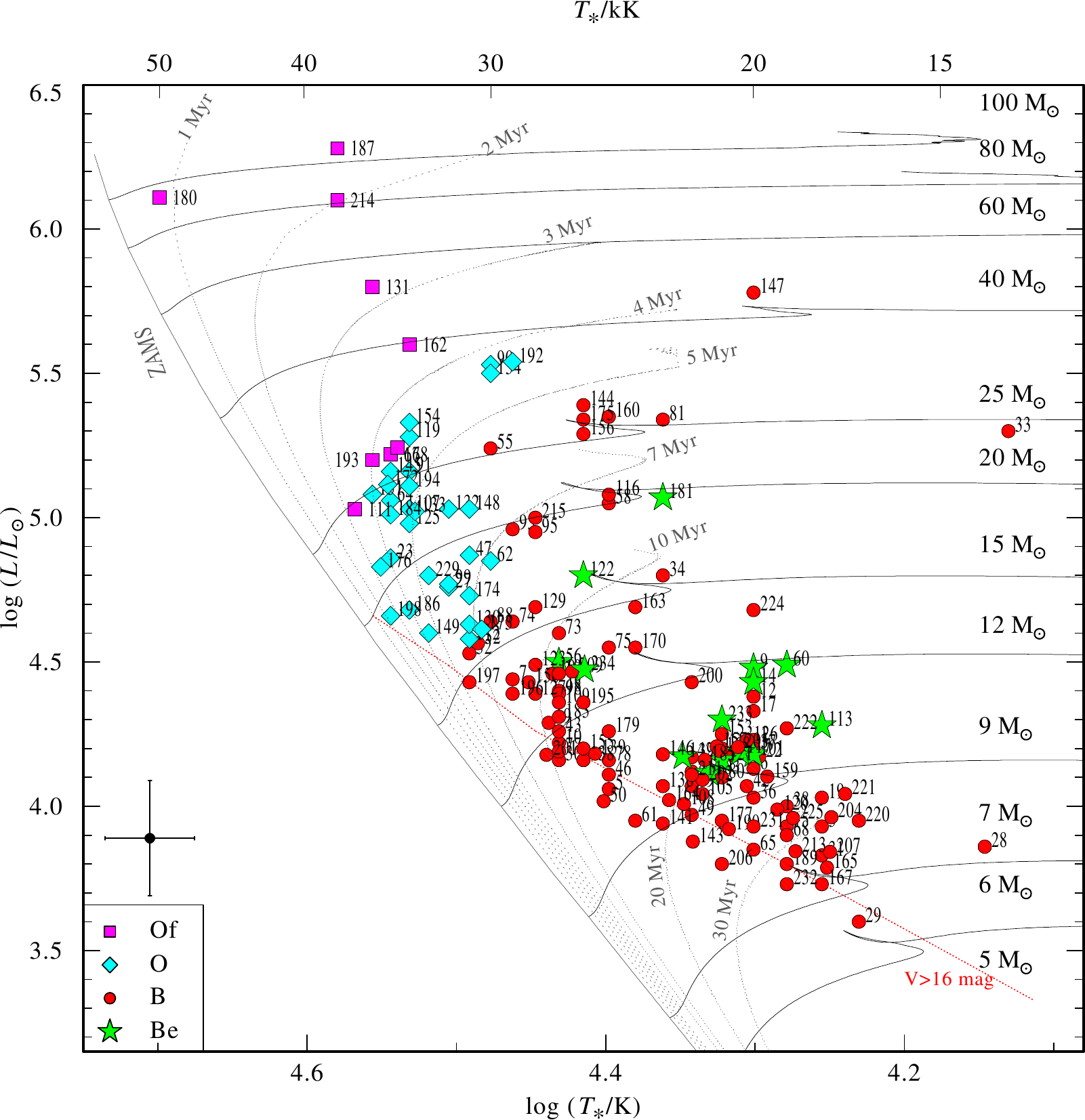}
\caption{Hertzsprung-Russell diagram for OB stars in the N\,206 superbubble in the LMC. The nine Of stars from Paper\,I are also included. The evolutionary tracks (thin solid lines) are labeled with their initial mass. Isochrones are represented by gray dotted lines. Tracks and isochrones are based on evolutionary models accounting for rotation with $\varv_{\mathrm{rot, init}} \,\sim$\,100 km\,s$^{-1} $ \citep{Brott2011,Kohler2015}.
Different spectral types are distinguished by different colors and shapes of the symbols (see legend). The typical uncertainties of the stellar temperatures and luminosities are indicated by the error bars in the lower left of the HRD. The red dotted line represents the visual-magnitude cut-off of the observed sample.}
\label{fig:hrd}
\end{figure*}


To identify runaway candidates in our sample, we compared the radial velocity estimates for each star with the mean radial velocity of all program stars and the corresponding standard deviation.
By accounting for a threshold of $|\delta \varv_{\rm rad}| > 3 \sigma$, we identify runaway candidates. Then we recalculate the mean velocity and standard deviation excluding these objects, continuing this process until no more stars with $|\delta \varv_{\rm rad}| > 3 \sigma$ remain. We find eight runaway candidates among the total sample. This includes  the Of binary N206-FS\,131 described in Paper\,I. All other runaways are early B-type stars. Their positions are marked in  Fig.\,\ref{fig:run}; their radial velocities as well as the deviation from the mean velocity are given in Table\,\ref{table:runstars}.

The mean radial velocity of the OB stars (excluding the runaway candidates) in this region is found to be $\approx$248$\pm$16\,km\,s$^{-1}$. This dispersion of $\pm 16\,$km\,s$^{-1} $ includes the $\pm 10\,$km\,s$^{-1} $ uncertainty of the $\varv_{\rm rad}$ measurement,  meaning that the actual velocity dispersion is smaller. The radial velocities do not show any obvious correlation with spatial structures in the complex.

\subsection{The wind momentum-luminosity relationship}
\label{sect:windML}

We quantitatively investigate the wind properties of our sample OB stars by plotting the modified wind momentum-luminosity relation (WLR) and compared our results to previous studies and theoretical predictions. Figure\,\ref{fig:wlr} depicts the modified stellar wind momentum, which is defined as $ D_{\mathrm{mom}} \equiv \dot{M} \varv_\infty R_\ast $ \citep{Kudritzki2000}, as a function of the stellar luminosity. The wind momentum of eight stars, which have available UV spectra, and nine OB stars with H$\alpha$ partially or completely in emission, are empirically determined and plotted in the figure. For all other stars, only an upper limit for the wind momentum can be estimated, as marked by  upside-down triangles. A linear regression to the logarithmic values of the modified wind momenta obtained in this work, accounting for the individual error bars, gives a slope of $1.61 \pm 0.22$, which is less steep than the theoretically predicted slope of 1.83 for LMC OB stars \citep{Vink2000} given in Fig.\,\ref{fig:wlr} (dashed line). The empirical WLR obtained by \citet{Mokiem2007A} for LMC OB stars with a slope of 1.81 is also marked in the figure (dotted line) for comparison. Figure\,\ref{fig:wlrfull} shows the WLR of the whole OB star sample 
including the Of stars from Paper\,I. This yields a linear regression of
\begin{equation}
\log\,D_{\mathrm{mom}}\, =\, (1.82 \pm 0.18)\, \log\,(L_\ast/L_{\odot})\,+\, (-12.25 \pm 0.95 )
\end{equation}
which is close to the theoretical and empirical relations shown. 

However, this agreement might be just fortuitous. First
of all, the empirical wind momenta show a large scatter. Moreover, they
are based on different diagnostics, which have their specific issues.
The Balmer line emission is fed by the recombination cascade and
therefore subject to the microclumping effect. The empirical mass-loss
rate derived from a given observed emission line scales roughly with the
square root of the adopted clumping contrast
\citep[e.g.][]{HK98}. For our analyses, we adopted a
clumping contrast of $D=10$, while the WLR from \citet{Mokiem2007A} shown
in Figs.\,\ref{fig:wlr} and \ref{fig:wlrfull} have been obtained with smooth-wind models.   

On the other hand, part of our empirical mass-loss rates have been
derived from fitting the P-Cygni profiles of UV lines (hatched symbol
filling in Figs.\,\ref{fig:wlr} and \ref{fig:wlrfull}). These resonance lines are not affected by
microclumping, but here  the usual neglect of macroclumping
(``porosity'') can lead to an  underestimation  of mass-loss
rates \citep{Oskinova2007}. 

Another uncertainty refers to the actual metallicity of the individual
stars.  The theoretical prediction from \cite{Vink2000} has been
calculated for a canonical LMC metallicity of 0.5 solar. Our young
stars, and especially the two most luminous Of stars in our sample,
might have formed already with higher metallicity. Our spectral data
allow only very limited abundance measurements, and the total
metallicity cannot be precisely determined. On the other hand, Vink
mass-loss rates have been often found to over-predict $\dot{M}$ by
about a factor of two for early OB-type stars \citep[e.g.][]{Surlan2013} and by more than an order of magnitude for late OB-type stars \citep{Martins2009,Shenar2017}.

\subsection{OB stars in the Hertzsprung-Russell diagram}
\label{sect:hrd}

\begin{figure}[!hbtp]
\centering
\includegraphics[scale=0.5]{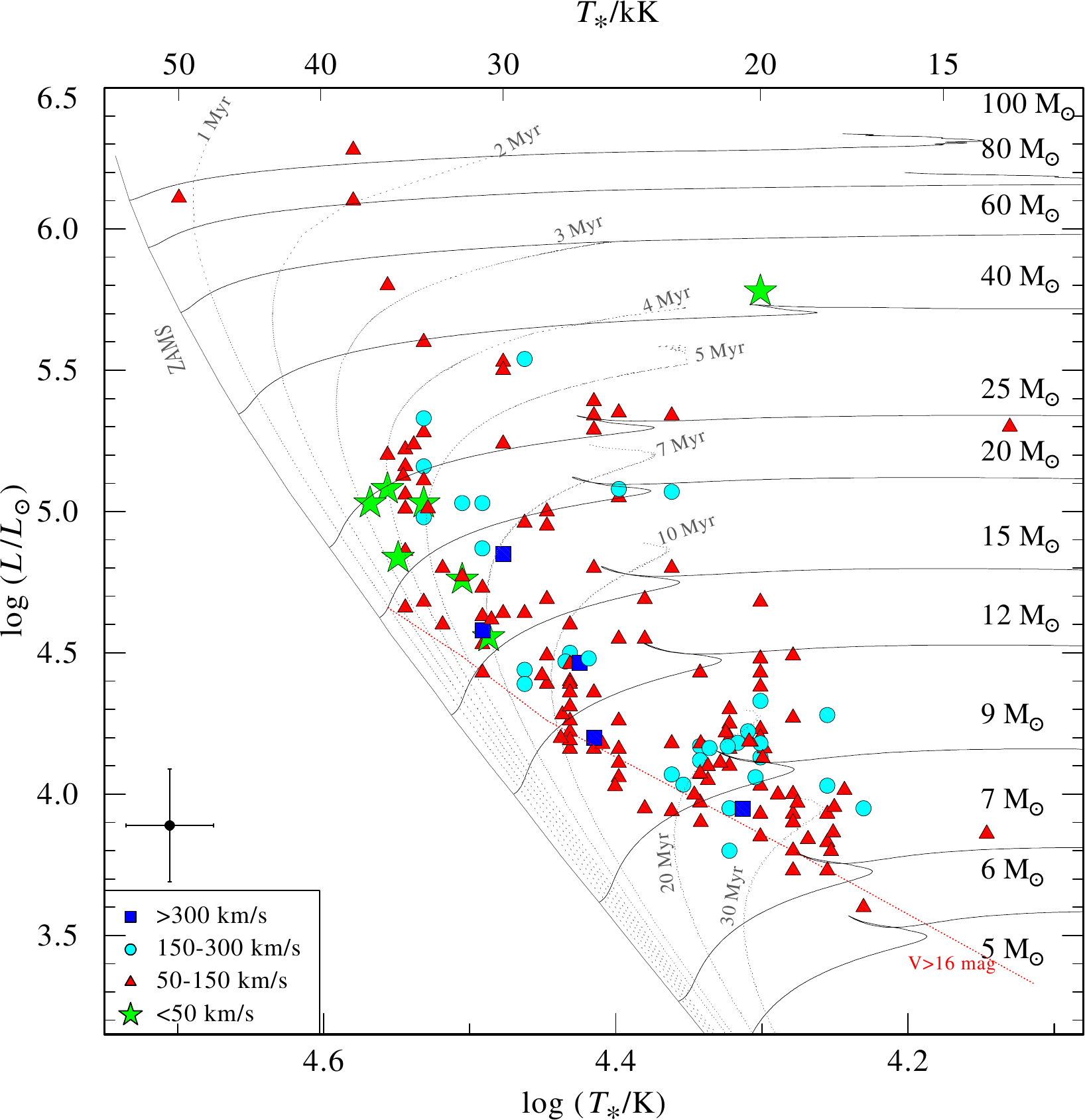}
\caption{Same as Fig.\,\ref{fig:hrd} with symbol shapes and colors referring to the projected rotational velocity (see legend).}
\label{fig:hrdrot}
\end{figure}

The evolutionary status of all hot OB stars in the N\,206 complex are investigated using the Hertzsprung-Russell diagram (HRD). Figure\,\ref{fig:hrd} shows  our sample of OB stars with the effective temperatures and luminosities as given in Table\,\ref{table:stellarparameters}. The nine `Of stars' from Paper\,I are also included.

The evolutionary tracks and isochrones are adapted from \citet{Brott2011} and \citet{Kohler2015}, which were calculated for an initial rotational velocity of $\sim$\,100\,km\,s$^{-1}$. This seems roughly adequate since the $\varv\sin i$ histogram in Fig.\,\ref{fig:hist} revealed an average of $\sim$\,126\,km\,s$^{-1}$. 
The evolutionary tracks are shown for stars with initial masses of 5 $ - $ 100 $M _{\odot} $, while the isochrones are shown for ages of  0, 1, 2, 3, 4\, 5, 7, 10, 20, and 30\,Myr, respectively.

Figure\,\ref{fig:hrd} shows that most of the O stars have ages between 1 and 7\,Myr, with initial evolutionary masses ranging from 15 to 40\,$M_{\odot}$.
In the case of B stars, ages extend from 5 to 30\,Myr. Interestingly, most of the Be stars are close to the terminal age main sequence as indicated by the loops in the evolutionary tracks.

The most massive stars  in our sample (N206-FS180, 187, and 214) are the youngest ones. This supports a scenario described in \citet{Bouret2013}, where the most massive stars of the cluster form last, and after their formation, they quench subsequent star formation. However, our finding could also be due to the V-magnitude ($V<16$\,mag) cut-off of the sample, since we might miss very young B-type stars.

Figure\,\ref{fig:hrdrot} shows also the HRD of the OB stars, but now color-coded with their respective projected rotational velocity. Five fast rotators with $\varv\sin i>$300\,km\,s$^{-1} $ (N206-FS\,62, 112, 155, 190, and 209) are found to have initial evolutionary masses less than 20\,$M _{\odot} $ (see dark-blue circles in the diagram). The most massive and youngest stars ($<$5\,Myr) are found to be slower rotators than the rest of the sample. Interestingly, one slow rotator (N206-FS\,52) with $\varv\sin i\approx$35\,km\,s$^{-1} $ and the fastest rotator (N206-FS\,112) with $\varv\sin i\approx$390\,km\,s$^{-1} $ fall on nearly the same position in the HRD.
However, both of these stars have similar spectral type and mass.

\begin{figure*}
\centering
\includegraphics[scale=0.49]{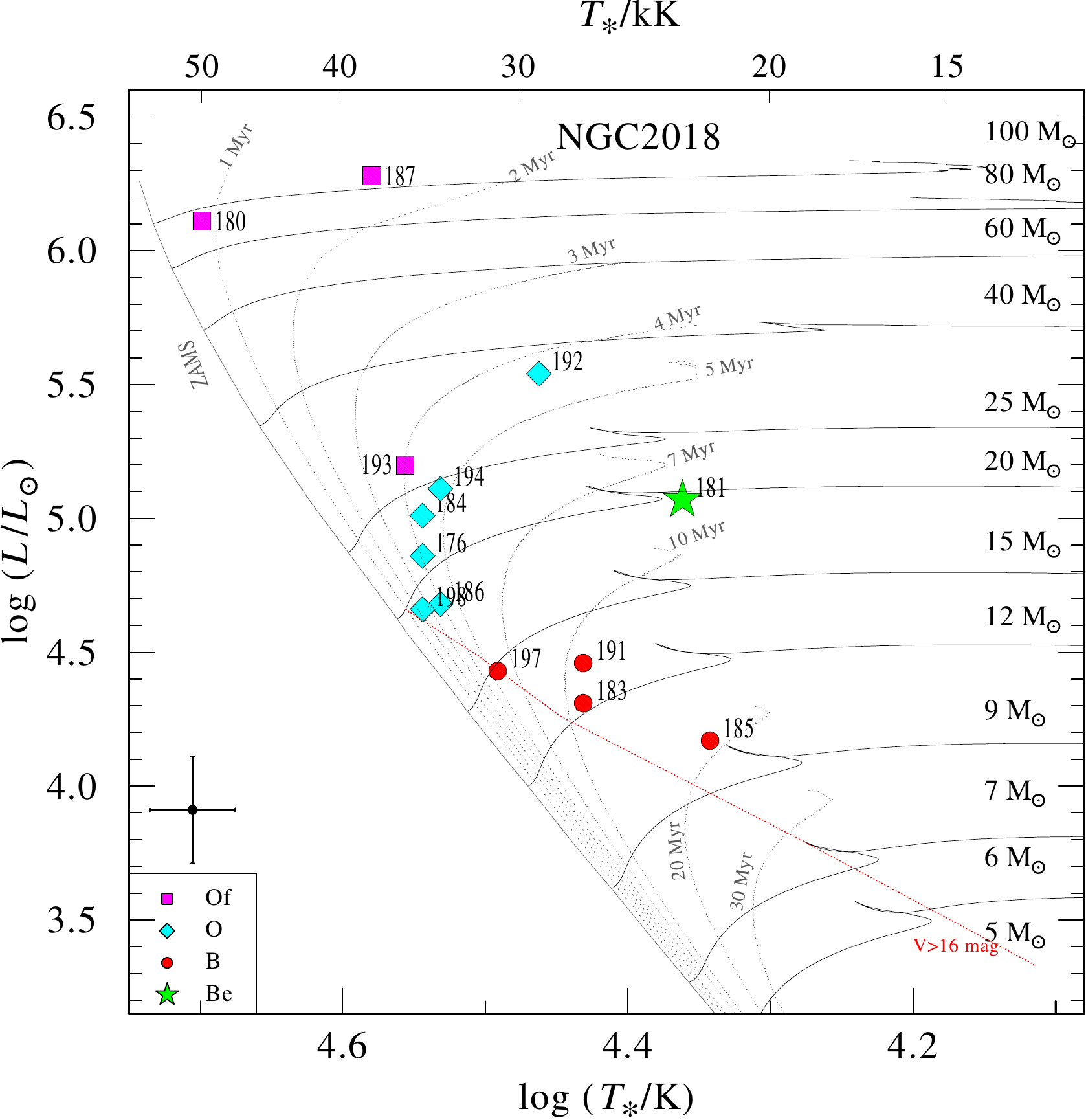}
\includegraphics[scale=0.49]{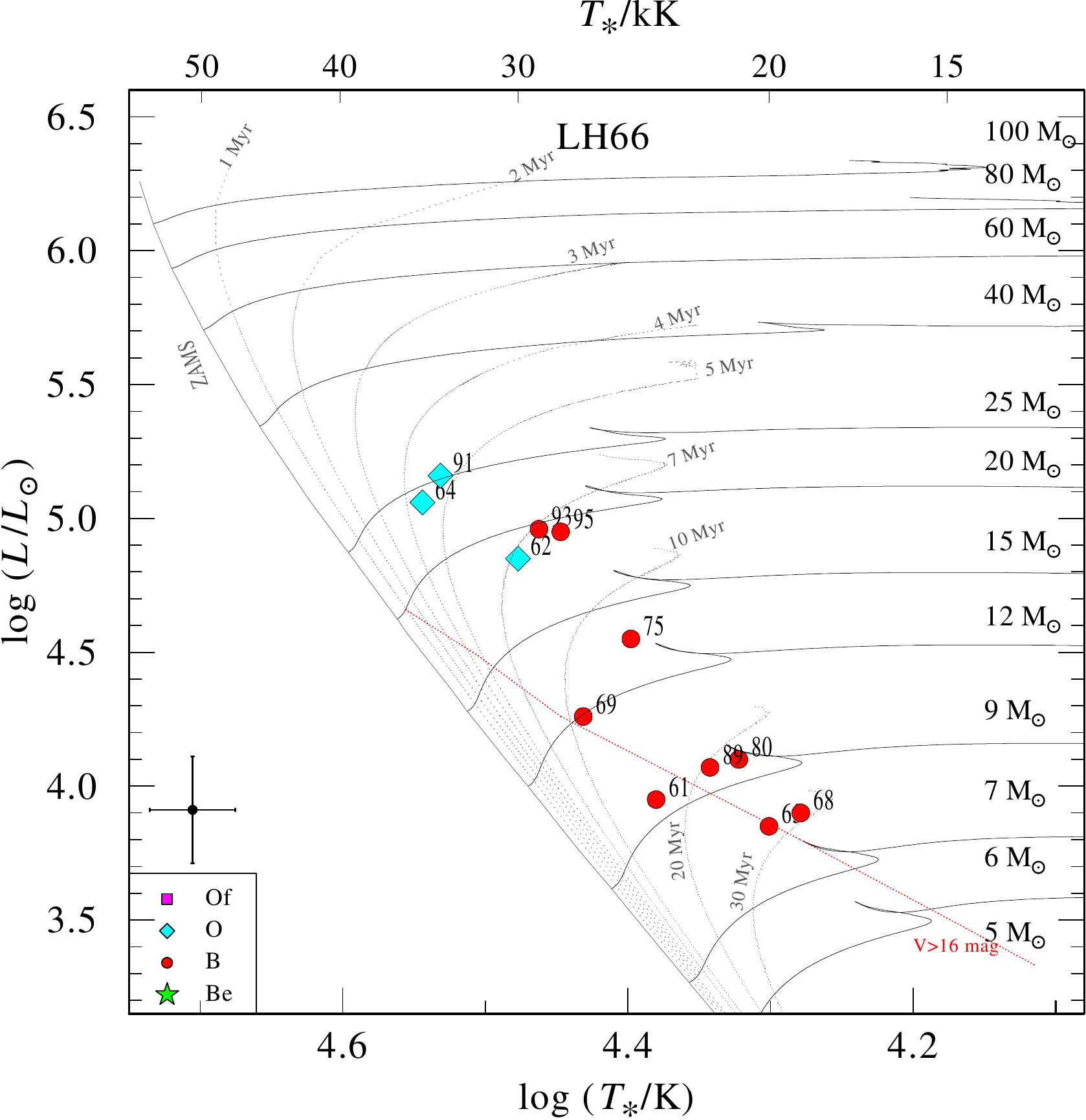}
\includegraphics[scale=0.49]{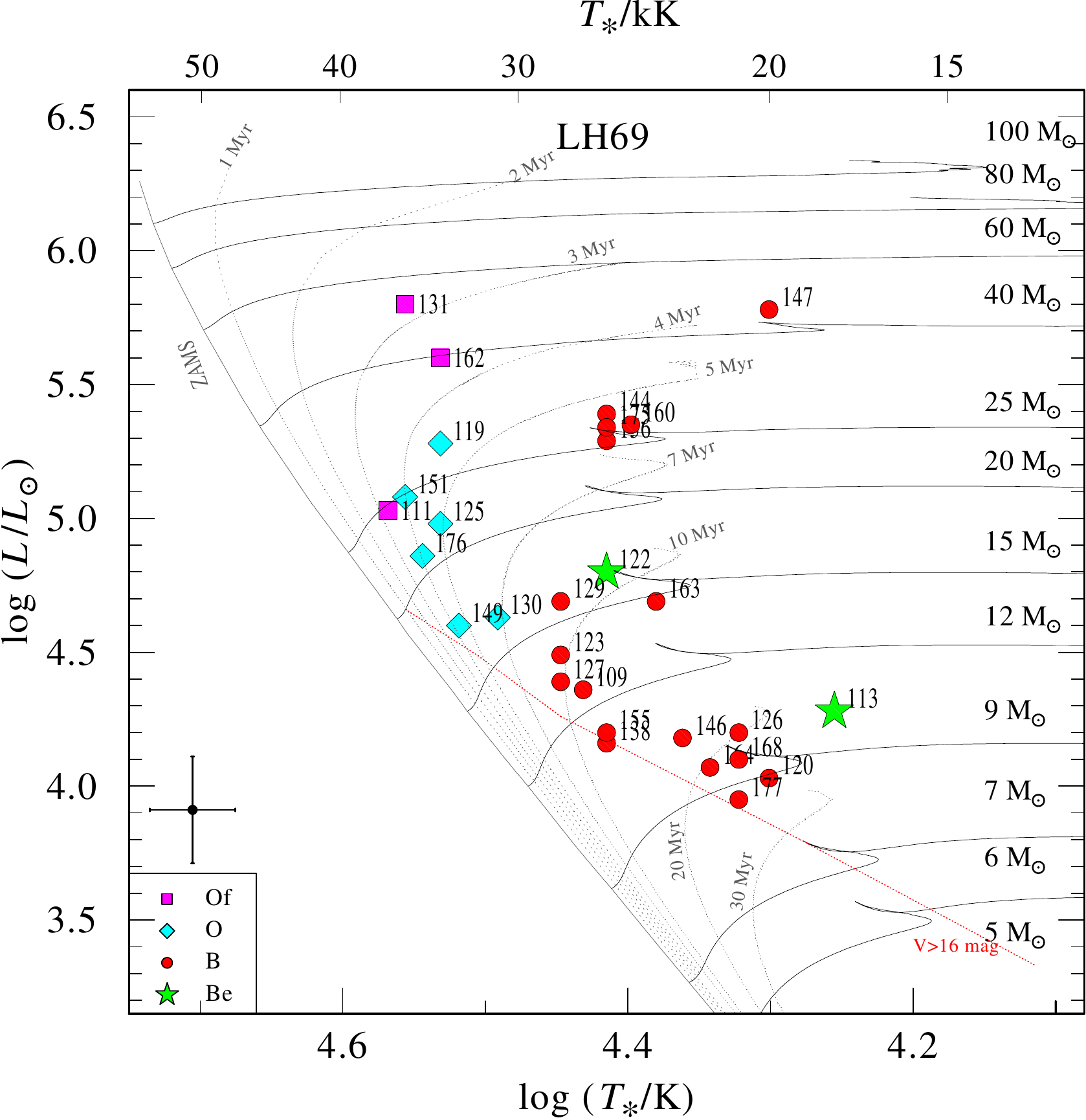}
\includegraphics[scale=0.49]{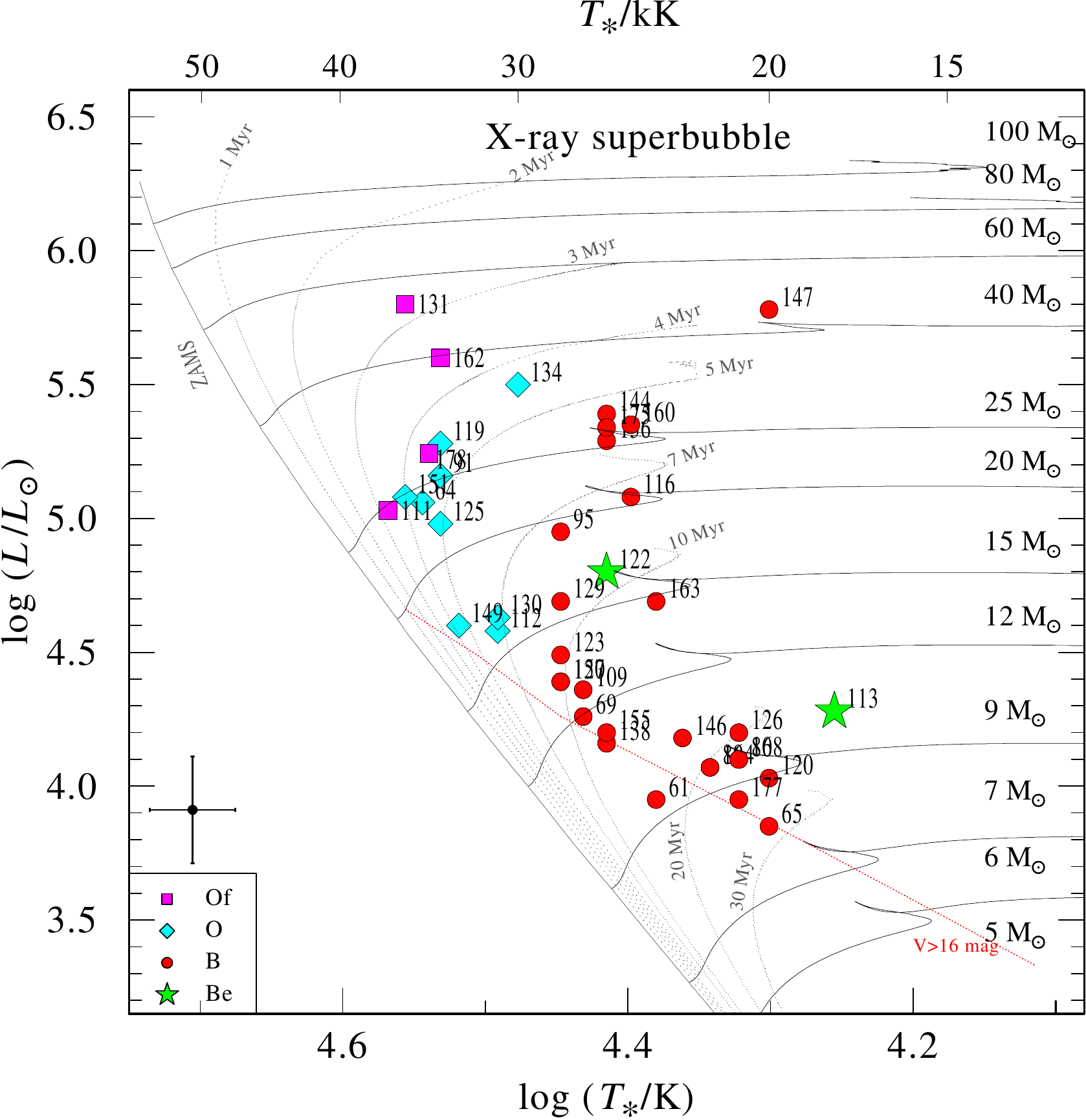}
\caption{Hertzsprung-Russell diagram for stars in subregions of N\,206 (circular regions shown for NGC\,2018, LH\,66, LH\,69, and X-ray superbubble in Fig.\,\ref{fig:bubble}). The evolutionary tracks, isochrones, and symbols have the same meaning as in Fig.\,\ref{fig:hrd}.  }
\label{fig:hrd_cl}
\end{figure*}
%

\subsubsection{HR diagram of substructures}
\label{subsect:hrdsub}

As discussed above, the empirical HRD indicates a spread of stellar ages rather than a coeval population of OB stars. This raises the question of multiple or progressing star formation throughout this large region. Therefore, we split our sample into different spatial regions as given in Fig.\,\ref{fig:bubble}, and plot their respective HRDs (Fig.\,\ref{fig:hrd_cl}).

The upper-left panel refers to the cluster NGC\,2018, which contains mostly O stars and especially the three most massive stars of the whole sample. Most of the stars in this cluster fall in the age range of $2-4$\,Myr. The cluster shows a large age dispersion from  $\leq$1\,Myr (N206-FS180) to $\sim $20\,Myr (N206-FS185). 

Most stars in the OB association LH\,66 (Fig.\,\ref{fig:hrd_cl}, upper-right panel) have ages in the range $7-10$\,Myr, with no star being younger than 5\,Myr. The OB association LH\,69 (lower-left panel) contains multiple populations with ages of 4, 5, 10, and 20\,Myr. The youngest star in this association is $\leq$3\,Myr old. 
Most of the stars in LH\,69 are part of the somewhat larger X-ray superbubble region, which shows an age dispersion in the range $3-30$\,Myr.

Summarizing this investigation revealed that each of the subregions of the N\,206 complex had multiple episodes of star formation over the last 30\,Myr. However, only the cluster NGC\,2018 formed stars as recently as 1\,Myr ago.

\subsubsection{The evolutionary status of the WC+O binary N206-FS\,128}
\label{subsect:wchrd}

\begin{figure}
\centering
\includegraphics[scale=0.51,trim={1cm 0 0 0}]{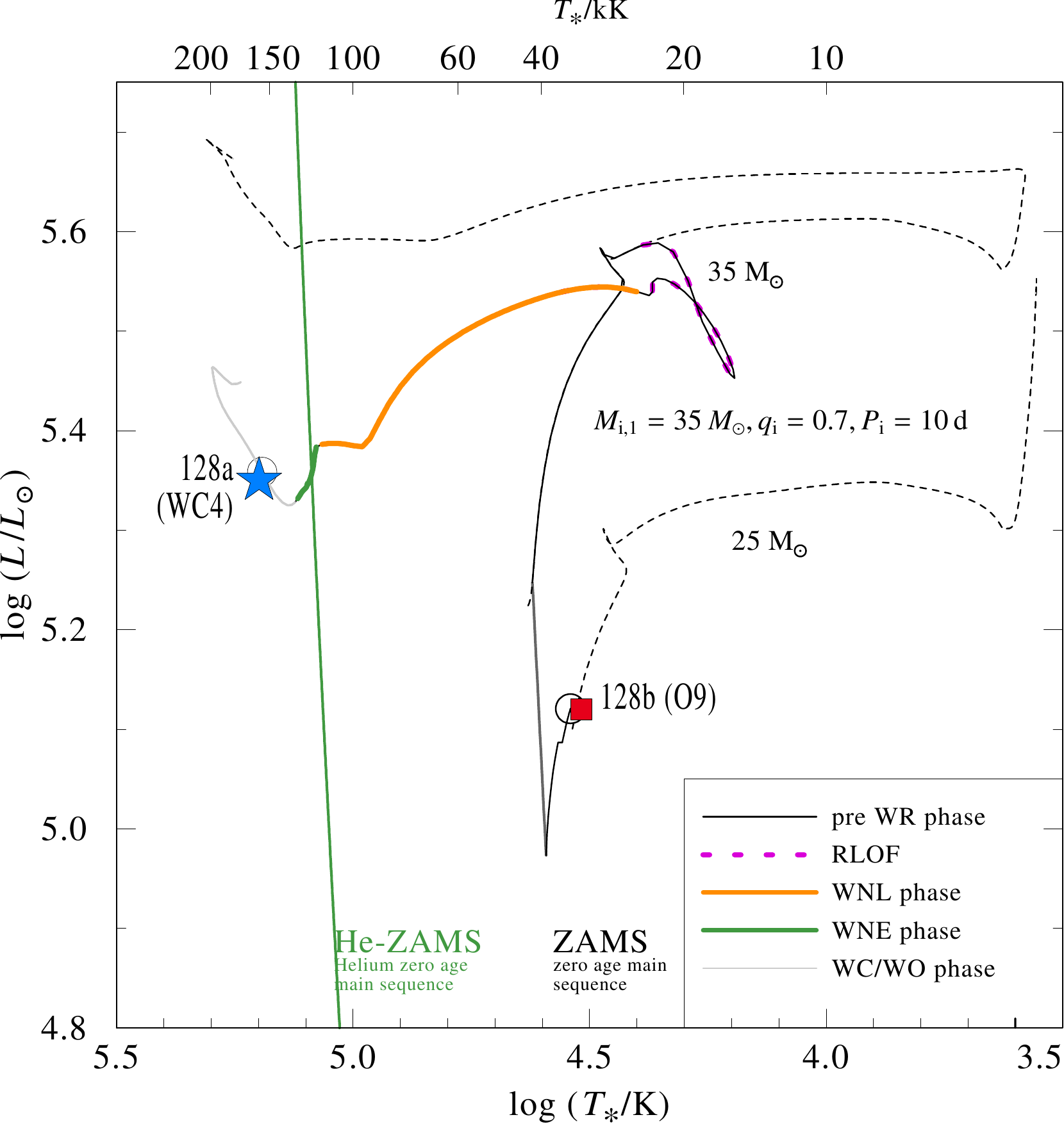}
\caption{HRD positions of N206-FS\,128a (WC4, blue asterisk) and N206-FS\,128b (O9, red square). The solid lines represent binary evolutionary tracks from \citet{Eldridge2008} for the respective components. The circles denote the best-fitting positions along the tracks after 6\,Myr of evolution. For comparison, the dashed lines show how the components would evolve without mutual interaction. See text for more details }
\label{fig:hrdwc}
\end{figure}

We compare the HRD positions of the binary components of N206-FS\,128 with evolutionary tracks from \citet{Eldridge2008}, which account for binary interaction (see Fig.\,\ref{fig:hrdwc}). These binary tracks are defined by three parameters : the initial mass of the primary $M_{i,1}$, the initial orbital period $P_{i}$, and the mass ratio $q_{i} = M_{i,2} /M_{i,1}$. Using the $\chi^{2}$ minimization algorithm described in \citet{Shenar2016}, we found the best fitting parameters for the binary track to be $M_{i,1}=35 M_{\odot}$, $P_{i}=10$ days, and $q_{i} =0.7$. After 6\,Myr of evolution, this binary track  reproduces not only the observed HRD positions but also the current surface hydrogen and carbon abundances of both components very well.
The current period of the binary system predicted by the model is $\approx 5$ days, and thus slightly higher than the observed value of 3.23 days measured by \citet{Mowlavi2017}.

According to the best-fitting binary track, the WC component N206-FS\,128a started its evolution with an initial mass of $35M_{\odot}$ and stayed on the main sequence (pre WR phase) for 5.4\,Myr. Before it entered into the WR phase, it had undergone a Roche lobe overflow (RLOF) phase (dashed magenta lines in Fig.\,\ref{fig:hrdwc}) over $\sim$20,000 years. During this phase, the primary looses more than half of its mass, and the secondary accretes a few solar masses. After this, N206-FS\,128a entered the WNL stage with a low hydrogen mass fraction $0< X_{\rm H} < 0.2$ at the surface.  The WNE phase is reached  when the surface hydrogen has vanished. When the helium-burning products appear at the surface, the star proceeds to the WC/WO phase. 

During the last 0.6\,Myr, the primary experienced very high  mass loss, which decreased its mass to about 10\,$M_{\odot}$, while the mass of the secondary remained roughly constant. This is the current state of the system, at an estimated age of roughly 6\,Myr. 

\begin{figure*}[htb]
\centering
\includegraphics[scale=0.76,trim={0 17cm 0 1cm}]{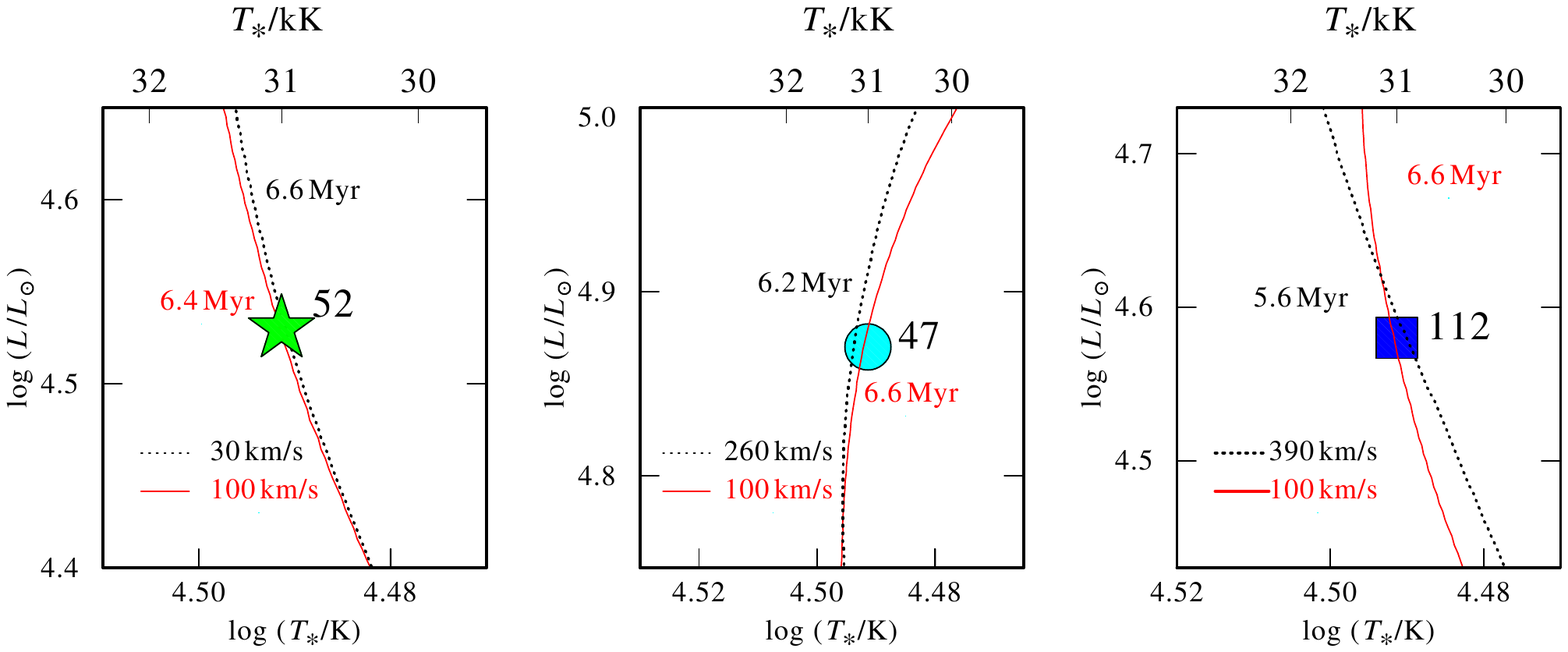}
\caption{Effect of adopted initial rotation velocity on age determination. The HRDs of N206-FS\,52 ( $\varv\sin i \approx 35 \rm{\,km\,s^{-1}}$), N206-FS\,47 ($\varv\sin i \approx 264 \rm{\,km\,s^{-1}}$), and N206-FS\,112 ($\varv\sin i \approx 390 \rm{\,km\,s^{-1}}$) are plotted from left to right. The black dashed lines represent the isochrones with $\varv\sin i $ corresponding to each star, and solid red lines are the isochrones with $\varv_{\rm rot} \sim$\,100\,km\,s$^{-1}$. 
}
\label{fig:rotdiff}
\end{figure*}

If, hypothetically, the star N206-FS\,128a had evolved without mutual interaction with its companion, it would have evolved as represented in Fig.\,\ref{fig:hrdwc} by the black dashed line. In that case, the star would end up as a WC star with a higher luminosity.

The secondary, N206-FS\,128b, is an O9 star with an initial mass of $25M_{\odot}$, which is still in the hydrogen burning stage. The binary track of the secondary does not continue to further evolutionary stages, since the primary explodes as a supernova in a few ten thousand years. After this, the secondary may evolve like a single star of $M=25M_{\odot}$ (see black dashed line in Fig.\,\ref{fig:hrdwc}).

\subsubsection{Evidence for sequential star formation?}
\label{subsect:sequential}
In the HRD (Fig.\,\ref{fig:hrd}), we have included the isochrones for evolutionary tracks with an initial rotational velocity of $\sim$\,100\,km\,s$^{-1} $\citep{Brott2011,Kohler2015}.
The same models have been used to estimate the ages and evolutionary masses for most of our sample stars as given in Table\,\ref{table:App_age}. However, the adopted initial rotation affects the isochrones and thus the age determination of the individual stars. In order to test this effect, we consider three stars with different measured  $\varv\sin i$ = 35, 264, and 390 $\rm{\,km\,s^{-1}}$, respectively, but similar HRD positions. (Fig.\,\ref{fig:rotdiff}).
In each of the HRDs we show the matching isochrone for $\varv_{\rm rot} \sim$\,100\,km\,s$^{-1}$. Additionally, we plot an isochrone that is selected from the set of tracks that is more suitable for the respective star according to its measured $\varv\sin i$.
In the first two cases, the difference in the determined age is relatively small ($<0.4$\,Myr). However, in the case of the rapidly rotating star N206-FS\,112, the isochrone corresponding  to  $\varv_{\rm rot}=100$\,km\,s$^{-1}$ yields an age that is higher by about 1\,Myr. Therefore, we finally applied
 isochrones for an initial rotational velocity of $\geq$\,300\,km\,s$^{-1}$ for those stars with measured $\varv\sin i>300$\,km\,s$^{-1} $ for determining the age and evolutionary mass. The uncertainties in the age are approximately 20-40\%, which are due to the uncertainties of the derived parameters (temperature, luminosity, $\varv\sin i$) as well as uncertainty in choosing the adequate isochrones.

\begin{figure}
\centering
\includegraphics[scale=0.48]{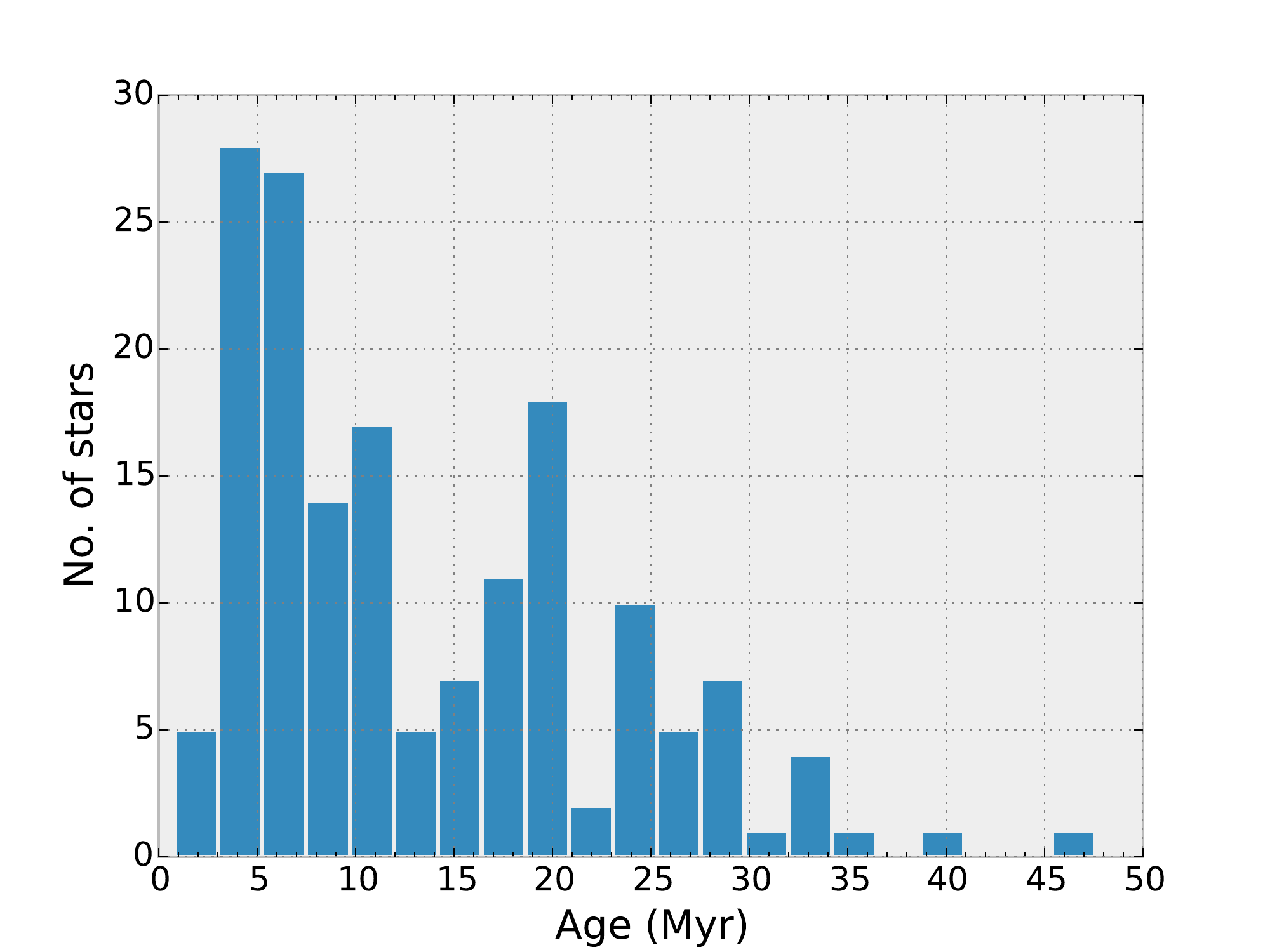}
\caption{Histogram of ages of the OB stars in the N\,206 complex, with bin width of $\sim$2\,Myr.}
\label{fig:histage}
\end{figure}

The distribution of ages of all OB stars in our sample is displayed in Fig.\,\ref{fig:histage}. It clearly shows multiple peaks across stellar ages, with a maximum in the age range 3-7\,Myr. From the HRD we can see that the stars in this age range are mostly O stars and B supergiants. So, massive star formation in the N\,206 complex must have peaked in this time period.
There are also local maxima at ages 10, 20, and 30\,Myr populated by B stars.

\begin{figure}
\centering
\includegraphics[scale=0.3,trim={1cm 0 0cm 0}]{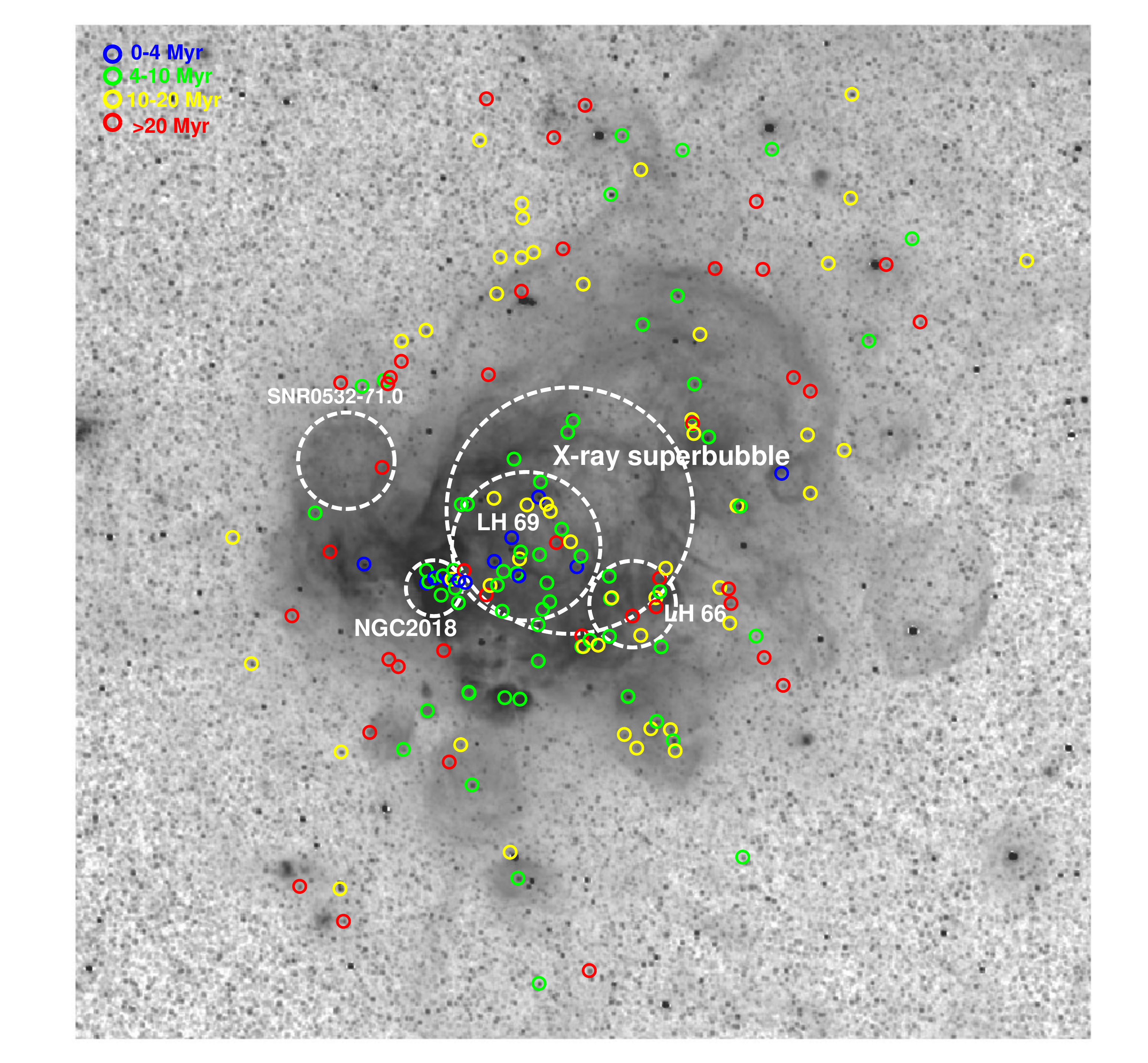}
\caption{Distribution of the OB stars in the N\,206 complex, color-coded with their age }
\label{fig:age_distr}
\end{figure}

We checked our results for correlations between stellar ages and location. Figure \,\ref{fig:age_distr} shows the position of the OB stars, color-coded with their respective estimated age. For this purpose, we grouped the stars into four different age categories  0-4\,Myr, 4-10\,Myr, 10-20\,Myr, and above 20\,Myr, respectively. We can see that the youngest stars are concentrated in the central parts of the complex, especially in the NGC\,2018 cluster. Most of the older OB stars are located in the periphery of the N\,206 complex. 

\begin{figure}
\centering
\includegraphics[scale=0.47,trim={1cm 0 0 0}]{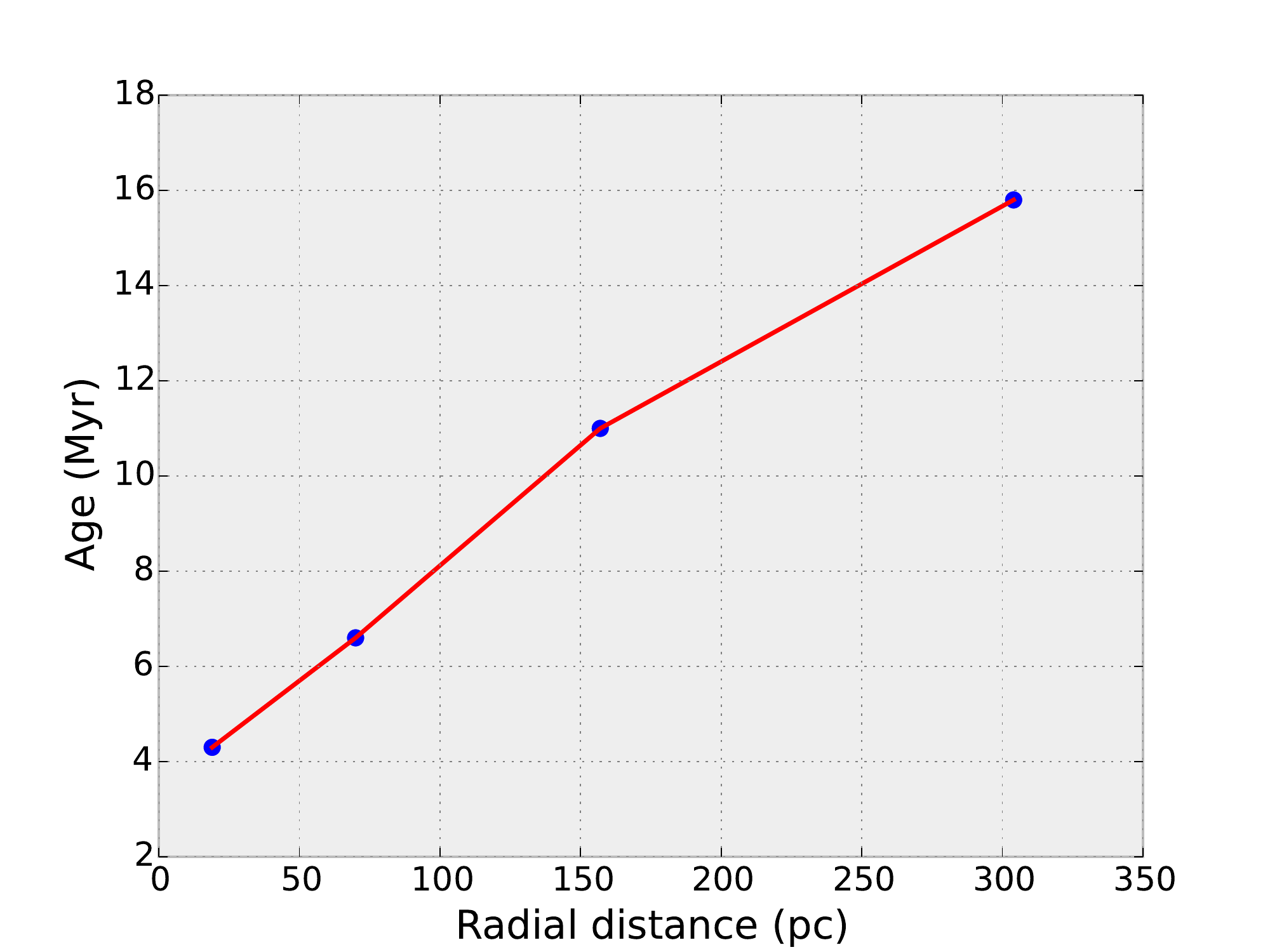}
\caption{Median age of the OB stars as function of their radial distance from the cluster NGC\,2018. }
\label{fig:age_pro}
\end{figure}


To investigating the variation of ages with location, we divided the region into four annuli  with radii of 80, 300, 670, and 1300$''$, respectively, centered at the young cluster NGC\,2018. The median of the ages of all OB stars inside each annulus is calculated and plotted in Fig.\,\ref{fig:age_pro}. The radial distances are converted from arc seconds to pc using the LMC distance. Interestingly, we can see a clear trend of the age increasing  with distance from the cluster center throughout the whole complex.

We propose two possible explanations for this correlation. One is that the star formation process began in the outer parts. The massive OB stars formed at this time period must have already cleared out their surrounding molecular cloud, and the star formation propagated inwards, where  dense molecular gas was left. Another possible scenario is that the star formation happened near the center of the complex, and these stars migrated outward during their lifetime. Given the average radial velocity dispersion of $16\,\rm{km\,s^{-1}}$, a star could travel a projected distance of 240\,pc ($\sim$ radius of the complex) in 15\,Myr. The dynamical ejection of massive stars from young clusters are also supported by  \citet{Oh2015} and \citet{Oh2016}.

\subsubsection{Mass discrepancy}
\label{subsect:discrepancy}
In Paper\,I we discussed the mass discrepancy noticed for the Of stars. The same discrepancy is also found here for the whole OB sample. The evolutionary masses of the individual stars are derived from the HRD as described above (see Sect.\,\ref{sect:hrd}). The evolutionary mass (see Table\,\ref{table:App_age}) represents the current stellar mass of the star as predicted by the track, while the spectroscopic mass (see Table\,\ref{table:stellarparameters}) is inferred from $\log\,g$.

\begin{figure}
\includegraphics[scale=0.5,trim={0cm 0 0cm 0}]{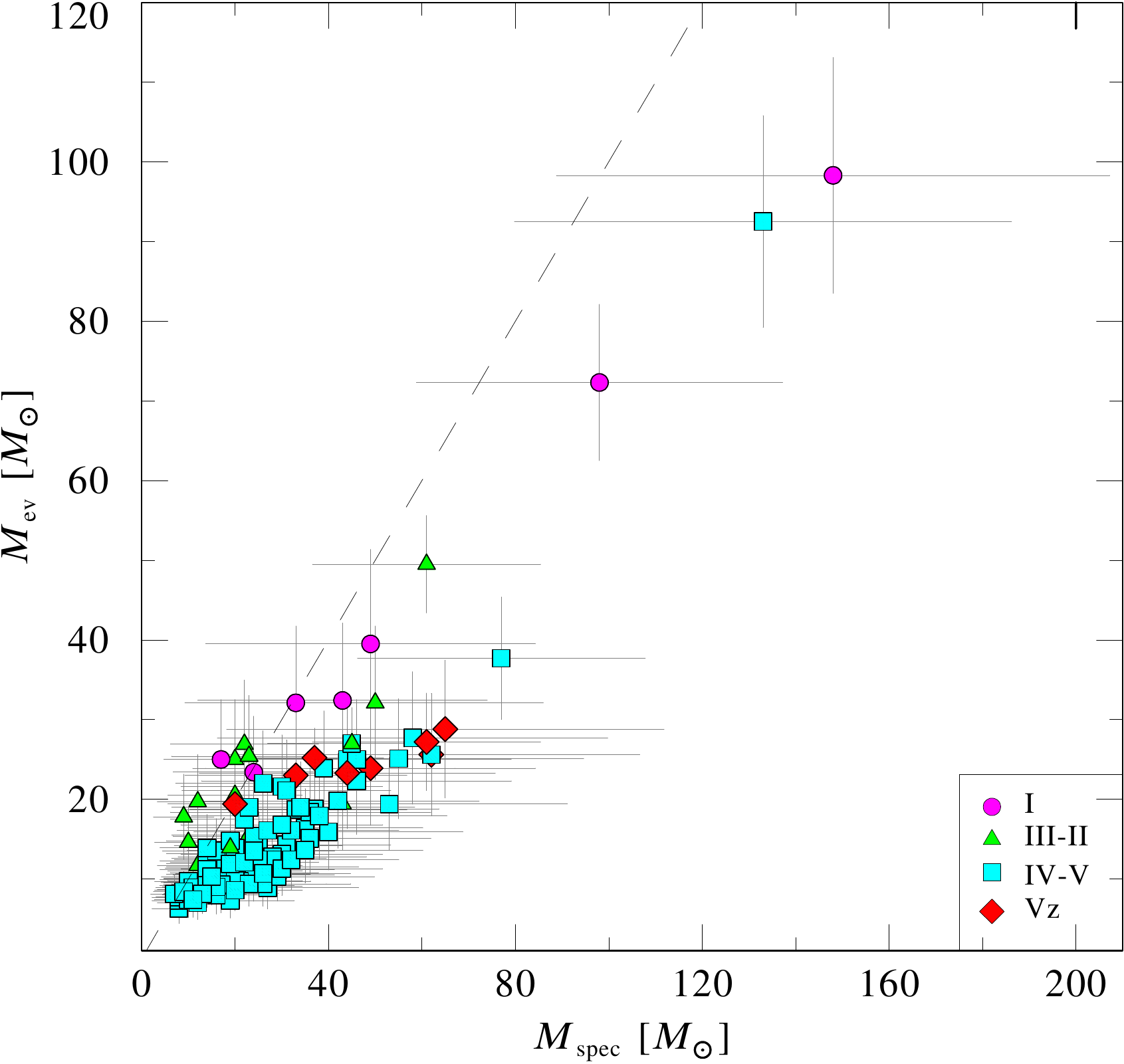}
\caption{Spectroscopic versus evolutionary masses. The 
one-to-one correlation of spectroscopic and evolutionary masses is indicated by 
the dashed line. The sample also includes the Of stars from Paper\,I. Different luminosity classes are denoted using different shapes and colors as given in the legend.}
\label{fig:massdis}
\end{figure}

In order to check for the discrepancy that we encountered in Paper\,I, we compared both masses for all OB stars (Fig.\,\ref{fig:massdis}). We can see a systematic difference between evolutionary and spectroscopic masses, even though the uncertainties are quite high. 
This mass discrepancy is a well known problem in astrophysics, especially in the case of OB stars in the Magellanic Clouds \citep{McEvoy2015,Bouret2013,Massey2009,Massey2013}. In our sample, the evolutionary masses are systematically lower than the spectroscopic masses. Moreover, the evolved stars (giants and supergiants) show much smaller discrepancies than the main sequence stars. It should be noted that spectroscopic masses have a much higher uncertainty than the evolutionary masses. The one-to-one correlation of the masses falls within this error limit. The study of Galactic B stars by \citet{Nieva2014} gives similar results, where the spectroscopic masses appear to be larger than corresponding evolutionary masses for $M_{\rm spec} > 15 M_{\odot}$.

Moreover, we cannot rule out the possibility of undetected companion(s) in the spectra, which would also affect the mass determinations. Also we know that a significant fraction of the OB stars in clusters or associations are binaries \citep{Sana2013b,Sana2011}. Some O stars studied by \citet{Bouret2013} also showed higher spectroscopic than evolutionary masses, and turned out to be binaries. They concluded that the bias toward higher spectroscopic masses is probably related to the binary status of the objects. However, it seems unlikely that all stars in our sample are biased this way.

\subsection{Present day mass function}
\label{subsect:imf}
\begin{figure}
\centering
\includegraphics[scale=0.56]{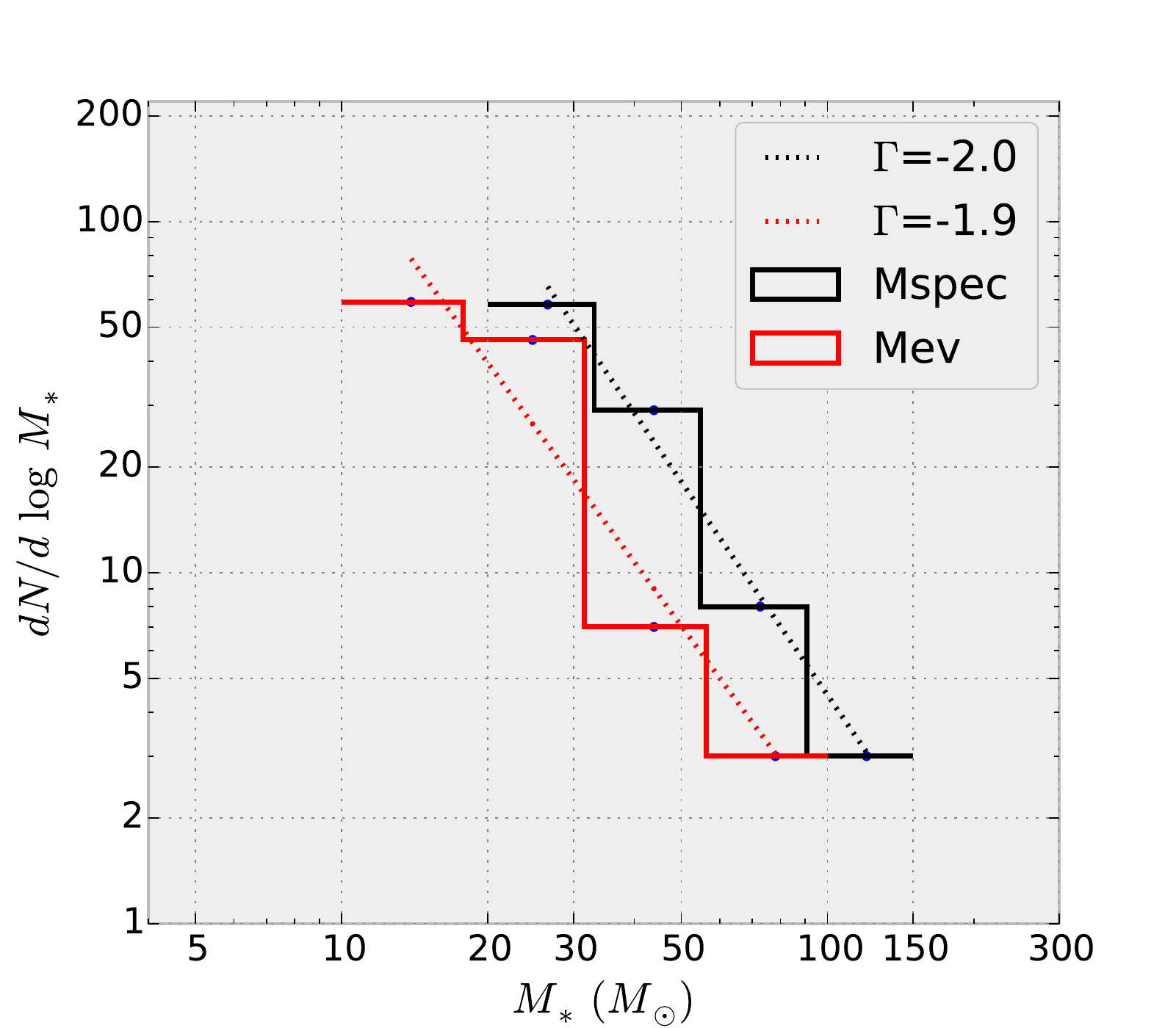}
\includegraphics[scale=0.56]{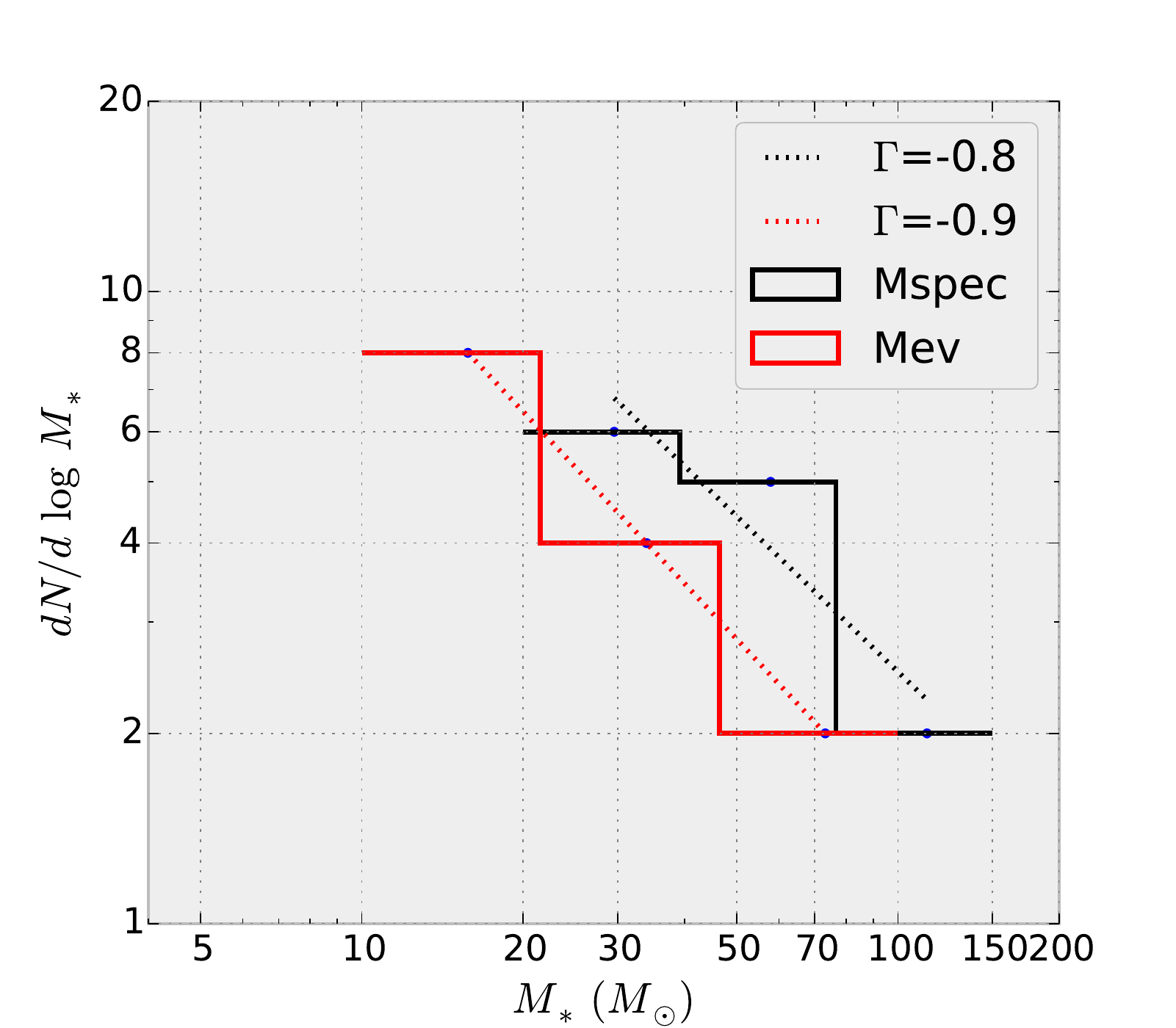}
\caption{Present day mass function of the whole OB stars in the sample (upper panel) and only for the NGC\,2018 cluster (lower panel). The histograms are for spectroscopic (black)  and evolutionary (red) masses, respectively. Power laws are fitted to the mass functions (dashed lines) with their slopes $\Gamma$ as given in the legends. }
\label{fig:imf}
\end{figure}

The spectral analysis of hot OB stars in the N\,206 region gives a good sample to reveal the present day mass function. Figure\,\ref{fig:imf} (upper panel) shows the present day mass function (PDMF) for the massive stars based on both spectroscopic and evolutionary masses. The corresponding power-law fits give a slope of $\Gamma \approx -2.0$ in both cases. Since the sample becomes incomplete at the lowest masses, the power-law fit is restricted to masses above 20\,$M_{\odot}$ in the case of spectroscopic masses, and 10\,$M_{\odot}$ for evolutionary masses. It should be noted that the PDMF plotted here might be biased due to unresolved binaries \citep{Weidne2009}. Moreover, the PDMF discussed here is distinct from the initial mass function (IMF), because stars lose mass over their lives, and some disappear after supernova explosions. Since multiple episodes of star formation have occurred in this region, the number of lower mass stars will increase over time and therefore lead to a steeper slope than in the IMF \citep{Kroupa2013}.

In Figure\,\ref{fig:imf} (lower panel) we plot the PDMF only for the young cluster NGC\,2018. For this subset of stars, the power-law fit gives a very shallow slope of $\approx0.9$. This is substantially flatter than the Salpeter slope of the IMF $\Gamma$ = -1.35, and could indicate a top-heavy IMF for this young cluster. In a recent work, \citet{Schneider2018} found a similarly flat IMF for the 30\,Doradus starburst in the LMC. Moreover, \cite{Marks2012}  
deduced a variation of the IMF as becoming top-heavy with decreasing metallicity.

Thus we find a substantial variation of the mass function for different parts of the N\,206 complex. To quantify this, we calculate the slopes of the PDMF as a function of distance from NGC\,1028. Similar to the analysis in Sect.\,\ref{subsect:sequential}, we divide the region into four annuli and determine the PDMF for each of the different areas. From the central young cluster to the outer parts of the complex, the slopes are -0.9, -1.4, -1.9, and -2.3, respectively. So, the slope of the PDMF gets gradually steeper with distance from the NGC\,2018 cluster. This could indicate mass segregation in the complex.  Similar trends are also observed in other clusters like  NGC\,346 \citep{Sabbi2008}, NGC\,3603 \citep{Harayama2008}, and Westerlund\,2 \citep{Zeidler2017}.

\subsection{Present star formation rate}

The distribution of stellar ages (Fig.\,\ref{fig:histage}) indicates a phase
of vivid star formation over the last 7.5\,Myr. We extract from our
results the spectroscopic masses of all OB stars younger than 7.5\,Myr.
Together with the two WR binaries, their total mass adds to
$2600\,M_\odot$. Neglecting the mass lost by winds and SNe, the mass
that has been assembled per year into massive stars is thus
$2600\,M_\odot / 7.5\,\mathrm{Myr} = 3.6 \times 10^{-4}\,M_\odot
\mathrm{yr}^{-1}$.

Our sample contains the stars with spectroscopic masses above $20\,M_\odot$ (see \ Fig.\,\ref{fig:imf}). For a rough estimate of the mass which went into any stars including those with masses lower than this limit, we adopt a Salpeter IMF and compare the integral from $20$ to $150\,M_\odot$ with the integral starting from a lower mass cut-off of $0.5\,M_\odot$ \citep[e.g.][]{Kroupa2002}. This calculation indicates a correction factor of about six. Thus, the total star formation
rate we obtain is about $2.2 \times 10^{-3}\,M_\odot\,\mathrm{yr}^{-1}$. This number compares very well with the estimate of $2.5 \times 10^{-3}\,M_\odot\,\mathrm{yr}^{-1}$ derived from infrared data on young stellar objects, as reported in the Introduction (Sect.\,\ref{sect:intro}).

\subsection{Stellar feedback in the N\,206 complex}
\label{Subsec:fb}
We quantify the total stellar feedback of the N\,206 complex from our spectroscopic study of its massive star population. The interstellar environment consists roughly of three phases: the cold, neutral, or molecular gas, the ionized \ion{H}{ii} regions, and the shock-heated plasma that emits X-rays. Neglecting mutual interactions, the \ion{H}{ii} regions are powered by the ionizing radiation of stars, while the X-ray plasma is created by the shocked stellar wind and supernova explosions.

The ionizing photon fluxes generated by the massive stars in our sample will be discussed in Sect.\,\ref{Subsubsec:Ifb}. These ionizing fluxes could be used for constructing photoionization models, and their predictions could be compared with the diffuse emission observed in this complex. In fact, we have placed a number of FLAMES fibers on the diffuse background and plan to do such modeling in the future, but this is beyond the scope of the present paper. 

In Sect.\,\ref{Subsubsec:Mfb}, we calculate the total mechanical luminosity produced by the stellar winds of all sample stars as well as the integrated kinetic energy of the winds over the life of the star. In Sect.\,\ref{Subsubsec:SNfb}, we discuss the possible supernova rate in the complex and estimate the input energy. Assuming no interaction between different phases of the ISM, we discuss the energy budget of the superbubble in Sect.\,\ref{subsec:budget} by comparing with the stored energy calculated by \citet{Kavanagh2012}.

\begin{figure}
\includegraphics[scale=0.57]{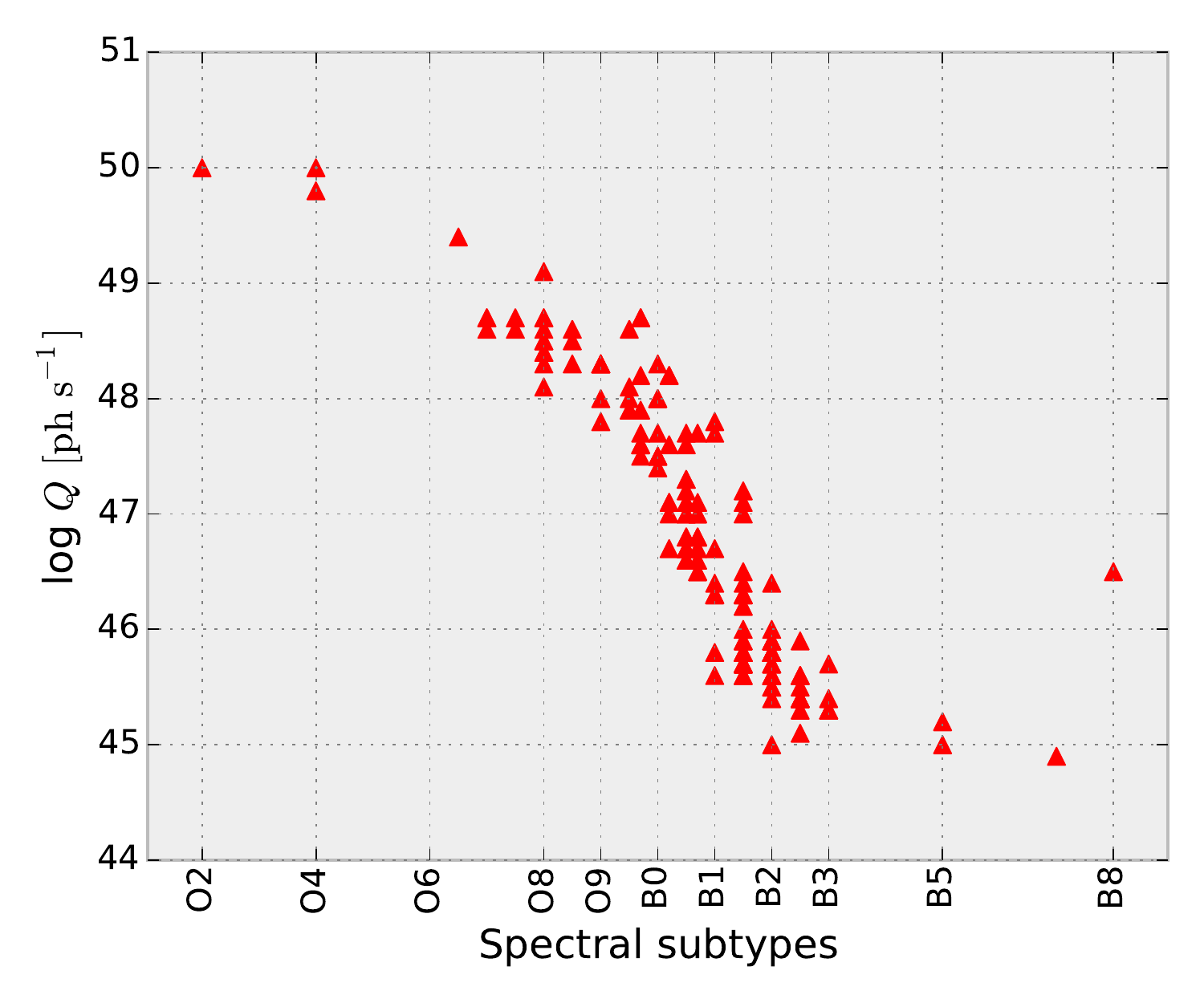}
\includegraphics[scale=0.57]{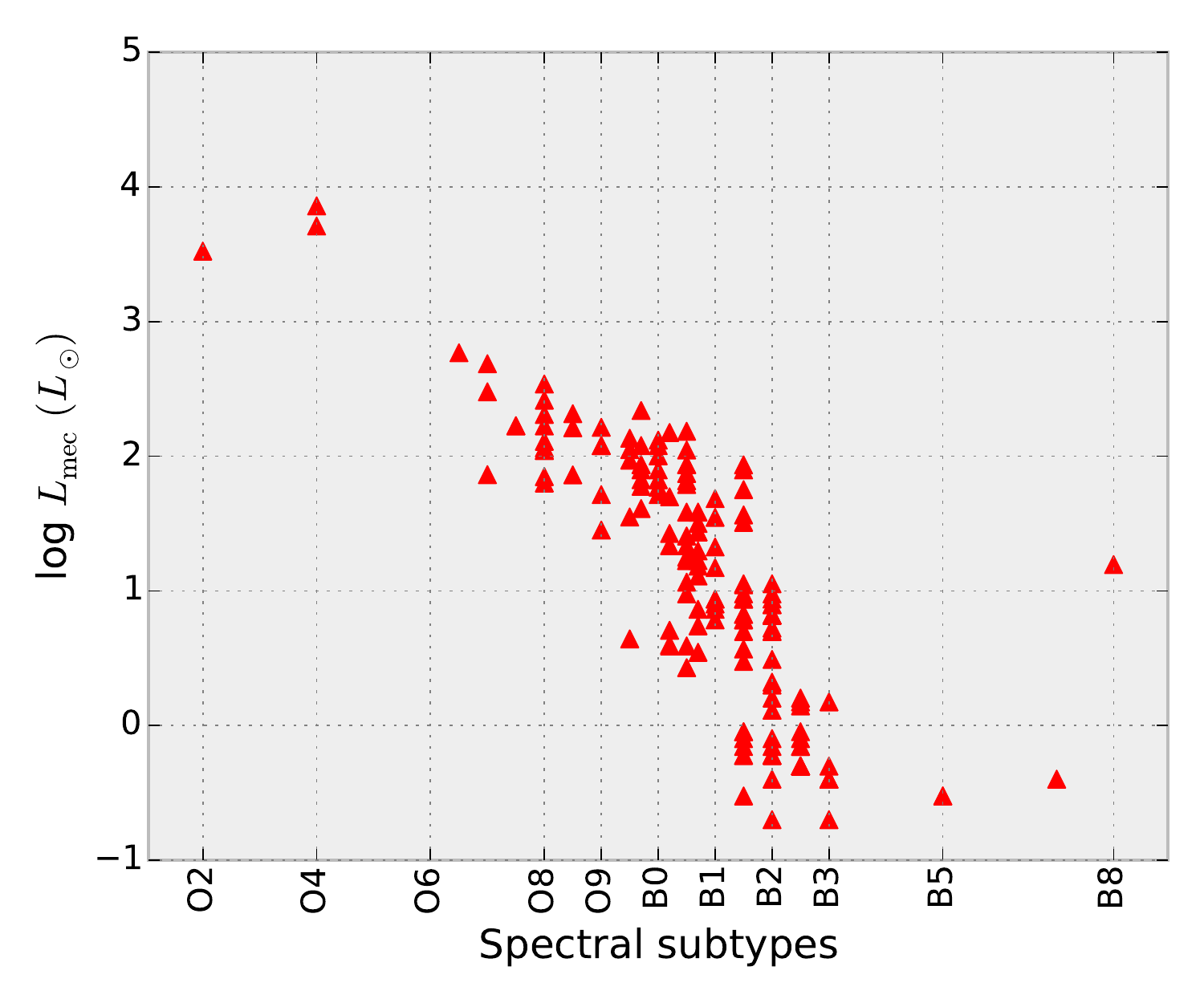}
\caption{Ionizing photon flux (upper panel) and mechanical luminosity (lower panel) as a function of spectral subtypes. }
\label{fig:qlmec}
\end{figure}

\subsubsection{Ionizing feedback from the WR and OB stars}
\label{Subsubsec:Ifb}
The rate of hydrogen ionizing photons ($Q_{0}$) for each OB star predicted by their final model is given in the Table.\,\ref{table:stellarparameters}. This ionizing photon flux strongly varies with the spectral subtypes of the OB stars as shown in Fig.\,\ref{fig:qlmec} (upper panel). The present-day total ionizing photon flux produced by all the OB stars in our sample is $Q_{0} \approx 4 \times 10^{50}\, \rm{s^{-1}}$. More than 65\% of this total ionizing photon flux is contributed by the three very massive Of stars (N206-FS180, 187, and 214).
The ionizing photon flux generated by the two WR binaries in the complex is $Q_{0} \approx 5 \times 10^{49}\, \rm{s^{-1}}$, which contributes about twelve percent to the combined ionizing feedback from WR and OB stars.

The three Of stars are the only stars in our sample that produce a significant number of \ion{He}{ii} ionizing photons. The number rate of such photons with $\lambda < 228$\,\AA\ is about $ \approx 6 \times 10^{45}\, \rm{s^{-1}}$. 
\subsubsection{Mechanical feedback from stellar winds}
\label{Subsubsec:Mfb}
The mechanical luminosities of the OB stars in the N\,206 complex as a function of their spectral subtype are shown in Fig.\,\ref{fig:qlmec} (lower panel). The total mechanical luminosity generated by all sample OB stars is estimated to be $ L_{\rm mec}=\,0.5\,\dot{M}\,\varv_\infty^{2}\approx\,2.34\times10^{4}\,L_{\odot}$. 
Similar to the ionizing photon flux, the total mechanical luminosity is also dominated ($\sim$62\%) by the three very massive Of stars.

\begin{figure}
\includegraphics[scale=0.63]{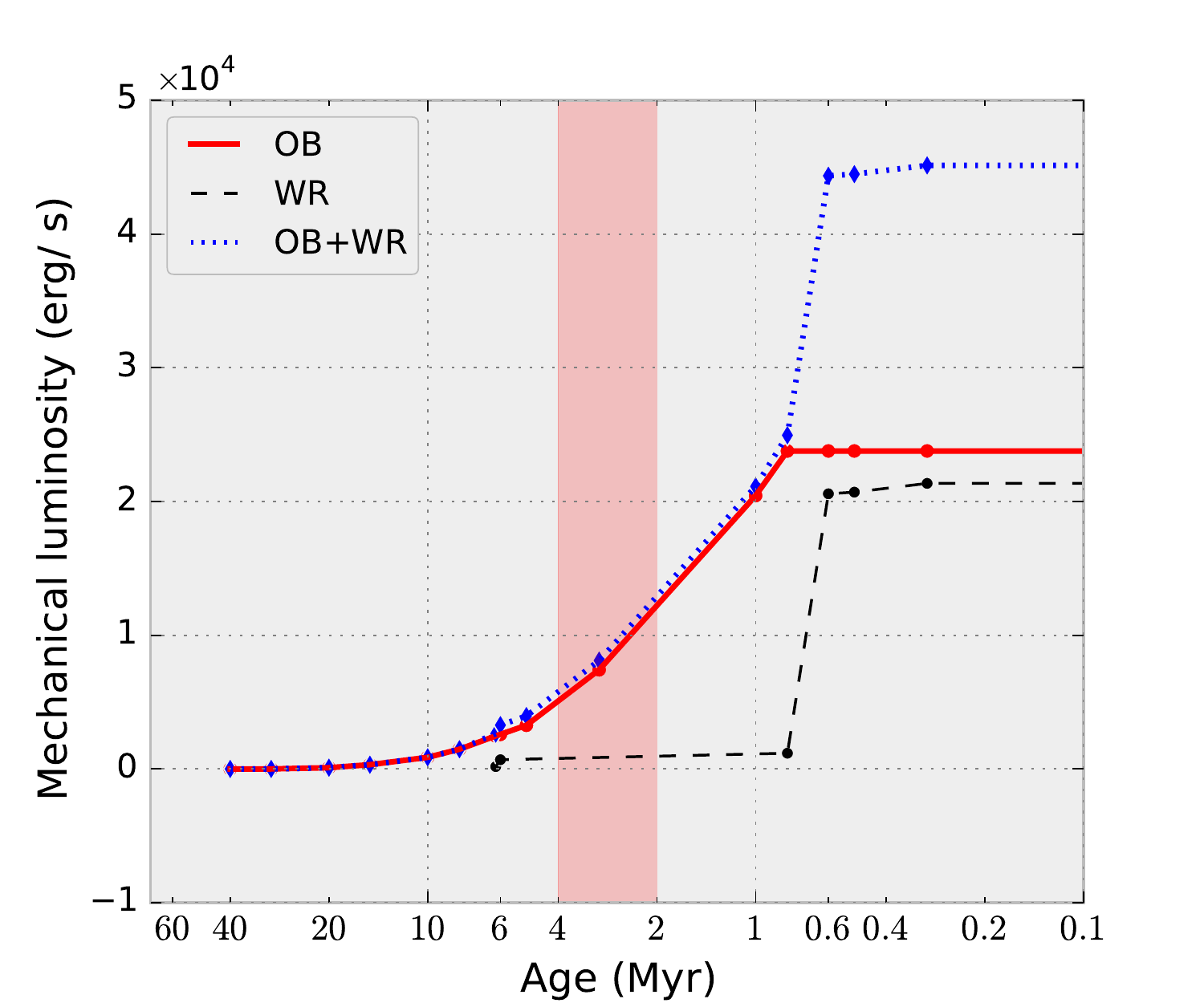}
\caption{Evolution of the mechanical luminosity in the N206\, complex, only considering the present stellar content. The mechanical luminosity from WR and OB stars is shown in black dashed and red solid lines, respectively. The combined mechanical luminosity is represented by the blue dotted line. The red shaded area indicates the time interval in which the superbubble formation has begun.}
\label{fig:lmecev}
\end{figure}

Since we know the ages (see Sect.\,\ref{subsect:sequential}) and the mechanical luminosity (see Table\,\ref{table:stellarparameters}) of the individual OB stars, we can construct a diagram for the evolution of the mechanical luminosity in the entire complex with time (see Fig.\,\ref{fig:lmecev}). 
The red line illustrates the total mechanical luminosity of all OB stars at a particular time. Here we assume that all our sample OB stars had a constant mechanical output since their formation. Since most of our sample OB stars are still in their hydrogen burning phase (before reaching the terminal age main sequence) as evident from the HRD, this assumption is justified for a rough estimate. Therefore, if only accounting for those OB stars which still exist today, the total mechanical luminosity would have increased over the past five Myr. Of course, this curve might change a lot due to previous generations of massive stars which are no longer present today.

In order to calculate the total mechanical energy released by all presently existing OB stars via stellar winds throughout their life, we must integrate their mechanical luminosity over their age. We obtained a total mechanical feedback from our sample OB stars of $E_{\rm mec} =\,0.5\,\dot{M}\,\varv_\infty^{2} t \,\approx 1.1 \times 10^{52}\,\rm{erg} $. 
Even if we are considering only the past five Myr, the accumulated mechanical feedback is $E_{\rm{mec}}\approx9.5\times10^{51}\, \rm{erg} $. However, these estimates of mechanical feedback only consider those massive stars that are still present in the N\,206 complex today, and does not account for those massive stars which have already disappeared. Therefore, these estimates are only lower limits to the stellar wind feedback. 

\begin{table}
\caption{Feedback from the WR binaries} 
\label{table:WRfeedback}     
\centering
\begin{tabular}{clcc}
\hline
\hline

\noalign{\vspace{1mm}}
& & N206-FS45 &N206-FS128 \\
\noalign{\vspace{1mm}}
\hline
\noalign{\vspace{1mm}}
\multirow{5}{=12mm}{pre-WR phase}& $\log \dot{M}$ [$M _{\odot}\,\mathrm{yr}^{-1} $]& -6.5&-6.1 \\
& $\varv_\infty $ [km\,s$^{-1}$]&2600&2800\\
&$L_{\rm mec}$ [$L _{\odot}$]&177 &514\\
& $t$ [Myr]&6.2 & 5.4\\
&$E_{\rm mec}$ [$10^{50}$erg]& 1.3 &3.3 \\
\hline
\noalign{\vspace{1mm}}
\multirow{5}{=12mm} {WR phase}& $\log \dot{M}$ [$M _{\odot}\,\mathrm{yr}^{-1} $]&-5.4 &-4.67  \\
& $\varv_\infty $ [km\,s$^{-1}$]&1700  &3400\\
&$L_{\rm mec}$ [$L _{\odot}$]& 950&20400\\
& $t$ [Myr]&0.3 &0.6\\
&$E_{\rm mec}$ [$10^{50}$erg]& 0.34 &14.75\\
\hline
\end{tabular}
\end{table}

The WR stars make a significant contribution to the mechanical feedback of the region. Here we only discuss the feedback from the primary (WR) component, since the contribution from the secondary component (late O star) is comparatively much smaller. The details of the feedback parameters of the WR stars N206-FS\,45 and 128 are given in Table\,\ref{table:WRfeedback}.

The evolutionary tracks predict (see Sect.\,\ref{subsect:wchrd}) the initial parameters of the star N206-FS\,128a at the ZAMS as $T_\ast=39$\,kK, $\log\,g_\ast=4.1$, $\log\,(L/L_{\odot}) = 5.2$,  $\log\dot{M} \sim -6.1\,[M_{\odot}$\,yr$^{-1}]$, and $\varv_\infty \sim 2800$\,km\,s$^{-1}$. Here,  $\varv_\infty $ is calculated theoretically from $\varv_{\rm esc}$. The mechanical energy produced by this star is about $3.3 \times 10^{50}\, \rm{erg} $ over the first 5.4\,Myr. The star then stayed in the WR stage for the last 0.6\,Myr, imparting a wind energy of approximately $1.5 \times 10^{51}\, \rm{erg} $ to the surroundings. Thus, the mechanical feedback from N206-FS\,128 throughout its life is $E_{\rm mec} \approx 1.8 \times 10^{51}\,\rm{erg} $.

The feedback contribution from the WN component of the binary N206-FS\,45 has been estimated by Shenar et al. (in prep.) on the basis of a spectral analysis.
The current period of the system is 36.9 days \citep{Foellmi2003b}. For the derived parameters (see Table\,\ref{table:stellarparametersWR}), the binary evolutionary tracks by \citet{Eldridge2009} suggest an age of $\approx$ 6.5\,Myr and an initial mass of the WN component of $30\, M_{\odot}$. The star spent $\approx$ 6.2\,Myr in the hydrogen burning phase. The mechanical luminosity  estimated for this time period is $\approx 177\, L_{\odot}$, and the integrated mechanical energy amounts to $ E_{\rm{mec}}\approx 1.3 \times 10^{50}\, \rm{erg} $. The star has stayed in the WR stage over the last 0.3\,Myr. The mechanical energy released by the WR wind during this time period is $\approx 3.4\times 10^{49}\, \rm{erg} $. Therefore, the model predicts the accumulated mechanical energy output of this star in both stages to be $ E_{\rm{mec}}\approx 1.6\times 10^{50}\, \rm{erg} $.

Table\,\ref{table:WRfeedback} summarizes the feedback from N206-FS\,45 and N206-FS\,128 in both the pre-WR and WR phase. Now we add these contributions to the mechanical luminosity over time plotted in (Fig.\,\ref{fig:lmecev}). From the diagram, we can see that the two WR stars  contribute to the feedback about as much as the whole sample of hundreds of OB stars. The total mechanical luminosity  is about $ L_{\rm{mec}} = 4.5 \times 10^{4}\, L_{\odot} = 1.7 \times 10^{38}\, \rm{erg \,s^{-1}}$.

The integrated mechanical energy over the lifetime of both WR stars in our sample is $\approx 2 \times 10^{51} \,\rm{erg} $. So, the total accumulated mechanical feedback over the lifetimes of the current WR and OB stars becomes $ E_{\rm{mec}}\approx 1.3 \times 10^{52}\, \rm{erg} $. Even if we consider only the past five Myr, the total integrated mechanical feedback is $\approx 1.2 \times 10^{52}\, \rm{erg}$.

From the size and expansion velocity of the X-ray superbubble, \citet{Dunne2001} suggest an age of $ \sim $2\,Myr. According to \citet{Kavanagh2012}, the expansion of the superbubble must have undergone induced acceleration in its history, and therefore it is likely to be older than 2\,Myr. The HRD of the stars in the X-ray superbubble region (Fig.\,\ref{fig:hrd_cl}, lower right) reveals many O stars within the age range $4-5$\,Myr. Therefore we speculate that the age of the superbubble is about $2-4$\,Myr. This time interval is marked in Fig.\,\ref{fig:lmecev} by a shaded area. As seen in Fig.\,\ref{fig:histage}, the star formation in the N\,206 complex peaked $3-5$\,Myr ago. In this period, the stellar winds from this population must have caused a rapid increase in the overall mechanical luminosity. 

Regarding the times more than five Myr ago, it should be noted that we are not aware of the full star formation history. There might have been other time intervals of peak star formation. However, considering only the present stellar content, $ L_{\rm{mec}}$ was negligible before 5\,Myr, then it started to grow and led to the formation of the superbubble. The huge increase in $ L_{\rm{mec}}$ over the last 0.6\,Myr is due to the two WR stars.

\subsubsection{Feedback from supernovae}
\label{Subsubsec:SNfb}
A single supernova (SN) injects typically $10^{51}$\,erg of mechanical energy into the ISM \citep{Woosley1995}. Therefore, for our study we need to estimate the number of supernovae that have already occurred in the N\,206 complex. 

Currently, there is only one SNR situated in this complex, which suggests that a SN that occurred within the last few ten thousand years. Older SNRs would have already faded below the observational limit. 

\begin{figure}
\centering
\includegraphics[scale=0.44]{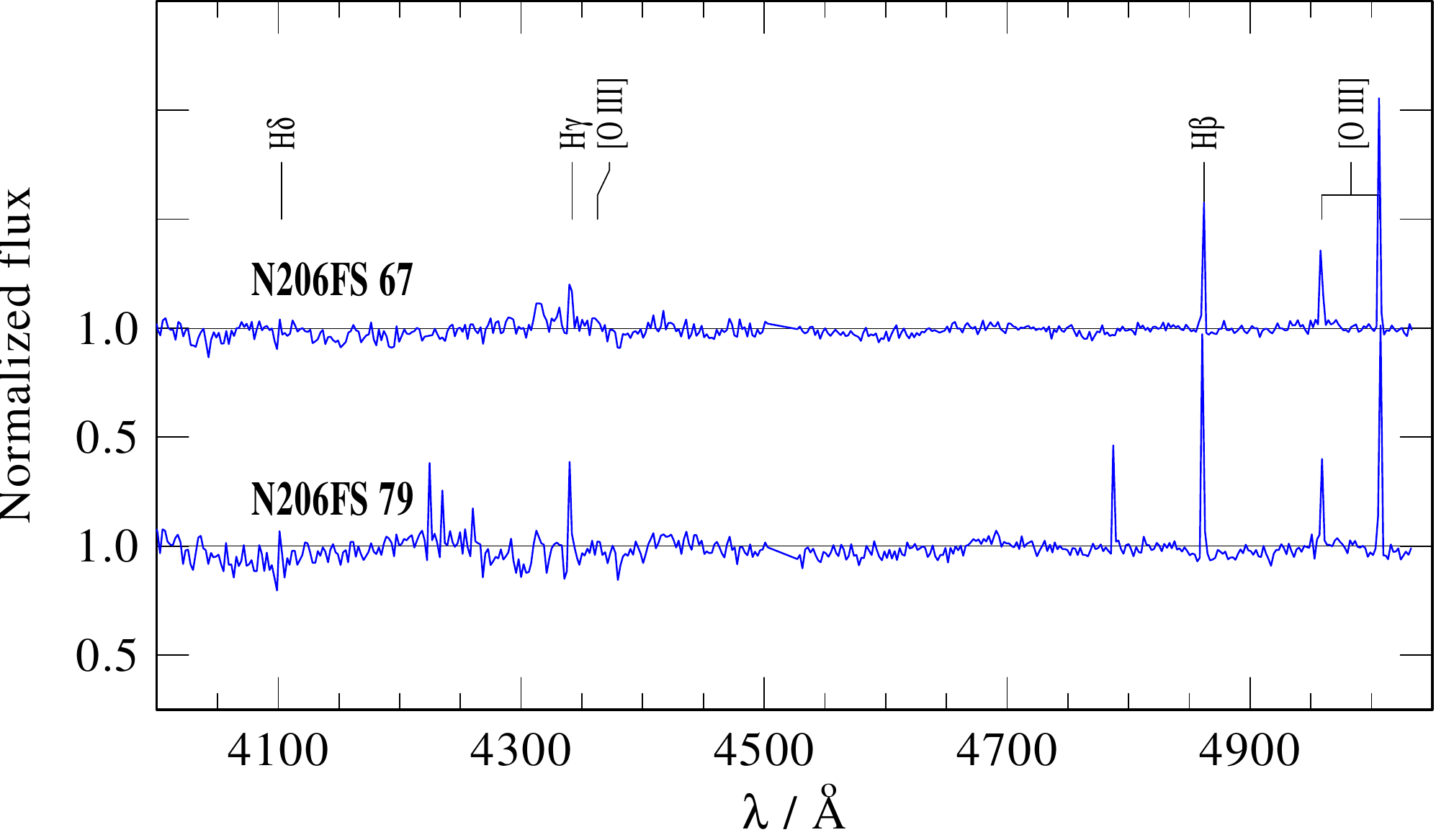}
\caption{Normalized VLT-FLAMES LR02+LR03 spectra taken at the positions of the X-ray sources XMMU J052952.9-705850 (upper panel) and USNO-B1.0 0189-00079195 (lower panel)}
\label{fig:spec_XB}
\end{figure}

Indirect evidence for previous SNe could come from the detection of X-ray binaries. The formation of their compact component, a black hole or neutron star, might have been accompanied by a SN explosion. 
Previous studies by  \citet{Shtykovskiy2005} and \citet{Kavanagh2012} have identified two HMXB candidates, namely XMMU J052952.9-705850 and USNO-B1.0 0189-00079195, by the X-ray analysis of the region using \textit{XMM-Newton} telescope data. In order to check  the nature of these objects, we have taken VLT-FLAMES spectra at the positions of both X-ray sources (Fig.\,\ref{fig:spec_XB}). In both cases, no stellar features can be detected, neither lines nor a continuum. The data show only nebular emission from the diffuse background in both cases. Hence we do not support the identification of these X-ray sources with HMXBs. Our conclusion is supported by the Million Optical - Radio/X-ray Associations (MORX) Catalog \citep{Flesch2016}, which also suggests that these X-ray sources are most probably contaminations from background quasars or external galaxies.

Further indirect evidence for previous SN occurrences is given by OB runaways. These stars might have been ejected from a binary system, when the SN explosion of the primary disrupted the system. In Sect.\,\ref{sect:sparameters} we identified eight runaway candidates with radial velocities in excess of $\pm$48\,km\,s$^{-1}$. Interestingly, most of these stars are located inside or at the periphery of the X-ray superbubble as shown in Fig.\,\ref{fig:run}.

Considering the evolutionary masses of the OB stars, there are about 36 stars in the $20-40\,M_{\odot}$  bins  and five stars with masses $\geq40\,M_{\odot}$, which must become SNe within the next ten and five Myr, respectively. Additionally, there are two WR binaries in the complex. These numbers are higher than predicted by \citet{Kavanagh2012} from the IMF. If the SFR was roughly constant in time, there should have been a similar number of massive stars in the past.

In \citet{Ferrand2010}, the rate of supernovae in massive clusters or OB associations is expressed to first approximation as
\begin{equation}
\frac{{\rm d}n}{{\rm d}t_{\rm SN}} \approx \frac{N_{*}}{\Delta t^{*}_{\rm OB}}.
\end{equation}
Here $\Delta t^{*}_{\rm OB} = t_{\rm SN}(m_{\rm min}) - t_{\rm SN}(m_{\rm max})$ is the active lifetime of the cluster, where $m_{\rm min}$ and $m_{\rm max}$ are the minimum and maximum initial mass of the stars in the region that explode as SNe.  The symbol $N_{*}$ is the number of stars with masses between $m_{\rm min}$ and $m_{\rm max}$.
Regarding the SN rate in the N\,206 complex, we can take $m_{\rm min} = 8\,M_{\odot}$ and $m_{\rm max} = 100\, M_{\odot}$. This gives $t_{\rm SN}(m_{\rm min}) \approx 37$\,Myr and $t_{\rm SN}(m_{\rm max}) \approx 3.5$\,Myr.  From the HRD, 162 stars have $M_{\rm ini} \geq 8\, M_{\odot}$ (including binary companions). So, the supernova rate is $\approx$ 4.8 per Myr or one SN per 210,000 years. Even if we consider only stars with initial masses below $40\,M_{\odot}$, because more massive stars might collapse silently \citep{Heger2003}, the rate is $\approx$ five supernovae per Myr. This number is in statistical agreement with the presence of just one SNR today. 

If we do the same calculation only for massive stars within or close to the X-ray superbubble (including stars in NGC\,2018, LH\,66, and LH\,69), the estimated SN rate is $\approx$2.2 per Myr. So, if we consider only the past five Myr and massive stars close to the X-ray superbubble, the accumulated mechanical input by supernova explosions is estimated to be $ E_{\rm{mec}}\approx 1.1 \times 10^{52} \,\rm{erg} $, and hence contributes approximately the same amount to the mechanical feedback as the massive star winds. 

\subsubsection{Superbubble energy budget}
\label{subsec:budget}

We compare our energy feedback results with the X-ray analysis of the superbubble by \citet{Kavanagh2012}. Using XMM-\textit{Newton} data, they estimated the X-ray luminosity of the N\,206 superbubble as $\approx 7 \times 10^{35}$\,erg\,s$^{-1}$. \changed{In young star clusters, the unresolved X-ray emission from the population of active pre-main sequence stars could be significant \citep{Nayakshin2005,Wang2006}, also at sub-solar metallicities \citep{oskinova2013}.
\citet{Kavanagh2012} estimated the contribution from the unresolved low-mass population to the observed X-ray emission as $\sim 6 $\%. Moreover, the distribution of YSO candidates \citep{Carlson2012} shows that they are not concentrated in the X-ray superbubble area. }

\changed{ To estimate the contribution of young active stars of low mass to the extended X-ray emission in N\,206, we use the scaling provided in \citet{Oskinova2005}. Their figure\,5 compares the relative contributions to the X-ray emission from the hot cluster wind with the emission from the unresolved population of low-mass stars. As can be seen in this figure, in clusters younger than $\sim 2.5$\,Myr (assuming an instantaneous burst of star formation), the collective X-ray luminosity of low-mass stars dominates the cluster X-ray luminosity. However, for older clusters, the input of mechanical energy from massive stars dominantly powers the diffuse X-ray emission. We estimated the age of the X-ray superbubble in N\,206 as higher than 2\,Myr (see Fig.\,\ref{fig:lmecev}). Moreover, we found that star formation has been ongoing in this region for a longer time. Hence, it is most likely that the observed soft diffuse X-ray emission is dominated by the mechanical energy input from massive stars.}

The observed X-ray luminosity is only 0.4\% of the combined mechanical luminosity of the WR and OB stellar winds (see Sect.\,\ref{Subsubsec:Mfb}). \changed{The uncertainty in $ L_{\rm{mec}}$ is about 30\% reflecting the error margins of the stellar wind parameters ($\dot{M}, \varv_\infty$).}
Previous studies on other massive star forming regions such as the Omega Nebula, the Rosette Nebula \citep{Townsley2003}, 30 Doradus \citep{Lopez2010,Doran2013}, and the Carina Nebula \citep{Smith2006,Townsley2011} also revealed similar trends, where only a small portion of the wind energy is converted into X-ray emission. \citet{Rosen2014} published a detailed study about missing stellar wind energy from massive star clusters. They suggest that turbulent mixing at the hot-cold interface or physical leakage of the shock heated gas can efficiently remove the kinetic energy injected by the massive star winds and supernovae.

 \citet{Kavanagh2012} also estimated the total energy stored in the superbubble in the form of thermal energy of the shock-heated gas and the kinetic energies of the expanding H$\alpha$ and \ion{H}{i} shells to be $(4.7 \pm 1.3)\,\times 10^{51}$\,erg. The feedback calculations discussed in the previous sections reveal a much higher mechanical energy input by the massive stars to the N\,206 complex. Adding the mechanical energy input during the last five Myr from the OB and WR star winds (Sect.\,\ref{Subsubsec:Mfb}) and the SNe (Sect.\,\ref{Subsubsec:SNfb}), we obtain $\approx 2.3 \times 10^{52}$\,erg, which is almost five times higher than the stored energy content observed by \citet{Kavanagh2012}. Even more, the feedback estimates are only a lower limit, as discussed in Sect.\,\ref{Subsubsec:Mfb}. \changed{With the SN rate estimated in Sect.\,\ref{Subsubsec:SNfb}, about 25 supernovae might have occurred in the entire complex during the past five Myr. Also, there would be a contribution from stellar winds of their progenitors. Considering these numbers, the total $ E_{\rm{mec}}$ can increase by a factor of roughly two. However, this does not include the contribution from stars which already died as SNe between 5 to 50\,Myr ago (Fig.\,\ref{fig:lmecev}). According to the SNe estimation given in Sect.\,\ref{Subsubsec:SNfb}, about 250 stars might have already exploded as SNe. Here we assumed a constant star formation rate over the past 50\,Myr, as high as the current rate. During their evolutionary stages, stellar winds (OB/WR) of these stars would have provided a significant amount of mechanical feedback. Supernovae and stellar winds together might have released about 15 times more mechanical energy than our estimated value of $ 2.3 \times 10^{52}$\,erg.}  As we have seen above, the radiative cooling by X-ray emission is negligible. Hence, this comparison shows that most of the mechanical energy input has been consumed by other processes or escaped from the X-ray superbubble.

\begin{figure}
 \centering
 \includegraphics[scale=0.33,trim={0cm 0cm 0cm 0}]{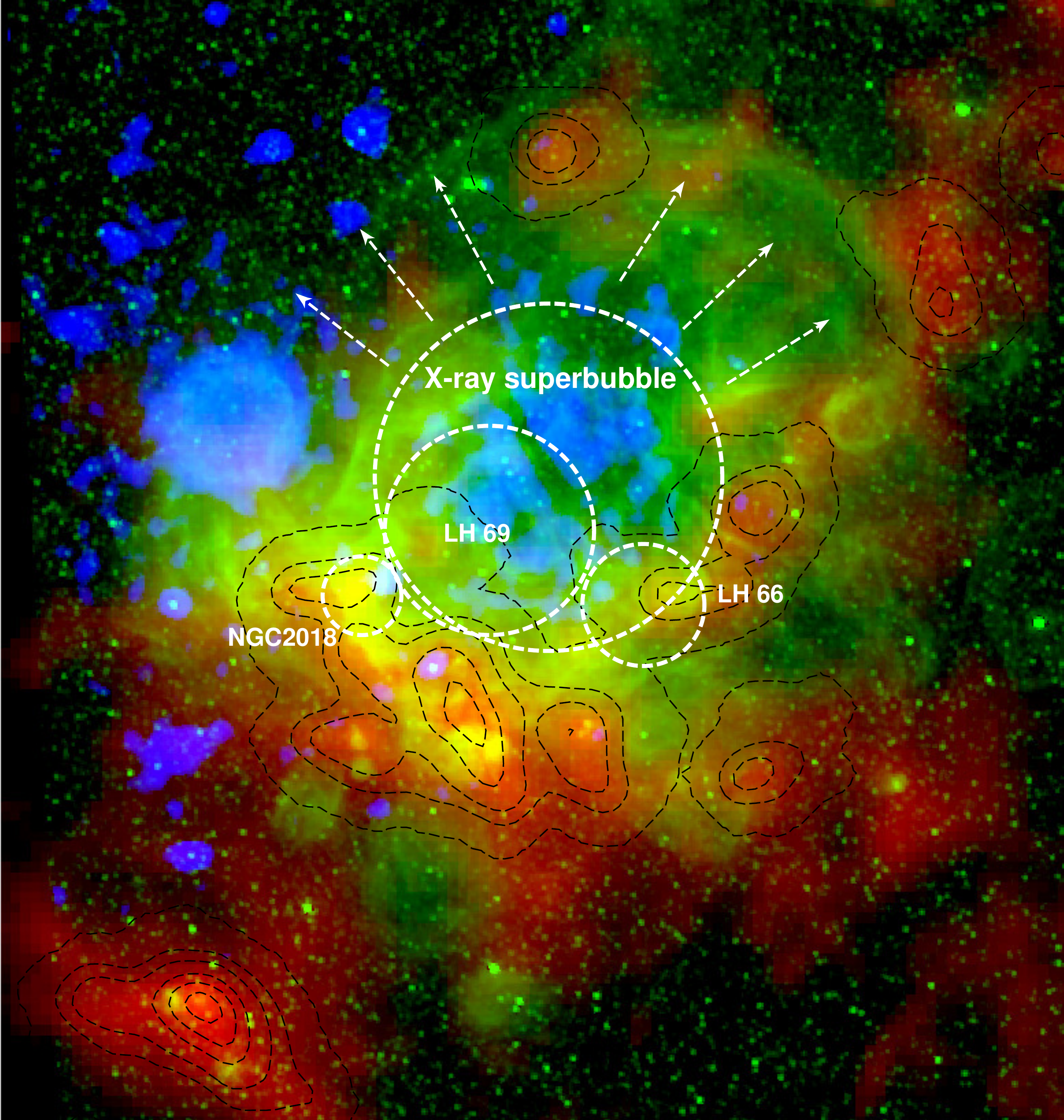}
 \caption{Color composite image of the N\,206 complex, constructed from the Herschel spire data (red: 500\,$\mu m$), H$\alpha$ \citep[green: MCELS,][]{Smith2005}, and X-rays (blue: XMM-\textit{Newton} $0.3-1$\,keV). CO intensity contours taken from the MAGMA survey are over-plotted to the image. The arrows suggest the possible leakage of hot gas from the X-ray superbubble.
 } 
 \label{fig:CO}
 \end{figure}

The hot X-ray emitting gas could not escape, if it were fully confined by surrounding cold and dense molecular material. Figure \ref{fig:CO} shows a multiwavelength image of the N\,206 complex at far-infrared, optical, and X-ray wavelengths, over-plotted with CO intensity contours. The CO intensity map taken from the Magellanic Mopra Assessment (MAGMA) \citep{Wong2011} traces the dense molecular gas. This map reveals that the X-ray superbubble is only partially confined by the dense molecular cloud. In the 2D projection, it seems that the  X-ray emitting gas leaks out in northern directions. 
Figure\,\ref{fig:CO} also gives a general picture of ongoing star formation in this region. When comparing this with Fig.\,\ref{fig:age_distr}, we can see that now the star formation is presently progressing at the rim of the cloud. This could be an indication of a triggered star formation in the complex by the expansion of the superbubble when it hits the rim of the cloud. The young cluster NGC\,2018 is also located in this dense part of the molecular cloud. At the other side of the superbubble, the cold  molecular gas is already removed, which might be due to the energetic winds of massive stars or SNe that occurred in the past. So, mostly older stars are left there, and there is not enough material available for further star formation.

\section{Summary and conclusions}
\label{sect:summary}

We have identified and analyzed the OB stars and WR binaries in the N\,206 complex located in the LMC. The stellar and wind parameters of the individual stars are determined using PoWR model atmospheres. We also present a detailed binary analysis of the WC4+O9 system, and discuss its evolutionary status.  
Our main conclusions are:

\begin{itemize}
\item Approximately 12\% of the whole sample are found to be emission-line stars with a disk. However, our analysis shows that the projected rotational velocities of most of these Oe/Be stars are far below their critical velocity.
\item From the radial velocity distribution of the OB stars, we suspect eight stars to be runaways.
\item The wind momentum-luminosity relation of most of the sample OB stars is consistent with expectations within uncertainties.

\item The HRD of our OB star sample in the complex reveals multiple populations of different ages. Moreover, each of the clusters and OB associations inside the region also shows large age dispersions, indicating multiple star formation events.

\item The current stellar masses as derived from evolutionary tracks are systematically lower than the masses that we obtain from our spectroscopic analyses. This discrepancy is higher for main sequence stars than for evolved stars.

\item We find indications of sequential star formation in the region. The youngest stars are mostly distributed inside the young clusters and the inner part of the complex, whereas outer parts contain slightly older populations.

\item The current star formation is taking place at the rim of the superbubble, where it hits the molecular cloud.

\item The present day mass function of the entire region shows a steeper slope than the Salpeter IMF. However, the cluster NGC\,2018 shows a top-heavy mass function.

\item We derive the current star formation rate of the complex as $2.2 \times 10^{-3} M_\odot\,\mathrm{yr}^{-1}$.

\item The total ionizing photon flux produced by all massive stars in the N\,206 complex is $Q_{0} \approx 5 \times 10^{50}\, \rm{s^{-1}}$.

 \item Three very massive Of stars out of 164 OB stars contribute $>$60\%  of the total ionizing photon flux and  the stellar wind mechanical luminosity.

\item The integrated mechanical energy feedback from all OB stars over the past five Myr is estimated to be  $\approx 1.3 \times 10^{52}\, \rm{erg}$. This is more than the energy that is found to be stored in the superbubble.

\item The two WR binaries contribute about as much mechanical luminosity as all the 164 OB stars in the sample.

\item The rate of supernova explosions in the complex is estimated to be about five per Myr.
\item The X-ray luminosity of the N\,206 superbubble is only 0.4\% of the current mechanical luminosity from the stellar winds of the OB + WR stars.
\item The energy input accumulated over time from the OB stars, WR stars, and SNe is found to be more than five times higher than the energy stored in the X-ray superbubble.

\item The morphology of the complex, with the X-ray superbubble not being fully enclosed by the cold molecular gas, gives the impression that the hot gas has the possibility to escape from the complex.

\item The fact that a few of the most massive stars dominate the energetics of the N\,206 complex would not have been revealed without spectroscopic information.
\end{itemize}
\begin{acknowledgements}
We wish to thank the referee,  Daniel Q.  Wang, for very constructive suggestions to improve the paper. Based on observations collected at the European Organization for
Astronomical Research in the Southern Hemisphere
under ESO program 096.D-0218(A) (P.I.:  L. Oskinova).
V.R. is grateful for financial support from Deutscher Akademischer Austauschdienst (DAAD), as a part of Graduate School Scholarship Program. LMO acknowledges support by the DLR grant 50 OR 1508.  A.A.C.S. is supported by the Deutsche Forschungsgemeinschaft (DFG) under grant HA 1455/26. T.S. acknowledges support from the German ``Verbundforschung'' (DLR) grant 50 OR 1612. This research made use of the VizieR catalog access tool, CDS,  
Strasbourg, France. The original description of the VizieR service was  
published in A\&AS 143, 23. Some data presented in this paper were retrieved from 
the Mikulski Archive for Space Telescopes (MAST). STScI is operated by the 
Association of Universities for Research in Astronomy, Inc., under NASA contract 
NAS5-26555. Support for MAST for non-HST data is provided by the NASA Office of 
Space Science via grant NNX09AF08G and by other grants and contracts.
\end{acknowledgements}

\bibliographystyle{aa}
\bibliography{paper}

\begin{thebibliography}{112}
\expandafter\ifx\csname natexlab\endcsname\relax\def\natexlab#1{#1}\fi

\bibitem[{{Bartzakos} {et~al.}(2001){Bartzakos}, {Moffat}, \&
  {Niemela}}]{Bartzakos2001}
{Bartzakos}, P., {Moffat}, A.~F.~J., \& {Niemela}, V.~S. 2001, \mnras, 324, 18

\bibitem[{{Blaauw}(1961)}]{Blaauw1961}
{Blaauw}, A. 1961, \bain, 15, 265

\bibitem[{{Bonanos} {et~al.}(2009){Bonanos}, {Massa}, {Sewilo}, {Lennon},
  {Panagia}, {Smith}, {Meixner}, {Babler}, {Bracker}, {Meade}, {Gordon},
  {Hora}, {Indebetouw}, \& {Whitney}}]{Bonanos2009}
{Bonanos}, A.~Z., {Massa}, D.~L., {Sewilo}, M., {et~al.} 2009, \aj, 138, 1003

\bibitem[{{Bouret} {et~al.}(2013){Bouret}, {Lanz}, {Martins}, {Marcolino},
  {Hillier}, {Depagne}, \& {Hubeny}}]{Bouret2013}
{Bouret}, J.-C., {Lanz}, T., {Martins}, F., {et~al.} 2013, \aap, 555, A1

\bibitem[{{Brott} {et~al.}(2011){Brott}, {de Mink}, {Cantiello}, {Langer}, {de
  Koter}, {Evans}, {Hunter}, {Trundle}, \& {Vink}}]{Brott2011}
{Brott}, I., {de Mink}, S.~E., {Cantiello}, M., {et~al.} 2011, \aap, 530, A115

\bibitem[{{Carlson} {et~al.}(2012){Carlson}, {Sewi{\l}o}, {Meixner}, {Romita},
  \& {Lawton}}]{Carlson2012}
{Carlson}, L.~R., {Sewi{\l}o}, M., {Meixner}, M., {Romita}, K.~A., \& {Lawton},
  B. 2012, \aap, 542, A66

\bibitem[{{Castor} {et~al.}(1975){Castor}, {Abbott}, \& {Klein}}]{CAK1975}
{Castor}, J.~I., {Abbott}, D.~C., \& {Klein}, R.~I. 1975, \apj, 195, 157

\bibitem[{{Chen} {et~al.}(2007){Chen}, {Chu}, {Gruendl}, \&
  {Heitsch}}]{Chen2007}
{Chen}, C.-H.~R., {Chu}, Y.-H., {Gruendl}, R.~A., \& {Heitsch}, F. 2007, in IAU
  Symposium, Vol. 237, Triggered Star Formation in a Turbulent ISM, ed. B.~G.
  {Elmegreen} \& J.~{Palous}, 401--401

\bibitem[{{Chu}(2008)}]{Chu2008}
{Chu}, Y.-H. 2008, in IAU Symposium, Vol. 250, Massive Stars as Cosmic Engines,
  ed. F.~{Bresolin}, P.~A. {Crowther}, \& J.~{Puls}, 341--354

\bibitem[{{Cranmer}(2005)}]{Cranmer2005}
{Cranmer}, S.~R. 2005, \apj, 634, 585

\bibitem[{{Crowther} {et~al.}(2002){Crowther}, {Dessart}, {Hillier}, {Abbott},
  \& {Fullerton}}]{Crowther2002}
{Crowther}, P.~A., {Dessart}, L., {Hillier}, D.~J., {Abbott}, J.~B., \&
  {Fullerton}, A.~W. 2002, \aap, 392, 653

\bibitem[{{Cutri} {et~al.}(2012){Cutri}, {Skrutskie}, {van Dyk}, {Beichman},
  {Carpenter}, {Chester}, {Cambresy}, {Evans}, {Fowler}, {Gizis}, {Howard},
  {Huchra}, {Jarrett}, {Kopan}, {Kirkpatrick}, {Light}, {Marsh}, {McCallon},
  {Schneider}, {Stiening}, {Sykes}, {Weinberg}, {Wheaton}, {Wheelock}, \&
  {Zacharias}}]{Cutri2012}
{Cutri}, R.~M., {Skrutskie}, M.~F., {van Dyk}, S., {et~al.} 2012, VizieR Online
  Data Catalog, 2281, 0

\bibitem[{{Doran} {et~al.}(2013){Doran}, {Crowther}, {de Koter}, {Evans},
  {McEvoy}, {Walborn}, {Bastian}, {Bestenlehner}, {Grafener}, {Herrero},
  {Kohler}, {Maiz Apellaniz}, {Najarro}, {Puls}, {Sana}, {Schneider}, {Taylor},
  {van Loon}, \& {Vink}}]{Doran2013}
{Doran}, E.~I., {Crowther}, P.~A., {de Koter}, A., {et~al.} 2013, ArXiv
  e-prints

\bibitem[{{Dunne} {et~al.}(2001){Dunne}, {Points}, \& {Chu}}]{Dunne2001}
{Dunne}, B.~C., {Points}, S.~D., \& {Chu}, Y.-H. 2001, \apjs, 136, 119

\bibitem[{{Eldridge} {et~al.}(2008){Eldridge}, {Izzard}, \&
  {Tout}}]{Eldridge2008}
{Eldridge}, J.~J., {Izzard}, R.~G., \& {Tout}, C.~A. 2008, \mnras, 384, 1109

\bibitem[{{Eldridge} \& {Stanway}(2009)}]{Eldridge2009}
{Eldridge}, J.~J. \& {Stanway}, E.~R. 2009, \mnras, 400, 1019

\bibitem[{{Evans} {et~al.}(2015){Evans}, {Kennedy}, {Dufton}, {Howarth},
  {Walborn}, {Markova}, {Clark}, {de Mink}, {de Koter}, {Dunstall},
  {H{\'e}nault-Brunet}, {Ma{\'{\i}}z Apell{\'a}niz}, {McEvoy}, {Sana},
  {Sim{\'o}n-D{\'{\i}}az}, {Taylor}, \& {Vink}}]{Evans2015}
{Evans}, C.~J., {Kennedy}, M.~B., {Dufton}, P.~L., {et~al.} 2015, \aap, 574,
  A13

\bibitem[{{Ferrand} \& {Marcowith}(2010)}]{Ferrand2010}
{Ferrand}, G. \& {Marcowith}, A. 2010, \aap, 510, A101

\bibitem[{{Flesch}(2016)}]{Flesch2016}
{Flesch}, E.~W. 2016, \pasa, 33, e052

\bibitem[{{Foellmi} {et~al.}(2003){Foellmi}, {Moffat}, \&
  {Guerrero}}]{Foellmi2003b}
{Foellmi}, C., {Moffat}, A.~F.~J., \& {Guerrero}, M.~A. 2003, \mnras, 338, 1025

\bibitem[{{Girard} {et~al.}(2011){Girard}, {van Altena}, {Zacharias}, {Vieira},
  {Casetti-Dinescu}, {Castillo}, {Herrera}, {Lee}, {Beers}, {Monet}, \&
  {L{\'o}pez}}]{Girard2011}
{Girard}, T.~M., {van Altena}, W.~F., {Zacharias}, N., {et~al.} 2011, \aj, 142,
  15

\bibitem[{{Gorjian} {et~al.}(2004){Gorjian}, {Werner}, {Mould}, {Gordon},
  {Muzzerole}, {Morrison}, {Surace}, {Rebull}, {Hurt}, {Smith}, {Points},
  {Aguilera}, {De Buizer}, \& {Packham}}]{Gorjian2004}
{Gorjian}, V., {Werner}, M.~W., {Mould}, J.~R., {et~al.} 2004, \apjs, 154, 275

\bibitem[{{Gouliermis} {et~al.}(2008){Gouliermis}, {Chu}, {Henning},
  {Brandner}, {Gruendl}, {Hennekemper}, \& {Hormuth}}]{Gouliermis2008}
{Gouliermis}, D.~A., {Chu}, Y.-H., {Henning}, T., {et~al.} 2008, \apj, 688,
  1050

\bibitem[{{Hainich} {et~al.}(2015){Hainich}, {Pasemann}, {Todt}, {Shenar},
  {Sander}, \& {Hamann}}]{Hainich2015}
{Hainich}, R., {Pasemann}, D., {Todt}, H., {et~al.} 2015, \aap, 581, A21

\bibitem[{{Hainich} {et~al.}(2014){Hainich}, {R{\"u}hling}, {Todt}, {Oskinova},
  {Liermann}, {Gr{\"a}fener}, {Foellmi}, {Schnurr}, \& {Hamann}}]{Hainich2014}
{Hainich}, R., {R{\"u}hling}, U., {Todt}, H., {et~al.} 2014, \aap, 565, A27

\bibitem[{{Hamann} {et~al.}(2006){Hamann}, {Gr{\"a}fener}, \&
  {Liermann}}]{Hamann2006}
{Hamann}, W.-R., {Gr{\"a}fener}, G., \& {Liermann}, A. 2006, \aap, 457, 1015

\bibitem[{{Hamann} \& {Koesterke}(1998)}]{HK98}
{Hamann}, W.-R. \& {Koesterke}, L. 1998, \aap, 335, 1003

\bibitem[{{Harayama} {et~al.}(2008){Harayama}, {Eisenhauer}, \&
  {Martins}}]{Harayama2008}
{Harayama}, Y., {Eisenhauer}, F., \& {Martins}, F. 2008, \apj, 675, 1319

\bibitem[{{Haschke} {et~al.}(2011){Haschke}, {Grebel}, \&
  {Duffau}}]{Haschke2011}
{Haschke}, R., {Grebel}, E.~K., \& {Duffau}, S. 2011, \aj, 141, 158

\bibitem[{{Heger} {et~al.}(2003){Heger}, {Fryer}, {Woosley}, {Langer}, \&
  {Hartmann}}]{Heger2003}
{Heger}, A., {Fryer}, C.~L., {Woosley}, S.~E., {Langer}, N., \& {Hartmann},
  D.~H. 2003, \apj, 591, 288

\bibitem[{{Henize}(1956)}]{Henize1956}
{Henize}, K.~G. 1956, \apjs, 2, 315

\bibitem[{{Howarth}(1983)}]{Howarth1983}
{Howarth}, I.~D. 1983, \mnras, 203, 301

\bibitem[{{Kato} {et~al.}(2007){Kato}, {Nagashima}, {Nagayama}, {Kurita},
  {Koerwer}, {Kawai}, {Yamamuro}, {Zenno}, {Nishiyama}, {Baba}, {Kadowaki},
  {Haba}, {Hatano}, {Shimizu}, {Nishimura}, {Nagata}, {Sato}, {Murai},
  {Kawazu}, {Nakajima}, {Nakaya}, {Kandori}, {Kusakabe}, {Ishihara},
  {Kaneyasu}, {Hashimoto}, {Tamura}, {Tanab{\'e}}, {Ita}, {Matsunaga},
  {Nakada}, {Sugitani}, {Wakamatsu}, {Glass}, {Feast}, {Menzies}, {Whitelock},
  {Fourie}, {Stoffels}, {Evans}, \& {Hasegawa}}]{Kato2007}
{Kato}, D., {Nagashima}, C., {Nagayama}, T., {et~al.} 2007, \pasj, 59, 615

\bibitem[{{Kavanagh} {et~al.}(2012){Kavanagh}, {Sasaki}, \&
  {Points}}]{Kavanagh2012}
{Kavanagh}, P.~J., {Sasaki}, M., \& {Points}, S.~D. 2012, \aap, 547, A19

\bibitem[{{K{\"o}hler} {et~al.}(2015){K{\"o}hler}, {Langer}, {de Koter}, {de
  Mink}, {Crowther}, {Evans}, {Gr{\"a}fener}, {Sana}, {Sanyal}, {Schneider}, \&
  {Vink}}]{Kohler2015}
{K{\"o}hler}, K., {Langer}, N., {de Koter}, A., {et~al.} 2015, \aap, 573, A71

\bibitem[{{Kroupa}(2002)}]{Kroupa2002}
{Kroupa}, P. 2002, Science, 295, 82

\bibitem[{{Kroupa} {et~al.}(2013){Kroupa}, {Weidner}, {Pflamm-Altenburg},
  {Thies}, {Dabringhausen}, {Marks}, \& {Maschberger}}]{Kroupa2013}
{Kroupa}, P., {Weidner}, C., {Pflamm-Altenburg}, J., {et~al.} 2013, {The
  Stellar and Sub-Stellar Initial Mass Function of Simple and Composite
  Populations}, ed. T.~D. {Oswalt} \& G.~{Gilmore}, 115

\bibitem[{{Kudritzki} \& {Puls}(2000)}]{Kudritzki2000}
{Kudritzki}, R.-P. \& {Puls}, J. 2000, \araa, 38, 613

\bibitem[{{Lamers} {et~al.}(1995){Lamers}, {Snow}, \& {Lindholm}}]{Lamers1995}
{Lamers}, H.~J.~G.~L.~M., {Snow}, T.~P., \& {Lindholm}, D.~M. 1995, \apj, 455,
  269

\bibitem[{{Lamers} {et~al.}(1998){Lamers}, {Zickgraf}, {de Winter}, {Houziaux},
  \& {Zorec}}]{Lamers1998}
{Lamers}, H.~J.~G.~L.~M., {Zickgraf}, F.-J., {de Winter}, D., {Houziaux}, L.,
  \& {Zorec}, J. 1998, \aap, 340, 117

\bibitem[{{Leitherer} {et~al.}(1992){Leitherer}, {Robert}, \&
  {Drissen}}]{Leitherer1992}
{Leitherer}, C., {Robert}, C., \& {Drissen}, L. 1992, \apj, 401, 596

\bibitem[{{Lopez} {et~al.}(2010){Lopez}, {Krumholz}, {Bolatto}, {Prochaska}, \&
  {Ramirez-Ruiz}}]{Lopez2010}
{Lopez}, L.~A., {Krumholz}, M., {Bolatto}, A., {Prochaska}, J.~X., \&
  {Ramirez-Ruiz}, E. 2010, in Bulletin of the American Astronomical Society,
  Vol.~42, American Astronomical Society Meeting Abstracts \#215, 261

\bibitem[{{Lucke} \& {Hodge}(1970)}]{Lucke1970}
{Lucke}, P.~B. \& {Hodge}, P.~W. 1970, \aj, 75, 171

\bibitem[{{Madore} \& {Freedman}(1998)}]{Madore1998}
{Madore}, B.~F. \& {Freedman}, W.~L. 1998, \apj, 492, 110

\bibitem[{{Marks} {et~al.}(2012){Marks}, {Kroupa}, {Dabringhausen}, \&
  {Pawlowski}}]{Marks2012}
{Marks}, M., {Kroupa}, P., {Dabringhausen}, J., \& {Pawlowski}, M.~S. 2012,
  \mnras, 422, 2246

\bibitem[{{Martayan} {et~al.}(2010){Martayan}, {Baade}, \&
  {Fabregat}}]{Martayan2010}
{Martayan}, C., {Baade}, D., \& {Fabregat}, J. 2010, \aap, 509, A11

\bibitem[{{Martayan} {et~al.}(2006){Martayan}, {Fr{\'e}mat}, {Hubert},
  {Floquet}, {Zorec}, \& {Neiner}}]{Martayan2006}
{Martayan}, C., {Fr{\'e}mat}, Y., {Hubert}, A.-M., {et~al.} 2006, \aap, 452,
  273

\bibitem[{{Martins} {et~al.}(2009){Martins}, {Hillier}, {Bouret}, {Depagne},
  {Foellmi}, {Marchenko}, \& {Moffat}}]{Martins2009}
{Martins}, F., {Hillier}, D.~J., {Bouret}, J.~C., {et~al.} 2009, \aap, 495, 257

\bibitem[{{Massey}(2002)}]{Massey2002}
{Massey}, P. 2002, \apjs, 141, 81

\bibitem[{{Massey} {et~al.}(2013){Massey}, {Neugent}, {Hillier}, \&
  {Puls}}]{Massey2013}
{Massey}, P., {Neugent}, K.~F., {Hillier}, D.~J., \& {Puls}, J. 2013, \apj,
  768, 6

\bibitem[{{Massey} {et~al.}(2009){Massey}, {Zangari}, {Morrell}, {Puls},
  {DeGioia-Eastwood}, {Bresolin}, \& {Kudritzki}}]{Massey2009}
{Massey}, P., {Zangari}, A.~M., {Morrell}, N.~I., {et~al.} 2009, \apj, 692, 618

\bibitem[{{Mathewson} \& {Clarke}(1973)}]{Mathewson1973}
{Mathewson}, D.~S. \& {Clarke}, J.~N. 1973, \apj, 180, 725

\bibitem[{{McEvoy} {et~al.}(2015){McEvoy}, {Dufton}, {Evans}, {Kalari},
  {Markova}, {Sim{\'o}n-D{\'{\i}}az}, {Vink}, {Walborn}, {Crowther}, {de
  Koter}, {de Mink}, {Dunstall}, {H{\'e}nault-Brunet}, {Herrero}, {Langer},
  {Lennon}, {Ma{\'{\i}}z Apell{\'a}niz}, {Najarro}, {Puls}, {Sana},
  {Schneider}, \& {Taylor}}]{McEvoy2015}
{McEvoy}, C.~M., {Dufton}, P.~L., {Evans}, C.~J., {et~al.} 2015, \aap, 575, A70

\bibitem[{{McSwain} {et~al.}(2008){McSwain}, {Huang}, {Gies}, {Grundstrom}, \&
  {Townsend}}]{McSwain2008}
{McSwain}, M.~V., {Huang}, W., {Gies}, D.~R., {Grundstrom}, E.~D., \&
  {Townsend}, R.~H.~D. 2008, \apj, 672, 590

\bibitem[{{Mokiem} {et~al.}(2007){Mokiem}, {de Koter}, {Evans}, {Puls},
  {Smartt}, {Crowther}, {Herrero}, {Langer}, {Lennon}, {Najarro}, {Villamariz},
  \& {Vink}}]{Mokiem2007A}
{Mokiem}, M.~R., {de Koter}, A., {Evans}, C.~J., {et~al.} 2007, \aap, 465, 1003

\bibitem[{{Mowlavi} {et~al.}(2017){Mowlavi}, {Lecoeur-Ta{\"i}bi}, {Holl},
  {Rimoldini}, {Barblan}, {Pr{\v s}a}, {Kochoska}, {S{\"u}veges}, {Eyer},
  {Nienartowicz}, {Jevardat}, {Charnas}, {Guy}, \& {Audard}}]{Mowlavi2017}
{Mowlavi}, N., {Lecoeur-Ta{\"i}bi}, I., {Holl}, B., {et~al.} 2017, \aap, 606,
  A92

\bibitem[{{Nayakshin} \& {Sunyaev}(2005)}]{Nayakshin2005}
{Nayakshin}, S. \& {Sunyaev}, R. 2005, \mnras, 364, L23

\bibitem[{{Nieva} \& {Przybilla}(2014)}]{Nieva2014}
{Nieva}, M.-F. \& {Przybilla}, N. 2014, \aap, 566, A7

\bibitem[{{Oey} {et~al.}(2001){Oey}, {Clarke}, \& {Massey}}]{Oey2001}
{Oey}, M.~S., {Clarke}, C.~J., \& {Massey}, P. 2001, in Dwarf galaxies and
  their environment, ed. K.~S. {de Boer}, R.-J. {Dettmar}, \& U.~{Klein}, 181

\bibitem[{{Oh} \& {Kroupa}(2016)}]{Oh2016}
{Oh}, S. \& {Kroupa}, P. 2016, \aap, 590, A107

\bibitem[{{Oh} {et~al.}(2015){Oh}, {Kroupa}, \& {Pflamm-Altenburg}}]{Oh2015}
{Oh}, S., {Kroupa}, P., \& {Pflamm-Altenburg}, J. 2015, \apj, 805, 92

\bibitem[{{Oskinova}(2005)}]{Oskinova2005}
{Oskinova}, L.~M. 2005, \mnras, 361, 679

\bibitem[{{Oskinova} {et~al.}(2007){Oskinova}, {Hamann}, \&
  {Feldmeier}}]{Oskinova2007}
{Oskinova}, L.~M., {Hamann}, W.-R., \& {Feldmeier}, A. 2007, \aap, 476, 1331

\bibitem[{{Oskinova} {et~al.}(2013){Oskinova}, {Sun}, {Evans},
  {H{\'e}nault-Brunet}, {Chu}, {Gallagher}, {Guerrero}, {Gruendl}, {G{\"u}del},
  {Silich}, {Chen}, {Naz{\'e}}, {Hainich}, \& {Reyes-Iturbide}}]{oskinova2013}
{Oskinova}, L.~M., {Sun}, W., {Evans}, C.~J., {et~al.} 2013, \apj, 765, 73

\bibitem[{{Oskinova} {et~al.}(2011){Oskinova}, {Todt}, {Ignace}, {Brown},
  {Cassinelli}, \& {Hamann}}]{Oskinova2011}
{Oskinova}, L.~M., {Todt}, H., {Ignace}, R., {et~al.} 2011, \mnras, 416, 1456

\bibitem[{{Penny}(1996)}]{Penny1996}
{Penny}, L.~R. 1996, \apj, 463, 737

\bibitem[{{Pietrzy{\'n}ski} {et~al.}(2013){Pietrzy{\'n}ski}, {Graczyk},
  {Gieren}, {Thompson}, {Pilecki}, {Udalski}, {Soszy{\'n}ski}, {Koz{\l}owski},
  {Konorski}, {Suchomska}, {Bono}, {Moroni}, {Villanova}, {Nardetto},
  {Bresolin}, {Kudritzki}, {Storm}, {Gallenne}, {Smolec}, {Minniti}, {Kubiak},
  {Szyma{\'n}ski}, {Poleski}, {Wyrzykowski}, {Ulaczyk}, {Pietrukowicz},
  {G{\'o}rski}, \& {Karczmarek}}]{Pietrzynski2013}
{Pietrzy{\'n}ski}, G., {Graczyk}, D., {Gieren}, W., {et~al.} 2013, \nat, 495,
  76

\bibitem[{{Puls} {et~al.}(2008){Puls}, {Vink}, \& {Najarro}}]{Puls2008}
{Puls}, J., {Vink}, J.~S., \& {Najarro}, F. 2008, \aapr, 16, 209

\bibitem[{{Ramachandran} {et~al.}(2018){Ramachandran}, {Hainich}, {Hamann},
  {Oskinova}, {Shenar}, {Sander}, {Todt}, \& {Gallagher}}]{Ramachandran2018}
{Ramachandran}, V., {Hainich}, R., {Hamann}, W.-R., {et~al.} 2018, \aap, 609,
  A7

\bibitem[{{Ram{\'{\i}}rez-Agudelo} {et~al.}(2013){Ram{\'{\i}}rez-Agudelo},
  {Sim{\'o}n-D{\'{\i}}az}, {Sana}, {de Koter}, {Sab{\'{\i}}n-Sanjul{\'{\i}}an},
  {de Mink}, {Dufton}, {Gr{\"a}fener}, {Evans}, {Herrero}, {Langer}, {Lennon},
  {Ma{\'{\i}}z Apell{\'a}niz}, {Markova}, {Najarro}, {Puls}, {Taylor}, \&
  {Vink}}]{Ramirez-Agudelo2013}
{Ram{\'{\i}}rez-Agudelo}, O.~H., {Sim{\'o}n-D{\'{\i}}az}, S., {Sana}, H.,
  {et~al.} 2013, \aap, 560, A29

\bibitem[{{Rivinius} {et~al.}(2013){Rivinius}, {Carciofi}, \&
  {Martayan}}]{Rivinius2013}
{Rivinius}, T., {Carciofi}, A.~C., \& {Martayan}, C. 2013, \aapr, 21, 69

\bibitem[{{Rolleston} {et~al.}(2002){Rolleston}, {Trundle}, \&
  {Dufton}}]{Rolleston2002}
{Rolleston}, W.~R.~J., {Trundle}, C., \& {Dufton}, P.~L. 2002, \aap, 396, 53

\bibitem[{{Romita} {et~al.}(2010){Romita}, {Carlson}, {Meixner}, {Sewi{\l}o},
  {Whitney}, {Babler}, {Indebetouw}, {Hora}, {Meade}, \& {Shiao}}]{Romita2010}
{Romita}, K.~A., {Carlson}, L.~R., {Meixner}, M., {et~al.} 2010, \apj, 721, 357

\bibitem[{{Rosen} {et~al.}(2014){Rosen}, {Lopez}, {Krumholz}, \&
  {Ramirez-Ruiz}}]{Rosen2014}
{Rosen}, A.~L., {Lopez}, L.~A., {Krumholz}, M.~R., \& {Ramirez-Ruiz}, E. 2014,
  \mnras, 442, 2701

\bibitem[{{Sabbi} {et~al.}(2008){Sabbi}, {Sirianni}, {Nota}, {Tosi},
  {Gallagher}, {Smith}, {Angeretti}, {Meixner}, {Oey}, {Walterbos}, \&
  {Pasquali}}]{Sabbi2008}
{Sabbi}, E., {Sirianni}, M., {Nota}, A., {et~al.} 2008, \aj, 135, 173

\bibitem[{{Sana} {et~al.}(2013{\natexlab{a}}){Sana}, {de Koter}, {de Mink},
  {Dunstall}, {Evans}, {H{\'e}nault-Brunet}, {Ma{\'{\i}}z Apell{\'a}niz},
  {Ram{\'{\i}}rez-Agudelo}, {Taylor}, {Walborn}, {Clark}, {Crowther},
  {Herrero}, {Gieles}, {Langer}, {Lennon}, \& {Vink}}]{Sana2013b}
{Sana}, H., {de Koter}, A., {de Mink}, S.~E., {et~al.} 2013{\natexlab{a}},
  \aap, 550, A107

\bibitem[{{Sana} {et~al.}(2011){Sana}, {James}, \& {Gosset}}]{Sana2011}
{Sana}, H., {James}, G., \& {Gosset}, E. 2011, \mnras, 416, 817

\bibitem[{{Sana} {et~al.}(2013{\natexlab{b}}){Sana}, {van Boeckel}, {Tramper},
  {Ellerbroek}, {de Koter}, {Kaper}, {Moffat}, {Schnurr}, {Schneider}, \&
  {Gies}}]{Sana2013}
{Sana}, H., {van Boeckel}, T., {Tramper}, F., {et~al.} 2013{\natexlab{b}},
  \mnras, 432, L26

\bibitem[{{Sander} {et~al.}(2012){Sander}, {Hamann}, \& {Todt}}]{Sander2012}
{Sander}, A., {Hamann}, W.-R., \& {Todt}, H. 2012, \aap, 540, A144

\bibitem[{{Sander} {et~al.}(2015){Sander}, {Shenar}, {Hainich},
  {G{\'{\i}}menez-Garc{\'{\i}}a}, {Todt}, \& {Hamann}}]{Sander2015}
{Sander}, A., {Shenar}, T., {Hainich}, R., {et~al.} 2015, \aap, 577, A13

\bibitem[{{Sasaki} {et~al.}(2011){Sasaki}, {Breitschwerdt}, {Baumgartner}, \&
  {Haberl}}]{Sasaki2011}
{Sasaki}, M., {Breitschwerdt}, D., {Baumgartner}, V., \& {Haberl}, F. 2011,
  \aap, 528, A136

\bibitem[{{Schneider} {et~al.}(2018){Schneider}, {Sana}, {Evans},
  {Bestenlehner}, {Castro}, {Fossati}, {Gr{\"a}fener}, {Langer},
  {Ram{\'{\i}}rez-Agudelo}, {Sab{\'{\i}}n-Sanjuli{\'a}n},
  {Sim{\'o}n-D{\'{\i}}az}, {Tramper}, {Crowther}, {de Koter}, {de Mink},
  {Dufton}, {Garcia}, {Gieles}, {H{\'e}nault-Brunet}, {Herrero}, {Izzard},
  {Kalari}, {Lennon}, {Ma{\'{\i}}z Apell{\'a}niz}, {Markova}, {Najarro},
  {Podsiadlowski}, {Puls}, {Taylor}, {van Loon}, {Vink}, \&
  {Norman}}]{Schneider2018}
{Schneider}, F.~R.~N., {Sana}, H., {Evans}, C.~J., {et~al.} 2018, Science, 359,
  69

\bibitem[{{Seaton}(1979)}]{Seaton1979}
{Seaton}, M.~J. 1979, \mnras, 187, 73P

\bibitem[{{Shenar} {et~al.}(2016){Shenar}, {Hainich}, {Todt}, {Sander},
  {Hamann}, {Moffat}, {Eldridge}, {Pablo}, {Oskinova}, \&
  {Richardson}}]{Shenar2016}
{Shenar}, T., {Hainich}, R., {Todt}, H., {et~al.} 2016, \aap, 591, A22

\bibitem[{{Shenar} {et~al.}(2015){Shenar}, {Oskinova}, {Hamann}, {Corcoran},
  {Moffat}, {Pablo}, {Richardson}, {Waldron}, {Huenemoerder}, {Ma{\'{\i}}z
  Apell{\'a}niz}, {Nichols}, {Todt}, {Naz{\'e}}, {Hoffman}, {Pollock}, \&
  {Negueruela}}]{Shenar2015}
{Shenar}, T., {Oskinova}, L., {Hamann}, W.-R., {et~al.} 2015, \apj, 809, 135

\bibitem[{{Shenar} {et~al.}(2017){Shenar}, {Oskinova}, {J{\"a}rvinen},
  {Luckas}, {Hainich}, {Todt}, {Hubrig}, {Sander}, {Ilyin}, \&
  {Hamann}}]{Shenar2017}
{Shenar}, T., {Oskinova}, L.~M., {J{\"a}rvinen}, S.~P., {et~al.} 2017, \aap,
  606, A91

\bibitem[{{Shtykovskiy} \& {Gilfanov}(2005)}]{Shtykovskiy2005}
{Shtykovskiy}, P. \& {Gilfanov}, M. 2005, \aap, 431, 597

\bibitem[{{Sim{\'o}n-D{\'{\i}}az} \& {Herrero}(2014)}]{Simon-diaz2014}
{Sim{\'o}n-D{\'{\i}}az}, S. \& {Herrero}, A. 2014, \aap, 562, A135

\bibitem[{{Smith}(2006)}]{Smith2006}
{Smith}, N. 2006, \mnras, 367, 763

\bibitem[{{Smith} {et~al.}(2010){Smith}, {Povich}, {Whitney}, {Churchwell},
  {Babler}, {Meade}, {Bally}, {Gehrz}, {Robitaille}, \& {Stassun}}]{Smith2010}
{Smith}, N., {Povich}, M.~S., {Whitney}, B.~A., {et~al.} 2010, \mnras, 406, 952

\bibitem[{{Smith} {et~al.}(2005){Smith}, {Points}, {Chu}, {Winkler}, {Leiton},
  \& {MCELS Team}}]{Smith2005}
{Smith}, R.~C., {Points}, S., {Chu}, Y.-H., {et~al.} 2005, in Bulletin of the
  American Astronomical Society, Vol.~37, American Astronomical Society Meeting
  Abstracts, \#145.01

\bibitem[{{Sota} {et~al.}(2014){Sota}, {Ma{\'{\i}}z Apell{\'a}niz}, {Morrell},
  {Barb{\'a}}, {Walborn}, {Gamen}, {Arias}, \& {Alfaro}}]{Sota2014}
{Sota}, A., {Ma{\'{\i}}z Apell{\'a}niz}, J., {Morrell}, N.~I., {et~al.} 2014,
  \apjs, 211, 10

\bibitem[{{Sota} {et~al.}(2011){Sota}, {Ma{\'{\i}}z Apell{\'a}niz}, {Walborn},
  {Alfaro}, {Barb{\'a}}, {Morrell}, {Gamen}, \& {Arias}}]{Sota2011}
{Sota}, A., {Ma{\'{\i}}z Apell{\'a}niz}, J., {Walborn}, N.~R., {et~al.} 2011,
  \apjs, 193, 24

\bibitem[{{Struve}(1931)}]{Struve1931}
{Struve}, O. 1931, \apj, 73, 94

\bibitem[{{Subramaniam}(2005)}]{Subramaniam2005}
{Subramaniam}, A. 2005, \aap, 430, 421

\bibitem[{{Tenorio-Tagle} \& {Bodenheimer}(1988)}]{Tenorio-Tagle1988}
{Tenorio-Tagle}, G. \& {Bodenheimer}, P. 1988, \araa, 26, 145

\bibitem[{{Townsley} {et~al.}(2011){Townsley}, {Broos}, {Chu}, {Gruendl},
  {Oey}, \& {Pittard}}]{Townsley2011}
{Townsley}, L.~K., {Broos}, P.~S., {Chu}, Y.-H., {et~al.} 2011, \apjs, 194, 16

\bibitem[{{Townsley} {et~al.}(2003){Townsley}, {Feigelson}, {Montmerle},
  {Broos}, {Chu}, \& {Garmire}}]{Townsley2003}
{Townsley}, L.~K., {Feigelson}, E.~D., {Montmerle}, T., {et~al.} 2003, \apj,
  593, 874

\bibitem[{{Trundle} {et~al.}(2007){Trundle}, {Dufton}, {Hunter}, {Evans},
  {Lennon}, {Smartt}, \& {Ryans}}]{Trundle2007}
{Trundle}, C., {Dufton}, P.~L., {Hunter}, I., {et~al.} 2007, \aap, 471, 625

\bibitem[{{{\v S}urlan} {et~al.}(2013){{\v S}urlan}, {Hamann}, {Aret},
  {Kub{\'a}t}, {Oskinova}, \& {Torres}}]{Surlan2013}
{{\v S}urlan}, B., {Hamann}, W.-R., {Aret}, A., {et~al.} 2013, \aap, 559, A130

\bibitem[{{Vink} {et~al.}(2000){Vink}, {de Koter}, \& {Lamers}}]{Vink2000}
{Vink}, J.~S., {de Koter}, A., \& {Lamers}, H.~J.~G.~L.~M. 2000, \aap, 362, 295

\bibitem[{{Walborn}(2006)}]{Walborn2006}
{Walborn}, N.~R. 2006, in IAU Joint Discussion, Vol.~4, IAU Joint Discussion

\bibitem[{{Walborn} {et~al.}(2014){Walborn}, {Sana}, {Sim{\'o}n-D{\'{\i}}az},
  {Ma{\'{\i}}z Apell{\'a}niz}, {Taylor}, {Evans}, {Markova}, {Lennon}, \& {de
  Koter}}]{Walborn2014}
{Walborn}, N.~R., {Sana}, H., {Sim{\'o}n-D{\'{\i}}az}, S., {et~al.} 2014, \aap,
  564, A40

\bibitem[{{Wang} {et~al.}(2006){Wang}, {Dong}, \& {Lang}}]{Wang2006}
{Wang}, Q.~D., {Dong}, H., \& {Lang}, C. 2006, \mnras, 371, 38

\bibitem[{{Weaver} {et~al.}(1977){Weaver}, {McCray}, {Castor}, {Shapiro}, \&
  {Moore}}]{Weaver1977}
{Weaver}, R., {McCray}, R., {Castor}, J., {Shapiro}, P., \& {Moore}, R. 1977,
  \apj, 218, 377

\bibitem[{{Weidner} {et~al.}(2009){Weidner}, {Kroupa}, \&
  {Maschberger}}]{Weidne2009}
{Weidner}, C., {Kroupa}, P., \& {Maschberger}, T. 2009, \mnras, 393, 663

\bibitem[{{Williams} {et~al.}(2005){Williams}, {Chu}, {Dickel}, {Gruendl},
  {Seward}, {Guerrero}, \& {Hobbs}}]{Williams2005}
{Williams}, R.~M., {Chu}, Y.-H., {Dickel}, J.~R., {et~al.} 2005, \apj, 628, 704

\bibitem[{{Wong} {et~al.}(2011){Wong}, {Hughes}, {Ott}, {Muller}, {Pineda},
  {Bernard}, {Chu}, {Fukui}, {Gruendl}, {Henkel}, {Kawamura}, {Klein},
  {Looney}, {Maddison}, {Mizuno}, {Paradis}, {Seale}, \& {Welty}}]{Wong2011}
{Wong}, T., {Hughes}, A., {Ott}, J., {et~al.} 2011, \apjs, 197, 16

\bibitem[{{Woosley} \& {Weaver}(1995)}]{Woosley1995}
{Woosley}, S.~E. \& {Weaver}, T.~A. 1995, \apjs, 101, 181

\bibitem[{{Zacharias} {et~al.}(2012){Zacharias}, {Finch}, {Girard}, {Henden},
  {Bartlett}, {Monet}, \& {Zacharias}}]{Zacharias2012}
{Zacharias}, N., {Finch}, C.~T., {Girard}, T.~M., {et~al.} 2012, VizieR Online
  Data Catalog, 1322

\bibitem[{{Zaritsky} {et~al.}(2004){Zaritsky}, {Harris}, {Thompson}, \&
  {Grebel}}]{Zaritsky2004}
{Zaritsky}, D., {Harris}, J., {Thompson}, I.~B., \& {Grebel}, E.~K. 2004, \aj,
  128, 1606

\bibitem[{{Zeidler} {et~al.}(2017){Zeidler}, {Nota}, {Grebel}, {Sabbi},
  {Pasquali}, {Tosi}, \& {Christian}}]{Zeidler2017}
{Zeidler}, P., {Nota}, A., {Grebel}, E.~K., {et~al.} 2017, \aj, 153, 122

\end{thebibliography}


\newpage

\appendix

\section{Additional tables}
\label{sect:appendixa}

\begin{center}

\tablehead{
\hline
\hline
\noalign{\vspace{1mm}}
 N206-FS& RA (J2000)   & DEC (J2000)  & Spectral type \\
 \# &(\degr)& (\degr)&  \\
\noalign{\vspace{1mm}}
\hline 
\noalign{\vspace{1mm}}}
\tabletail{\hline }
\topcaption{Coordinates and spectral types of OB stars in our sample (continuation of Table\,\ref{table:obstars})}
\label{table:App_coord}
\begin{supertabular}{cccl}
31 	&  82.307375 	&  -70.865306    &	B3  IV           \\
32 	&  82.325375 	&  -71.218750    &	B0 V             \\
33 	&  82.331500 	&  -71.029611    &	B8 Ia            \\
34 	&  82.337333 	&  -71.029389    &	B1.5 III Nwk     \\
36 	&  82.346625 	&  -71.082000    &	B2  IV           \\
37 	&  82.348583 	&  -71.092639    &	B1.5 IV          \\
38 	&  82.350667 	&  -71.074139    &	B2 V             \\
39 	&  82.365375 	&  -71.073444    &	B1.5 V           \\
42 	&  82.374917 	&  -70.901500    &	B2  IV           \\
43 	&  82.384500 	&  -70.992333    &	B0.5 V           \\
46 	&  82.399375 	&  -70.937000    &	B1.5 V           \\
47 	&  82.408458 	&  -70.963778    &	O9.7 V           \\
48 	&  82.409125 	&  -70.990472    &	B0.7 V           \\
49 	&  82.411208 	&  -70.985083    &	B1.5 V           \\
50 	&  82.412125 	&  -70.985778    &	B1 V             \\
51 	&  82.412458 	&  -70.982944    &	B2  IV           \\
52 	&  82.429292 	&  -70.837722    &	B0 V             \\
55 	&  82.436708 	&  -70.916278    &	B0.2 IV          \\
56 	&  82.438917 	&  -71.161500    &	B0.7 (IV)e       \\
58 	&  82.442250 	&  -71.156111    &	B0.5 II          \\
60 	&  82.446917 	&  -71.150111    &	B1.5 (IV)e       \\
61 	&  82.455208 	&  -71.063139    &	B1.5 V           \\
62 	&  82.462500 	&  -71.105500    &	O9.5 (III)e      \\
64 	&  82.464958 	&  -71.075722    &	O8 Vz            \\
65 	&  82.465500 	&  -71.068611    &	B2  IV           \\
68 	&  82.471250 	&  -71.083889    &	B2.5 IV          \\
69 	&  82.471583 	&  -71.079000    &	B0.2 V           \\
73 	&  82.479625 	&  -71.149639    &	B0.7 IV          \\
74 	&  82.494250 	&  -70.931778    &	B0.5 V           \\
75 	&  82.496542 	&  -71.099333    &	B0.7 IV          \\
77 	&  82.497500 	&  -70.848306    &	B0.5 V           \\
78 	&  82.503083 	&  -71.160111    &	B1 V             \\
80 	&  82.510292 	&  -71.088917    &	B1.5 V           \\
81 	&  82.517667 	&  -71.132278    &	B1 II            \\
84 	&  82.523792 	&  -71.152778    &	B1.5 IV          \\
88 	&  82.528583 	&  -70.829889    &	B0 V             \\
89 	&  82.544917 	&  -71.079056    &	B2  IV           \\
90 	&  82.546875 	&  -70.861667    &	O9.7 Iab         \\
91 	&  82.546958 	&  -71.079583    &	O8 IV            \\
93 	&  82.548625 	&  -71.099917    &	B0.2 V           \\
95 	&  82.549083 	&  -71.067472    &	B0.5 IV          \\
98 	&  82.567542 	&  -71.104611    &	B0.7 IV          \\
99 	&  82.581125 	&  -71.102000    &	O9.5 V           \\
101 	&  82.582333 	&  -71.280056    &	B1.5 (IV)e       \\
103 	&  82.589083 	&  -70.813639    &	B2  (IV)e        \\
104 	&  82.592167 	&  -70.910028    &	B0.5 V           \\
105 	&  82.592250 	&  -71.105389    &	B1.5 V           \\
107 	&  82.594583 	&  -71.105417    &	O8 Vz            \\
108 	&  82.594708 	&  -71.099611    &	B1.5 V           \\
109 	&  82.596000 	&  -71.056639    &	B0.5 V           \\
112 	&  82.609125 	&  -70.983639    &	O9.7 V           \\
113 	&  82.613167 	&  -71.048917    &	B3 (V)e          \\
116 	&  82.617792 	&  -70.989889    &	B0.7 III         \\
117 	&  82.625625 	&  -70.890972    &	B2 (IV)e         \\
119 	&  82.627875 	&  -71.042083    &	O8 (V)e          \\
120 	&  82.636708 	&  -71.049361    &	B2 IV            \\
121 	&  82.640458 	&  -70.831000    &	B1.5 (V)e        \\
122 	&  82.646792 	&  -71.032472    &	B1.5 (IV)e       \\
123 	&  82.647833 	&  -71.081194    &	B0.2 V           \\
125 	&  82.652583 	&  -71.071000    &	O9 V             \\
126 	&  82.653000 	&  -71.028583    &	B2 V             \\
127 	&  82.659750 	&  -71.085278    &	B0.2 V           \\
129 	&  82.663083 	&  -71.016611    &	B0.5 IV          \\
130 	&  82.664750 	&  -71.055778    &	O9.7 V           \\
132 	&  82.666750 	&  -71.286972    &	O9 IV            \\
133 	&  82.667125 	&  -71.113139    &	O9 V             \\
134 	&  82.667250 	&  -71.093611    &	O9.7 Iab         \\
136 	&  82.674125 	&  -70.892889    &	B1 V             \\
138 	&  82.685458 	&  -71.029000    &	B0.7 V           \\
139 	&  82.691667 	&  -70.874306    &	B0.7 V           \\
141 	&  82.693083 	&  -70.866500    &	B1.5 V           \\
142 	&  82.693750 	&  -70.895667    &	B2 (V)e          \\
143 	&  82.694000 	&  -70.913778    &	B1.5 V           \\
144 	&  82.696042 	&  -71.054333    &	B0 II Nwk        \\
145 	&  82.697958 	&  -71.133667    &	O8 Vz            \\
146 	&  82.697958 	&  -71.058083    &	B1.5 IV          \\
147 	&  82.699083 	&  -71.067278    &	B1 Ia Nwk        \\
148 	&  82.701792 	&  -71.230222    &	O9.7 IV          \\
149 	&  82.703083 	&  -71.065917    &	O9 IV            \\
150 	&  82.706750 	&  -71.004278    &	B0.5 V           \\
151 	&  82.710958 	&  -71.046639    &	O7.5 V           \\
153 	&  82.714833 	&  -71.216167    &	B2 IV            \\
154 	&  82.723083 	&  -71.132889    &	O7 Vz            \\
155 	&  82.724667 	&  -71.064833    &	B0.7 V           \\
156 	&  82.726875 	&  -71.086278    &	B0.2 I           \\
157 	&  82.728917 	&  -70.895361    &	B2 V             \\
159 	&  82.734792 	&  -70.915111    &	B2 V             \\
160 	&  82.734917 	&  -71.072167    &	B0 II Nwk        \\
163 	&  82.740167 	&  -71.025389    &	B0.5 III         \\
164 	&  82.747125 	&  -71.072389    &	B1.5 V           \\
165 	&  82.748792 	&  -70.958750    &	B3 V             \\
167 	&  82.750750 	&  -70.810028    &	B5 V             \\
168 	&  82.753417 	&  -71.077389    &	B2 V             \\
170 	&  82.762250 	&  -70.832333    &	B1 IV            \\
172 	&  82.778292 	&  -71.179861    &	O8.5 V           \\
173 	&  82.782458 	&  -71.130361    &	O9.7 V           \\
174 	&  82.783083 	&  -71.129889    &	O9.7 V           \\
175 	&  82.784208 	&  -71.028500    &	B0 III           \\
176 	&  82.788292 	&  -71.070694    &	O8.5 V           \\
177 	&  82.789958 	&  -71.064306    &	B2 V             \\
179 	&  82.796833 	&  -71.158167    &	B1 V             \\
181 	&  82.799750 	&  -71.081639    &	B1.5 (III)e      \\
183 	&  82.804833 	&  -71.073722    &	B0.5 V           \\
184 	&  82.807750 	&  -71.064167    &	O8 Vz            \\
185 	&  82.810333 	&  -71.068875    &	B2 V             \\
186 	&  82.813100 	&  -71.069500    &	O9 (V)e          \\
189 	&  82.816125 	&  -71.167472    &	B2.5 V           \\
190 	&  82.824708 	&  -71.107389    &	B2 V             \\
191 	&  82.825125 	&  -71.067111    &	B0.5 V           \\
192 	&  82.828458 	&  -71.077500    &	O9.5 (III)e      \\
194 	&  82.848833 	&  -71.070194    &	O8.5 V           \\
195 	&  82.851708 	&  -70.934694    &	B0.7 V           \\
196 	&  82.852250 	&  -71.139750    &	B0.2 IV          \\
197 	&  82.852792 	&  -71.064139    &	B0 V             \\
198 	&  82.852833 	&  -71.070889    &	O8 Vz            \\
200 	&  82.892125 	&  -70.940389    &	B1.5 IV          \\
201 	&  82.892458 	&  -71.160556    &	B0.5 V           \\
202 	&  82.892583 	&  -70.951361    &	B2 V             \\
204 	&  82.900333 	&  -71.115917    &	B2.5 V           \\
205 	&  82.910917 	&  -70.959972    &	B1.5 V           \\
206 	&  82.915042 	&  -70.963417    &	B2 V             \\
207 	&  82.916000 	&  -71.111972    &	B2.5 V           \\
209 	&  82.921000 	&  -70.961556    &	B0.7 V           \\
211 	&  82.925333 	&  -71.008500    &	B1 V             \\
213 	&  82.948583 	&  -71.151278    &	B3 V             \\
215 	&  82.957792 	&  -70.964722    &	B0 IV            \\
220 	&  82.993000 	&  -70.962694    &	B3 IV            \\
221 	&  82.994625 	&  -71.253000    &	B3 V             \\
222 	&  82.996583 	&  -71.161722    &	B1.5 IV          \\
224 	&  83.000167 	&  -71.235472    &	B2 III           \\
225 	&  83.012833 	&  -71.053778    &	B2.5 V           \\
229 	&  83.036750 	&  -71.032778    &	O9.5 V           \\
231 	&  83.067583 	&  -71.234000    &	B2 V             \\
232 	&  83.076792 	&  -71.088139    &	B2.5 V           \\
233 	&  83.145083 	&  -71.113722    &	B1.5 (V)e        \\
234 	&  83.174125 	&  -71.045583    &	B0.5 (V)e        \\
\end{supertabular}

\end{center}


\longtab{2}{     
\setlength{\tabcolsep}{3pt}
\begin{longtable}{llcccccrccccccr}
\caption{Stellar parameters of all OB stars in the N\,206 superbubble (continuation of Table\,\ref{table:stellarparameters})} \label{table:App_stellarparameters} \\


\hline 
\hline
\noalign{\vspace{1mm}}
N206-FS & Spectral type &	$T  _\ast$ 	& log\,$L$ &	log\,$g_\ast$ &	log $\dot{M}$ &	$E_{B-V} $&	$M_{\mathrm{V}}$&$R _\ast$  &	
$\varv_\infty$ 	& $\varv$\,sin\,$i$& $\varv_{\rm rad}$&$M  _\ast$ &	$\log Q_{0}$ & $L_{\mathrm{mec}}$\\ 

\#& & [kK] & [$L _{\odot}$]&[cm s$ ^{-2} $]& [$M _{\odot}\,\mathrm{yr}^{-1} $] & 
[mag] &[mag]& [$R _{\odot}$] & [km\,s$^{-1}$]&[km\,s$^{-1}$]&[km\,s$^{-1}$]&[$M _{\odot} $]&[s$ ^{-1} $] 
&[$L _{\odot}$]\\

\noalign{\vspace{1mm}}
\hline 
\noalign{\vspace{1mm}}
\endfirsthead

\caption{continued.}\\
\hline 
\hline
\noalign{\vspace{1mm}}
N206-FS & Spectral type &	$T  _\ast$ 	& log\,$L$ &	log\,$g_\ast$ &	log $\dot{M}$ &	$E_{B-V} $&	$M_{\mathrm{V}}$&$R _\ast$  &	
$\varv_\infty$ 	& $\varv$\,sin\,$i$& $\varv_{\rm rad}$&$M  _\ast$ &	$\log Q_{0}$ &	$L_{\mathrm{mec}}$\\ 

\#& & [kK] & [$L _{\odot}$]&[cm s$ ^{-2} $]& [$M _{\odot}\,\mathrm{yr}^{-1} $] & 
[mag] &[mag]& [$R _{\odot}$] & [km\,s$^{-1}$]&[km\,s$^{-1}$]&[km\,s$^{-1}$]&[$M _{\odot} $]&[s$ ^{-1} $] 
&[$L _{\odot}$]\\
\noalign{\vspace{1mm}} 

\hline 

\noalign{\vspace{1mm}}
\noalign{\vspace{1mm}}
\endhead
\noalign{\vspace{1mm}}
\hline
\endfoot

31 	  &	B3  IV              & 18.0 	& 3.83 	& 3.6 	& -8.07 	& 0.12 	& -3.03 	& 8.5  	& 800 	& 90 	& 290 	& 10 	& 45.3 	& 0.4 \\ 
32 	  &	B0 V                & 31.0 	& 4.53 	& 4.2 	& -6.85 	& 0.15 	& -3.18 	& 6.4  	& 2400 	& 150 	& 260 	& 24 	& 47.5 	& 67.1 \\ 
33 	  &	B8 Ia               & 13.5 	& 5.30 	& 2.0 	& -6.28 	& 0.12 	& -7.19 	& 81.9  	& 600 	& 65 	& 270 	& 24 	& 46.5 	& 15.68 \\ 
34 	  &	B1.5 III Nwk        & 23.0 	& 4.80 	& 3.4 	& -7.70 	& 0.12 	& -5.08 	& 15.9  	& 1500 	& 80 	& 220 	& 23 	& 47.0 	& 3.7 \\ 
36 	  &	B2  IV              & 20.0 	& 4.03 	& 3.6 	& -7.93 	& 0.15 	& -3.29 	& 8.6  	& 800 	& 150 	& 260 	& 11 	& 45.6 	& 0.6 \\ 
37 	  &	B1.5 IV             & 20.0 	& 4.20 	& 3.4 	& -8.00 	& 0.12 	& -3.76 	& 10.5  	& 600 	& 130 	& 260 	& 10 	& 45.7 	& 0.3 \\ 
38 	  &	B2 V                & 19.0 	& 4.00 	& 3.8 	& -7.70 	& 0.10 	& -3.47 	& 9.3  	& 1000 	& 100 	& 260 	& 20 	& 45.6 	& 1.6 \\ 
39 	  &	B1.5 V              & 22.0 	& 4.17 	& 4.0 	& -7.62 	& 0.12 	& -3.16 	& 8.4  	& 2400 	& 160 	& 240 	& 26 	& 45.9 	& 11.3 \\ 
42 	  &	B2  IV              & 20.0 	& 4.03 	& 4.0 	& -7.35 	& 0.12 	& -3.51 	& 8.6  	& 1200 	& 180 	& 250 	& 27 	& 45.9 	& 5.3 \\ 
43 	  &	B0.5 V              & 27.0 	& 4.26 	& 4.2 	& -6.78 	& 0.10 	& -3.03 	& 6.2  	& 2500 	& 75 	& 240 	& 22 	& 46.7 	& 86.6 \\ 
46 	  &	B1.5 V              & 25.0 	& 4.11 	& 4.2 	& -7.24 	& 0.11 	& -2.93 	& 6.1  	& 2600 	& 100 	& 260 	& 21 	& 46.3 	& 32.3 \\ 
47 	  &	O9.7 V              & 31.0 	& 4.87 	& 4.0 	& -6.78 	& 0.07 	& -4.39 	& 9.5  	& 2200 	& 264 	& 240 	& 33 	& 47.9 	& 67.0 \\ 
48 	  &	B0.7 V              & 27.0 	& 4.40 	& 4.0 	& -7.45 	& 0.10 	& -3.78 	& 7.3  	& 2100 	& 100 	& 180 	& 19 	& 47.0 	& 12.9 \\ 
49 	  &	B1.5 V              & 22.0 	& 3.97 	& 4.0 	& -7.78 	& 0.08 	& -2.97 	& 6.7  	& 2100 	& 80 	& 240 	& 16 	& 45.6 	& 6.1 \\ 
50 	  &	B1 V                & 25.0 	& 4.06 	& 4.2 	& -7.78 	& 0.08 	& -3.12 	& 5.7  	& 2500 	& 100 	& 250 	& 19 	& 46.3 	& 8.7 \\ 
51 	  &	B2  IV              & 21.0 	& 4.20 	& 3.8 	& -7.70 	& 0.05 	& -3.71 	& 9.5  	& 2000 	& 245 	& 240 	& 21 	& 45.8 	& 6.6 \\ 
52 	  &	B0 V                & 31.0 	& 4.53 	& 4.2 	& -6.85 	& 0.08 	& -3.68 	& 6.4  	& 2400 	& 35 	& 240 	& 24 	& 47.5 	& 67.1 \\ 
55 	  &	B0.2 IV             & 30.0 	& 5.24 	& 3.8 	& -6.43 	& 0.12 	& -5.05 	& 15.5  	& 2200 	& 110 	& 270 	& 55 	& 48.2 	& 150.1 \\ 
56 	  &	B0.7 (IV)e          & 27.0 	& 4.50 	& 4.0 	& -7.38 	& 0.22 	& -3.35 	& 8.1  	& 2200 	& 188 	& 260 	& 24 	& 47.1 	& 16.8 \\ 
58 	  &	B0.5 II             & 25.0 	& 5.05 	& 3.0 	& -7.95 	& 0.20 	& -5.34 	& 17.9  	& 1700 	& 100 	& 240 	& 12 	& 47.7 	& 2.69 \\ 
60 	  &	B1.5 (IV)e          & 19.0 	& 4.49 	& 3.4 	& -7.90 	& 0.23 	& -4.07 	& 16.3  	& 760 	& 86 	& 240 	& 24 	& 46.2 	& 0.6 \\ 
61 	  &	B1.5 V              & 24.0 	& 3.95 	& 4.2 	& -7.78 	& 0.03 	& -2.82 	& 5.5  	& 2500 	& 120 	& 180 	& 17 	& 46.0 	& 8.7 \\ 
62 	  &	O9.5 (III)e         & 30.0 	& 4.85 	& 3.4 	& -7.07 	& 0.18 	& -3.97 	& 9.9  	& 800 	& 383 	& 260 	& 9 	& 48.1 	& 4.4 \\ 
64 	  &	O8 Vz               & 35.0 	& 5.06 	& 4.2 	& -6.50 	& 0.09 	& -4.65 	& 9.2  	& 2800 	& 100 	& 240 	& 49 	& 48.5 	& 204.6 \\ 
65 	  &	B2  IV              & 20.0 	& 3.85 	& 3.8 	& -8.54 	& 0.10 	& -2.88 	& 7.0  	& 900 	& 100 	& 260 	& 11 	& 45.0 	& 0.2 \\ 
68 	  &	B2.5 IV             & 19.0 	& 3.90 	& 3.8 	& -7.78 	& 0.10 	& -2.94 	& 8.2  	& 1000 	& 80 	& 230 	& 16 	& 45.5 	& 1.4 \\ 
69 	  &	B0.2 V              & 27.0 	& 4.26 	& 4.2 	& -7.38 	& 0.09 	& -3.07 	& 6.2  	& 2500 	& 150 	& 210 	& 22 	& 46.7 	& 21.7 \\ 
73 	  &	B0.7 IV             & 27.0 	& 4.60 	& 4.0 	& -7.70 	& 0.12 	& -3.89 	& 9.1  	& 2100 	& 70 	& 240 	& 31 	& 47.0 	& 7.3 \\ 
74 	  &	B0.5 V              & 29.0 	& 4.64 	& 4.2 	& -6.62 	& 0.12 	& -3.80 	& 8.3  	& 2800 	& 100 	& 240 	& 40 	& 47.3 	& 153.4 \\ 
75 	  &	B0.7 IV             & 25.0 	& 4.55 	& 3.8 	& -7.78 	& 0.15 	& -4.06 	& 10.1  	& 2000 	& 150 	& 250 	& 23 	& 46.8 	& 5.5 \\ 
77 	  &	B0.5 V              & 27.0 	& 4.19 	& 4.2 	& -7.43 	& 0.07 	& -2.99 	& 5.7  	& 2400 	& 150 	& 240 	& 19 	& 46.6 	& 17.8 \\ 
78 	  &	B1 V                & 25.0 	& 4.16 	& 4.2 	& -7.20 	& 0.08 	& -3.07 	& 6.4  	& 2600 	& 80 	& 240 	& 24 	& 46.3 	& 35.2 \\ 
80 	  &	B1.5 V              & 21.0 	& 4.10 	& 3.8 	& -7.78 	& 0.12 	& -3.33 	& 8.5  	& 1900 	& 100 	& 220 	& 17 	& 45.7 	& 5.0 \\ 
81 	  &	B1 II               & 23.0 	& 5.34 	& 2.8 	& -5.95 	& 0.12 	& -6.06 	& 29.5  	& 400 	& 87 	& 160 	& 20 	& 47.8 	& 14.9 \\ 
84 	  &	B1.5 IV             & 22.0 	& 4.12 	& 4.0 	& -7.66 	& 0.14 	& -3.44 	& 7.9  	& 2300 	& 218 	& 230 	& 23 	& 45.8 	& 9.5 \\ 
88 	  &	B0 V                & 30.0 	& 4.64 	& 4.2 	& -6.70 	& 0.06 	& -4.08 	& 7.8  	& 2700 	& 70 	& 260 	& 35 	& 47.5 	& 120.0 \\ 
89 	  &	B2  IV              & 22.0 	& 4.07 	& 4.0 	& -7.70 	& 0.12 	& -3.07 	& 7.5  	& 2300 	& 130 	& 260 	& 20 	& 45.8 	& 8.7 \\ 
90 	  &	O9.7 Iab            & 30.0 	& 5.53 	& 3.4 	& -6.86 	& 0.12 	& -5.86 	& 21.6  	& 1891 	& 90 	& 260 	& 43 	& 48.7 	& 40.9 \\ 
91 	  &	O8 IV               & 34.0 	& 5.16 	& 4.0 	& -6.75 	& 0.26 	& -4.31 	& 11.0  	& 2200 	& 274 	& 260 	& 44 	& 48.5 	& 70.6 \\ 
93 	  &	B0.2 V              & 29.0 	& 4.96 	& 4.0 	& -7.78 	& 0.21 	& -4.60 	& 12.0  	& 1700 	& 80 	& 240 	& 53 	& 47.6 	& 3.98 \\ 
95 	  &	B0.5 IV             & 28.0 	& 4.95 	& 3.8 	& -6.70 	& 0.12 	& -4.80 	& 12.7  	& 2000 	& 80 	& 260 	& 37 	& 47.6 	& 65.9 \\ 
98 	  &	B0.7 IV             & 27.0 	& 4.39 	& 4.2 	& -7.28 	& 0.03 	& -3.65 	& 7.2  	& 2700 	& 100 	& 250 	& 30 	& 46.8 	& 31.7 \\ 
99 	  &	O9.5 V              & 32.0 	& 4.77 	& 4.2 	& -6.73 	& 0.20 	& -3.50 	& 7.9  	& 2700 	& 150 	& 240 	& 36 	& 47.9 	& 112.0 \\ 
101 	  &	B1.5 (IV)e          & 20.0 	& 4.18 	& 3.6 	& -7.81 	& 0.24 	& -3.30 	& 10.3  	& 800 	& 232 	& 230 	& 15 	& 45.7 	& 0.8 \\ 
103 	  &	B2  (IV)e           & 21.0 	& 4.10 	& 3.8 	& -7.78 	& 0.16 	& -3.41 	& 8.5  	& 1900 	& 150 	& 280 	& 17 	& 45.7 	& 5.0 \\ 
104 	  &	B0.5 V              & 27.0 	& 4.50 	& 4.0 	& -7.38 	& 0.04 	& -3.72 	& 8.1  	& 2200 	& 200 	& 230 	& 24 	& 47.1 	& 16.8 \\ 
105 	  &	B1.5 V              & 22.0 	& 4.07 	& 4.0 	& -7.70 	& 0.12 	& -3.00 	& 7.5  	& 2300 	& 100 	& 240 	& 20 	& 45.8 	& 8.7 \\ 
107 	  &	O8 Vz               & 34.0 	& 5.03 	& 4.0 	& -6.45 	& 0.14 	& -4.39 	& 9.5  	& 2100 	& 40 	& 250 	& 33 	& 48.4 	& 129.1 \\ 
108 	  &	B1.5 V              & 22.0 	& 3.97 	& 4.0 	& -7.78 	& 0.10 	& -2.99 	& 6.7  	& 2200 	& 140 	& 240 	& 16 	& 45.7 	& 6.7 \\ 
109 	  &	B0.5 V              & 27.0 	& 4.36 	& 4.2 	& -6.70 	& 0.15 	& -3.10 	& 6.9  	& 2600 	& 90 	& 250 	& 28 	& 46.8 	& 111.3 \\ 
112 	  &	O9.7 V              & 31.0 	& 4.58 	& 4.2 	& -6.81 	& 0.08 	& -3.68 	& 6.8  	& 2500 	& 390 	& 260 	& 27 	& 47.5 	& 79.4 \\ 
113 	  &	B3 (V)e             & 18.0 	& 4.28 	& 3.6 	& -7.74 	& 0.20 	& -3.51 	& 14.2  	& 1000 	& 230 	& 240 	& 29 	& 45.7 	& 1.5 \\ 
116 	  &	B0.7 III            & 25.0 	& 5.08 	& 3.0 	& -6.78 	& 0.12 	& -5.27 	& 18.5  	& 1200 	& 252 	& 260 	& 13 	& 47.7 	& 19.83 \\ 
117 	  &	B2 (IV)e            & 20.0 	& 4.18 	& 3.6 	& -7.81 	& 0.16 	& -3.73 	& 10.3  	& 800 	& 72 	& 260 	& 15 	& 45.7 	& 0.8 \\ 
119 	  &	O8 (V)e             & 34.0 	& 5.28 	& 4.0 	& -6.66 	& 0.17 	& -5.10 	& 12.6  	& 2470 	& 100 	& 240 	& 58 	& 48.7 	& 110.78 \\ 
120 	  &	B2 IV               & 20.0 	& 4.03 	& 3.6 	& -7.93 	& 0.08 	& -3.32 	& 8.6  	& 800 	& 110 	& 250 	& 11 	& 45.6 	& 0.6 \\ 
121 	  &	B1.5 (V)e           & 21.0 	& 4.16 	& 3.8 	& -7.73 	& 0.23 	& -3.42 	& 9.1  	& 2000 	& 140 	& 260 	& 19 	& 45.8 	& 6.1 \\ 
122 	  &	B1.5 (IV)e          & 26.0 	& 4.80 	& 3.8 	& -6.66 	& 0.28 	& -4.44 	& 12.4  	& 2100 	& 140 	& 240 	& 36 	& 47.2 	& 79.2 \\ 
123 	  &	B0.2 V              & 28.0 	& 4.49 	& 4.0 	& -7.85 	& 0.10 	& -3.49 	& 7.5  	& 2100 	& 90 	& 240 	& 20 	& 47.1 	& 5.1 \\ 
125 	  &	O9 V                & 34.0 	& 4.98 	& 4.2 	& -6.70 	& 0.13 	& -4.38 	& 8.9  	& 2700 	& 160 	& 290 	& 46 	& 48.3 	& 120.0 \\ 
126 	  &	B2 V                & 21.0 	& 4.20 	& 3.8 	& -7.70 	& 0.03 	& -3.76 	& 9.5  	& 2000 	& 251 	& 230 	& 21 	& 45.8 	& 6.6 \\ 
127 	  &	B0.2 V              & 28.0 	& 4.39 	& 4.0 	& -7.93 	& 0.10 	& -3.23 	& 6.7  	& 2000 	& 80 	& 260 	& 16 	& 47.0 	& 3.9 \\ 
129 	  &	B0.5 IV             & 28.0 	& 4.69 	& 4.0 	& -7.70 	& 0.10 	& -4.22 	& 9.4  	& 2400 	& 110 	& 200 	& 32 	& 47.3 	& 9.5 \\ 
130 	  &	O9.7 V              & 31.0 	& 4.63 	& 4.2 	& -6.78 	& 0.07 	& -3.67 	& 7.2  	& 2500 	& 80 	& 260 	& 30 	& 47.6 	& 86.6 \\ 
132 	  &	O9 IV               & 32.0 	& 5.03 	& 3.8 	& -6.93 	& 0.12 	& -4.61 	& 10.7  	& 1700 	& 160 	& 260 	& 26 	& 48.3 	& 28.3 \\ 
133 	  &	O9 V                & 34.0 	& 4.98 	& 4.2 	& -6.70 	& 0.40 	& -3.91 	& 8.9  	& 2700 	& 110 	& 260 	& 46 	& 48.3 	& 120.0 \\ 
134 	  &	O9.7 Iab            & 30.0 	& 5.50 	& 3.4 	& -6.09 	& 0.16 	& -6.22 	& 20.9  	& 1800 	& 90 	& 260 	& 40 	& 48.7 	& 218.58 \\ 
136 	  &	B1 V                & 23.0 	& 4.07 	& 4.0 	& -7.70 	& 0.07 	& -2.84 	& 6.8  	& 2200 	& 180 	& 250 	& 17 	& 45.8 	& 8.0 \\ 
138 	  &	B0.7 V              & 26.0 	& 4.16 	& 4.2 	& -7.28 	& 0.05 	& -2.75 	& 5.9  	& 2500 	& 140 	& 260 	& 20 	& 46.5 	& 27.4 \\ 
139 	  &	B0.7 V              & 26.0 	& 4.16 	& 4.2 	& -7.28 	& 0.08 	& -3.02 	& 5.9  	& 2500 	& 110 	& 250 	& 20 	& 46.5 	& 27.4 \\ 
141 	  &	B1.5 V              & 23.0 	& 3.94 	& 4.2 	& -7.70 	& 0.06 	& -2.74 	& 5.9  	& 2600 	& 80 	& 240 	& 20 	& 46.2 	& 11.1 \\ 
142 	  &	B2 (V)e             & 21.0 	& 4.20 	& 3.8 	& -7.70 	& 0.14 	& -3.72 	& 9.5  	& 2000 	& 120 	& 240 	& 21 	& 45.8 	& 6.6 \\ 
143 	  &	B1.5 V              & 22.0 	& 3.97 	& 4.0 	& -7.78 	& 0.10 	& -3.18 	& 6.7  	& 2200 	& 120 	& 270 	& 16 	& 45.7 	& 6.7 \\ 
144 	  &	B0 II Nwk           & 26.0 	& 5.39 	& 3.0 	& -6.55 	& 0.25 	& -5.84 	& 24.5  	& 1585 	& 75 	& 320 	& 22 	& 48.3 	& 58.76 \\ 
145 	  &	O8 Vz               & 35.0 	& 5.16 	& 4.2 	& -6.43 	& 0.18 	& -4.97 	& 10.4  	& 2900 	& 120 	& 250 	& 62 	& 48.6 	& 260.8 \\ 
146 	  &	B1.5 IV             & 23.0 	& 4.18 	& 3.8 	& -8.49 	& 0.10 	& -3.36 	& 7.8  	& 1800 	& 120 	& 220 	& 14 	& 46.3 	& 0.9 \\ 
147 	  &	B1 Ia Nwk           & 20.0 	& 5.78 	& 2.5 	& -6.13 	& 0.15 	& -7.88 	& 64.8  	& 588 	& 40 	& 260 	& 49 	& 47.7 	& 21.2 \\ 
148 	  &	O9.7 IV             & 31.0 	& 5.03 	& 3.8 	& -6.65 	& 0.12 	& -4.66 	& 11.4  	& 1800 	& 160 	& 220 	& 30 	& 48.2 	& 59.8 \\ 
149 	  &	O9 IV               & 33.0 	& 4.60 	& 4.2 	& -6.93 	& 0.12 	& -3.27 	& 6.1  	& 2300 	& 70 	& 260 	& 22 	& 47.8 	& 51.9 \\ 
150 	  &	B0.5 V              & 28.0 	& 4.39 	& 4.0 	& -7.93 	& 0.08 	& -3.48 	& 6.7  	& 2000 	& 90 	& 250 	& 16 	& 47.0 	& 3.9 \\ 
151 	  &	O7.5 V              & 36.0 	& 5.08 	& 4.2 	& -6.55 	& 0.10 	& -4.44 	& 8.9  	& 2700 	& 30 	& 240 	& 46 	& 48.6 	& 168.6 \\ 
153 	  &	B2 IV               & 21.0 	& 4.25 	& 3.8 	& -7.66 	& 0.20 	& -3.79 	& 10.1  	& 2100 	& 130 	& 230 	& 24 	& 45.9 	& 7.9 \\ 
154 	  &	O7 Vz               & 34.0 	& 5.33 	& 4.0 	& -6.03 	& 0.40 	& -4.91 	& 13.4  	& 2500 	& 280 	& 260 	& 65 	& 48.7 	& 486.8 \\ 
155 	  &	B0.7 V              & 26.0 	& 4.20 	& 4.0 	& -7.93 	& 0.08 	& -2.91 	& 6.2  	& 1900 	& 342 	& 220 	& 14 	& 46.6 	& 3.5 \\ 
156 	  &	B0.2 I              & 26.0 	& 5.29 	& 3.0 	& -6.62 	& 0.16 	& -5.57 	& 21.8  	& 1585 	& 90 	& 240 	& 17 	& 48.2 	& 50.0 \\ 
157 	  &	B2 V                & 21.0 	& 4.25 	& 3.8 	& -7.66 	& 0.08 	& -3.89 	& 10.1  	& 2100 	& 80 	& 230 	& 24 	& 45.9 	& 7.9 \\ 
159 	  &	B2 V                & 21.0 	& 4.20 	& 3.8 	& -7.70 	& 0.10 	& -3.93 	& 9.5  	& 2000 	& 100 	& 320 	& 21 	& 45.8 	& 6.6 \\ 
160 	  &	B0 II Nwk           & 25.0 	& 5.35 	& 3.0 	& -6.28 	& 0.12 	& -5.97 	& 25.3  	& 1522 	& 80 	& 240 	& 23 	& 48.0 	& 100.9 \\ 
163 	  &	B0.5 III            & 24.0 	& 4.69 	& 3.2 	& -7.22 	& 0.10 	& -4.18 	& 12.8  	& 1527 	& 75 	& 250 	& 10 	& 47.0 	& 11.6 \\ 
164 	  &	B1.5 V              & 22.0 	& 4.07 	& 4.0 	& -7.70 	& 0.08 	& -3.19 	& 7.5  	& 2300 	& 250 	& 240 	& 20 	& 45.8 	& 8.7 \\ 
165 	  &	B3 V                & 18.0 	& 3.83 	& 3.6 	& -8.07 	& 0.10 	& -2.93 	& 8.5  	& 800 	& 80 	& 280 	& 10 	& 45.3 	& 0.4 \\ 
167 	  &	B5 V                & 18.0 	& 3.73 	& 3.6 	& -8.15 	& 0.10 	& -2.95 	& 7.6  	& 700 	& 120 	& 250 	& 8 	& 45.2 	& 0.3 \\ 
168 	  &	B2 V                & 21.0 	& 4.10 	& 3.8 	& -7.78 	& 0.08 	& -3.53 	& 8.5  	& 1900 	& 180 	& 240 	& 17 	& 45.7 	& 5.0 \\ 
170 	  &	B1 IV               & 24.0 	& 4.55 	& 3.8 	& -7.70 	& 0.06 	& -4.30 	& 10.9  	& 2100 	& 80 	& 240 	& 28 	& 46.7 	& 7.3 \\ 
172 	  &	O8.5 V              & 35.0 	& 5.16 	& 4.2 	& -6.53 	& 0.12 	& -4.75 	& 10.4  	& 2900 	& 60 	& 230 	& 62 	& 48.6 	& 207.2 \\ 
173 	  &	O9.7 V              & 31.0 	& 4.63 	& 4.2 	& -6.78 	& 0.23 	& -3.55 	& 7.2  	& 2500 	& 90 	& 240 	& 30 	& 47.6 	& 86.6 \\ 
174 	  &	O9.7 V              & 31.0 	& 4.73 	& 4.2 	& -6.70 	& 0.15 	& -3.49 	& 8.1  	& 2700 	& 120 	& 250 	& 38 	& 47.7 	& 120.0 \\ 
175 	  &	B0 III              & 26.0 	& 5.34 	& 3.2 	& -6.22 	& 0.12 	& -5.87 	& 23.1  	& 1624 	& 90 	& 330 	& 31 	& 48.0 	& 131.9 \\ 
176 	  &	O8.5 V              & 35.0 	& 4.86 	& 4.2 	& -6.85 	& 0.12 	& -3.83 	& 7.3  	& 2500 	& 31 	& 260 	& 31 	& 48.3 	& 72.8 \\ 
177 	  &	B2 V                & 21.0 	& 3.95 	& 3.8 	& -7.89 	& 0.10 	& -2.81 	& 7.2  	& 1700 	& 160 	& 200 	& 12 	& 45.6 	& 3.1 \\ 
179 	  &	B1 V                & 25.0 	& 4.26 	& 4.2 	& -7.12 	& 0.08 	& -3.43 	& 7.2  	& 2800 	& 80 	& 240 	& 30 	& 46.4 	& 48.5 \\ 
181 	  &	B1.5 (III)e         & 23.0 	& 5.07 	& 3.4 	& -6.62 	& 0.34 	& -5.06 	& 21.6  	& 1700 	& 170 	& 250 	& 43 	& 47.1 	& 56.5 \\ 
183 	  &	B0.5 V              & 27.0 	& 4.31 	& 4.2 	& -7.34 	& 0.10 	& -3.44 	& 6.5  	& 2600 	& 140 	& 240 	& 25 	& 46.7 	& 25.6 \\ 
184 	  &	O8 Vz               & 35.0 	& 5.01 	& 4.2 	& -6.74 	& 0.26 	& -3.99 	& 8.7  	& 2700 	& 110 	& 240 	& 44 	& 48.4 	& 110.1 \\ 
185 	  &	B2 V                & 22.0 	& 4.17 	& 4.0 	& -7.62 	& 0.14 	& -3.57 	& 8.4  	& 2400 	& 170 	& 250 	& 26 	& 45.9 	& 11.3 \\ 
186 	  &	O9 (V)e             & 34.0 	& 4.68 	& 4.2 	& -6.43 	& 0.10 	& -6.41 	& 6.3  	& 2300 	& 90 	& 240 	& 23 	& 48.0 	& 164.0 \\ 
189 	  &	B2.5 V              & 19.0 	& 3.80 	& 3.8 	& -7.85 	& 0.12 	& -3.01 	& 7.4  	& 900 	& 60 	& 240 	& 12 	& 45.4 	& 0.9 \\ 
190 	  &	B2 V                & 20.0 	& 4.03 	& 3.6 	& -7.93 	& 0.06 	& -3.11 	& 8.6  	& 800 	& 351 	& 250 	& 11 	& 45.5 	& 0.6 \\ 
191 	  &	B0.5 V              & 27.0 	& 4.46 	& 4.2 	& -7.22 	& 0.28 	& -3.53 	& 7.8  	& 2800 	& 90 	& 250 	& 35 	& 46.8 	& 38.5 \\ 
192 	  &	O9.5 (III)e         & 29.0 	& 5.54 	& 3.4 	& -6.07 	& 0.50 	& -5.64 	& 23.4  	& 1400 	& 220 	& 260 	& 50 	& 48.6 	& 136.1 \\ 
194 	  &	O8.5 V              & 34.0 	& 5.11 	& 4.0 	& -6.39 	& 0.18 	& -4.62 	& 10.4  	& 2200 	& 75 	& 260 	& 39 	& 48.5 	& 162.7 \\ 
195 	  &	B0.7 V              & 26.0 	& 4.36 	& 4.2 	& -7.62 	& 0.06 	& -3.69 	& 7.5  	& 2800 	& 90 	& 220 	& 32 	& 46.7 	& 15.3 \\ 
196 	  &	B0.2 IV             & 29.0 	& 4.39 	& 4.0 	& -7.00 	& 0.12 	& -3.02 	& 6.2  	& 1800 	& 180 	& 240 	& 14 	& 47.1 	& 26.7 \\ 
197 	  &	B0 V                & 31.0 	& 4.43 	& 4.2 	& -6.93 	& 0.12 	& -2.96 	& 5.7  	& 2300 	& 75 	& 250 	& 19 	& 47.4 	& 51.9 \\ 
198 	  &	O8 Vz               & 35.0 	& 4.66 	& 4.2 	& -6.80 	& 0.12 	& -3.78 	& 5.8  	& 2200 	& 150 	& 200 	& 20 	& 48.1 	& 63.3 \\ 
200 	  &	B1.5 IV             & 22.0 	& 4.43 	& 3.6 	& -6.81 	& 0.08 	& -4.31 	& 11.3  	& 1700 	& 100 	& 240 	& 19 	& 46.3 	& 36.7 \\ 
201 	  &	B0.5 V              & 27.0 	& 4.26 	& 4.2 	& -7.38 	& 0.08 	& -3.11 	& 6.2  	& 2500 	& 150 	& 240 	& 22 	& 46.7 	& 21.7 \\ 
202 	  &	B2 V                & 22.0 	& 4.12 	& 4.0 	& -7.66 	& 0.20 	& -3.47 	& 7.9  	& 2300 	& 120 	& 260 	& 23 	& 45.8 	& 9.5 \\ 
204 	  &	B2.5 V              & 18.0 	& 3.93 	& 3.6 	& -8.00 	& 0.10 	& -3.39 	& 9.5  	& 800 	& 140 	& 240 	& 13 	& 45.4 	& 0.5 \\ 
205 	  &	B1.5 V              & 20.0 	& 4.13 	& 3.6 	& -7.85 	& 0.16 	& -3.54 	& 9.7  	& 800 	& 80 	& 240 	& 14 	& 45.6 	& 0.7 \\ 
206 	  &	B2 V                & 21.0 	& 3.80 	& 3.8 	& -8.00 	& 0.07 	& -3.01 	& 6.0  	& 1600 	& 160 	& 230 	& 8 	& 45.4 	& 2.1 \\ 
207 	  &	B2.5 V              & 18.0 	& 3.93 	& 3.6 	& -8.00 	& 0.12 	& -3.06 	& 9.5  	& 800 	& 110 	& 240 	& 13 	& 45.4 	& 0.5 \\ 
209 	  &	B0.7 V              & 27.0 	& 4.46 	& 4.2 	& -7.22 	& 0.10 	& -3.75 	& 7.8  	& 2800 	& 366 	& 260 	& 35 	& 46.8 	& 38.5 \\ 
211 	  &	B1 V                & 22.0 	& 3.97 	& 4.0 	& -7.78 	& 0.12 	& -2.97 	& 6.7  	& 2100 	& 90 	& 240 	& 16 	& 45.6 	& 6.1 \\ 
213 	  &	B3 V                & 18.0 	& 3.83 	& 3.6 	& -8.07 	& 0.10 	& -3.17 	& 8.5  	& 800 	& 80 	& 250 	& 10 	& 45.3 	& 0.4 \\ 
215 	  &	B0 IV               & 28.0 	& 5.00 	& 3.8 	& -6.66 	& 0.10 	& -4.90 	& 13.5  	& 2100 	& 110 	& 260 	& 42 	& 47.7 	& 79.2 \\ 
220 	  &	B3 IV               & 17.0 	& 3.95 	& 3.2 	& -8.07 	& 0.10 	& -3.39 	& 10.9  	& 500 	& 233 	& 240 	& 7 	& 45.3 	& 0.2 \\ 
221 	  &	B3 V                & 18.0 	& 3.93 	& 3.6 	& -8.00 	& 0.08 	& -3.53 	& 9.5  	& 800 	& 80 	& 250 	& 13 	& 45.4 	& 0.5 \\ 
222 	  &	B1.5 IV             & 19.0 	& 4.27 	& 3.4 	& -7.85 	& 0.12 	& -4.10 	& 12.6  	& 700 	& 90 	& 240 	& 15 	& 45.9 	& 0.6 \\ 
224 	  &	B2 III              & 20.0 	& 4.68 	& 3.2 	& -7.65 	& 0.05 	& -4.76 	& 18.3  	& 600 	& 70 	& 250 	& 19 	& 46.4 	& 0.7 \\ 
225 	  &	B2.5 V              & 19.0 	& 4.00 	& 3.8 	& -7.70 	& 0.10 	& -3.28 	& 9.3  	& 1000 	& 150 	& 240 	& 20 	& 45.6 	& 1.6 \\ 
229 	  &	O9.5 V              & 33.0 	& 4.80 	& 4.2 	& -6.78 	& 0.08 	& -4.12 	& 7.7  	& 2600 	& 150 	& 230 	& 34 	& 48.0 	& 93.6 \\ 
231 	  &	B2 V                & 20.0 	& 3.93 	& 3.6 	& -8.00 	& 0.10 	& -3.33 	& 7.7  	& 700 	& 150 	& 240 	& 9 	& 45.5 	& 0.4 \\ 
232 	  &	B2.5 V              & 19.0 	& 3.73 	& 3.8 	& -8.15 	& 0.09 	& -2.92 	& 6.8  	& 900 	& 130 	& 240 	& 11 	& 45.1 	& 0.5 \\ 
233 	  &	B1.5 (V)e           & 21.0 	& 4.30 	& 3.8 	& -6.62 	& 0.17 	& -3.61 	& 10.7  	& 2100 	& 130 	& 260 	& 26 	& 46.4 	& 86.3 \\ 
234 	  &	B0.5 (V)e           & 27.0 	& 4.50 	& 4.0 	& -7.38 	& 0.14 	& -3.18 	& 8.1  	& 2200 	& 240 	& 260 	& 24 	& 47.1 	& 16.8 \\

\end{longtable}
}


\begin{center}

\tablehead{
\hline
\hline
\noalign{\vspace{1mm}}
 N206-FS& Age&$M_{\rm ev}$   \\
 \# &[Myr]& [$M_{\odot}$]  \\
\noalign{\vspace{1mm}}
\hline 
\noalign{\vspace{1mm}}}
\tabletail{\hline }
\topcaption{Ages and evolutionary masses of the OB stars determined from isochrones. See Sect.\,\ref{sect:hrd} for details}
\label{table:App_age}
\begin{supertabular}{cccl}
1	&  18.5	  &    10        \\
3	&  28.2	  &    8.1       \\
5	&  7.3	  &    10.6      \\
6	&  23.2	  &    9.2       \\
7	&  7.8	  &    14.2      \\
9	&  16.4	  &    11.7      \\
10	&  11.0	  &    11.7      \\
11	&  19.4	  &    9.7       \\
12	&  16.8	  &    11.2      \\
14	&  14.8	  &    11.5      \\
15	&  26.8	  &    8.2       \\
17	&  17.2	  &    10.8      \\
19	&  26.6	  &    8.6       \\
22	&  23.2	  &    9.2       \\
23	&  4.0	  &    21.1      \\
27	&  6.2	  &    18.4      \\
28	&  35.8	  &    7.3       \\
29	&  47.6	  &    6.3       \\
30	&  9.2	  &    11.4      \\
31	&  33.0	  &    7.4       \\
32	&  7.0	  &    15.4      \\
33	&  6.7	  &    23.4      \\
34	&  10.3	  &    15.5      \\
36	&  24.2	  &    8.9       \\
37	&  19.5	  &    9.7       \\
38	&  25.8	  &    8.6       \\
39	&  19.2	  &    9.4       \\
42	&  24.2	  &    8.9       \\
43	&  9.5	  &    12.1      \\
46	&  10.8	  &    10.6      \\
47	&  6.4	  &    18.7      \\
48	&  11.6	  &    13.2      \\
49	&  20.3	  &    9         \\
50	&  7.3	  &    10.6      \\
51	&  19.3	  &    9.7       \\
52	&  6.6	  &    15.4      \\
55	&  5.3	  &    25.1      \\
56	&  10.4	  &    13.5      \\
58	&  7.8	  &    19.7      \\
60	&  16.7	  &    11.7      \\
61	&  14.6	  &    9.5       \\
62	&  6.7	  &    17.8      \\
64	&  4.2	  &    23.9      \\
65	&  28.2	  &    7.8       \\
68	&  27.6	  &    8         \\
69	&  11.3	  &    12.1      \\
73	&  10.2	  &    14.3      \\
74	&  8.1	  &    15.9      \\
75	&  13.9	  &    12.6      \\
77	&  10.7	  &    11.5      \\
78	&  13.0	  &    10.5      \\
80	&  20.5	  &    9.3       \\
81	&  6.2	  &    25        \\
84	&  18.0	  &    9.4       \\
88	&  7.3	  &    16.4      \\
89	&  14.2	  &    9.8       \\
90	&  4.2	  &    32.4      \\
91	&  4.5	  &    25        \\
93	&  6.9	  &    19.4      \\
95	&  7.4	  &    18.8      \\
98	&  11.9	  &    13.1      \\
99	&  5.7	  &    18.4      \\
101	&  20.5	  &    9.6       \\
103	&  20.9	  &    9.3       \\
104	&  10.5	  &    13.5      \\
105	&  14.2	  &    9.8       \\
107	&  4.3	  &    23        \\
108	&  20.3	  &    9         \\
109	&  8.5	  &    12.8      \\
112	&  5.6	  &    16.1      \\
113	&  18.8	  &    10.3      \\
116	&  6.9	  &    19.9      \\
117	&  20.4	  &    9.6       \\
119	&  4.2	  &    27.7      \\
120	&  24.2	  &    8.9       \\
121	&  23.3	  &    9.2       \\
122	&  11.9	  &    15.1      \\
123	&  9.1	  &    13.9      \\
125	&  4.9	  &    22.3      \\
126	&  19.6	  &    9.7       \\
127	&  10.0	  &    13.5      \\
129	&  6.6	  &    16.1      \\
130	&  6.8	  &    16.8      \\
132	&  4.9	  &    22        \\
133	&  5.0	  &    22.3      \\
134	&  4.3	  &    31.8      \\
136	&  11.6	  &    10.1      \\
138	&  10.9	  &    11        \\
139	&  10.9	  &    11        \\
141	&  20.0	  &    9.1       \\
142	&  19.4	  &    9.7       \\
143	&  20.3	  &    9         \\
144	&  5.6	  &    26.9      \\
145	&  4.3	  &    25.6      \\
146	&  18.3	  &    9.9       \\
147	&  4.0	  &    39.5      \\
148	&  5.3	  &    21.6      \\
149	&  4.4	  &    17.5      \\
150	&  9.6	  &    13.5      \\
151	&  4.0	  &    25        \\
153	&  17.5	  &    10.3      \\
154	&  4.3	  &    28.8      \\
155	&  9.6	  &    11.2      \\
156	&  5.9	  &    25        \\
157	&  17.5	  &    10.3      \\
159	&  19.4	  &    9.7       \\
160	&  6.1	  &    25.4      \\
163	&  11.1	  &    14.6      \\
164	&  17.0	  &    9.8       \\
165	&  33.0	  &    7.4       \\
167	&  33.2	  &    7.3       \\
168	&  20.9	  &    9.3       \\
170	&  14.4	  &    12.4      \\
172	&  4.2	  &    25.6      \\
173	&  7.0	  &    16.8      \\
174	&  4.5	  &    17.9      \\
175	&  6.3	  &    24.7      \\
176	&  3.2	  &    21.1      \\
177	&  23.3	  &    8.7       \\
179	&  13.2	  &    11.3      \\
181	&  8.1	  &    19.4      \\
183	&  8.9	  &    12.4      \\
184	&  4.2	  &    23.3      \\
185	&  19.5	  &    9.4       \\
186	&  3.9	  &    19        \\
189	&  39.6	  &    7         \\
190	&  23.2	  &    8.9       \\
191	&  8.9	  &    13.6      \\
192	&  4.3	  &    32.1      \\
194	&  4.7	  &    23.9      \\
195	&  12.4	  &    12.4      \\
196	&  7.2	  &    13.9      \\
197	&  5.3	  &    14.8      \\
198	&  2.3	  &    19.4      \\
200	&  15.3	  &    11.9      \\
201	&  9.5	  &    12.1      \\
202	&  20.2	  &    9.4       \\
204	&  28.2	  &    8.1       \\
205	&  23.2	  &    9.2       \\
206	&  26.7	  &    7.7       \\
207	&  28.2	  &    8.1       \\
209	&  9.7	  &    13.6      \\
211	&  20.3	  &    9         \\
213	&  33.0	  &    7.4       \\
215	&  7.1	  &    19.8      \\
220	&  31.0	  &    8.1       \\
221	&  28.2	  &    8.1       \\
222	&  18.3	  &    10.3      \\
224	&  12.6	  &    13.9      \\
225	&  25.8	  &    8.6       \\
229	&  4.9	  &    19        \\
231	&  25.2	  &    8.4       \\
232	&  29.5	  &    7.4       \\
233	&  16.8	  &    10.7      \\
234	&  10.4	  &    13.5      \\
\end{supertabular}

\end{center}

\newpage

\section{Spectral fits of the WR binary N206-FS\,128}
\label{sect:appendixb}

\begin{figure*}
\centering
\includegraphics[scale=0.83]{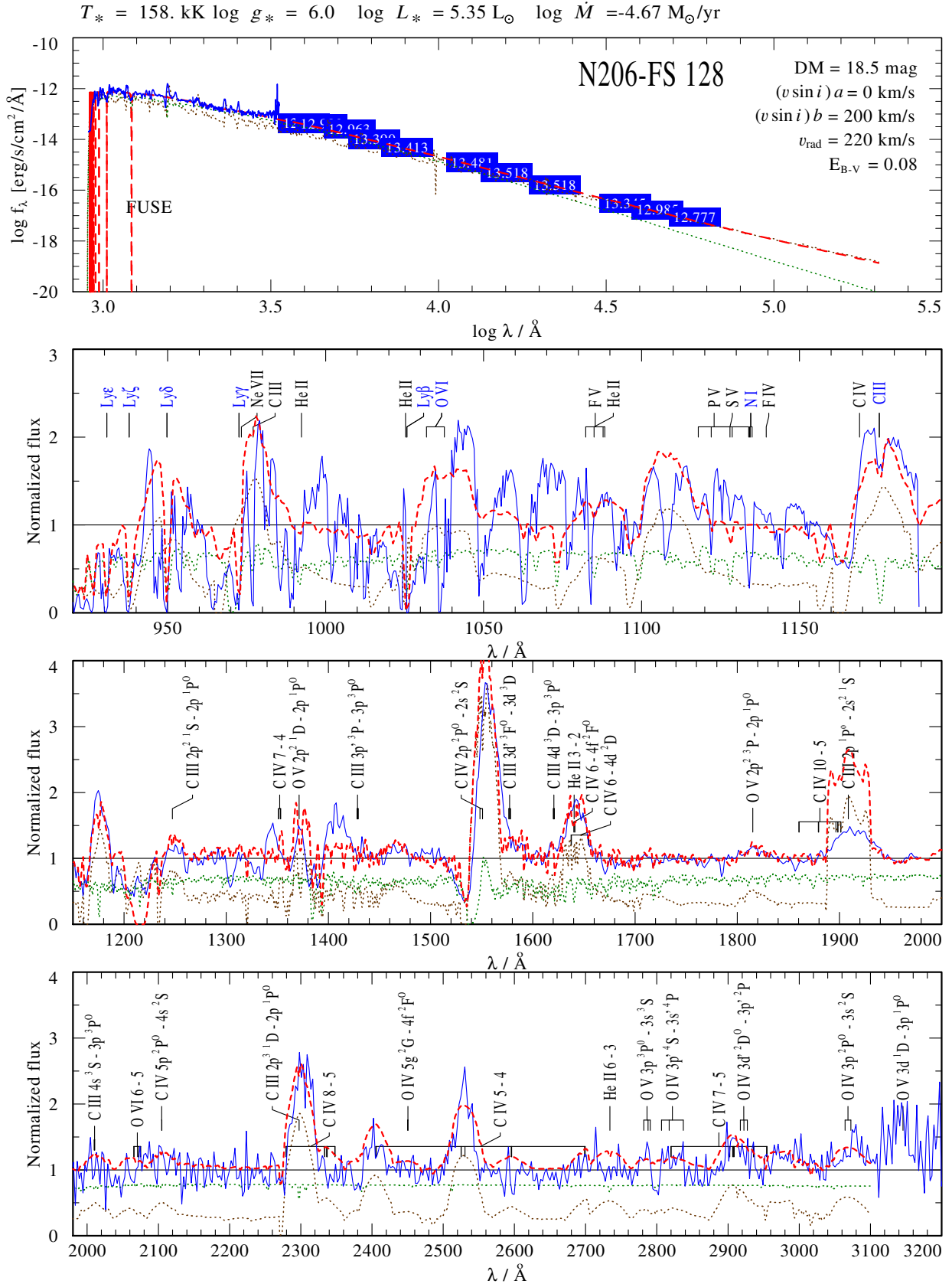}
\caption{Spectral fit for N206-FS\,128. The upper panel shows the spectral energy distribution with photometry from UV, optical, and infrared bands as well as the flux calibrated UV spectra. Lower panels show the normalized UV spectra from FUSE, IUE-short, and IUE-long (blue solid lines), fitted by composite model (red dashed lines). The WC and O star models are represented with brown and green dotted lines respectively. }
\label{fig:N206FS128_1}
\end{figure*}

\begin{figure*}
\centering
\includegraphics[scale=0.55,trim={7cm 6cm 7cm 0}]{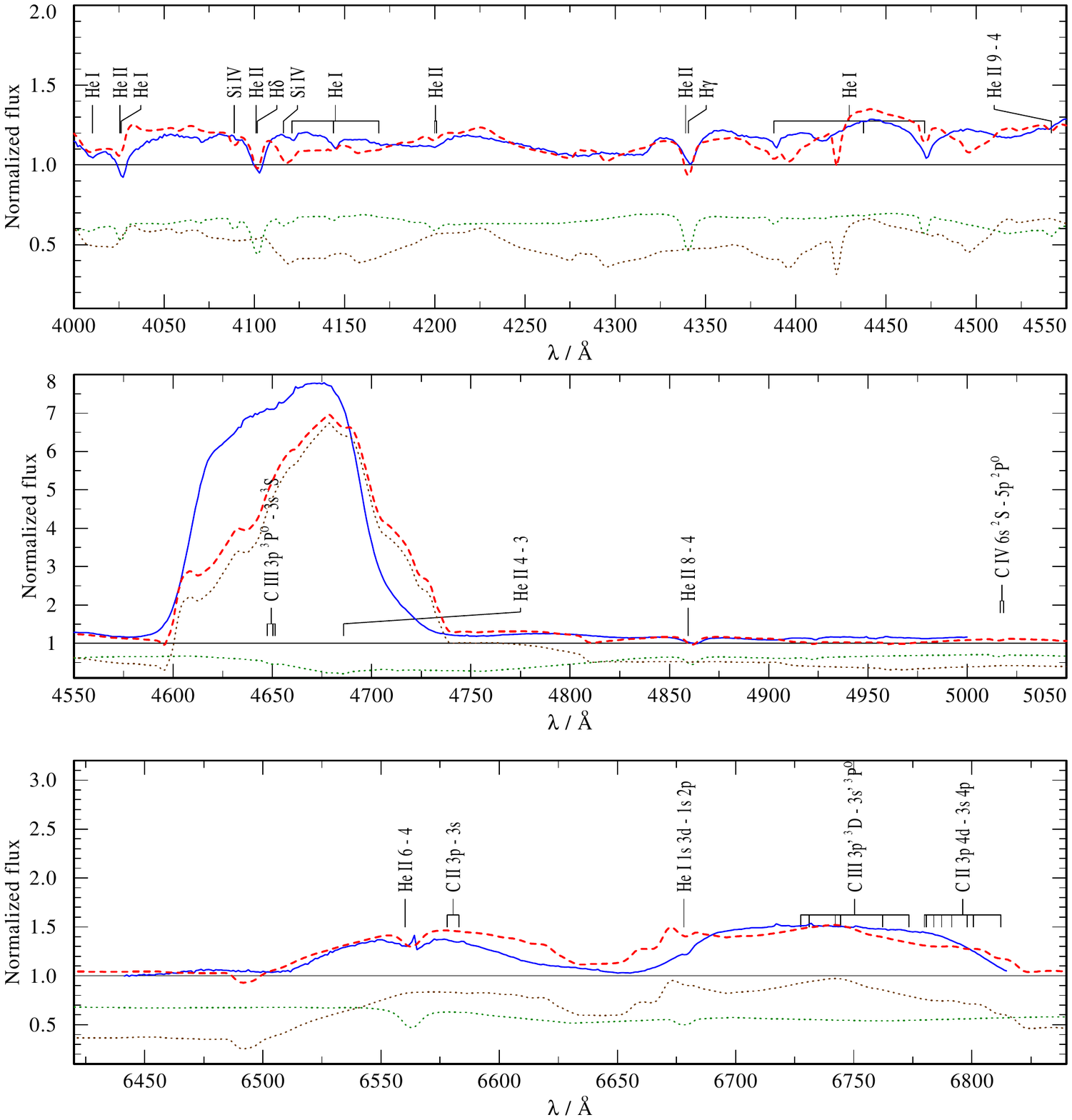}

\includegraphics[scale=0.55,trim={7cm 6cm 7cm 0}]{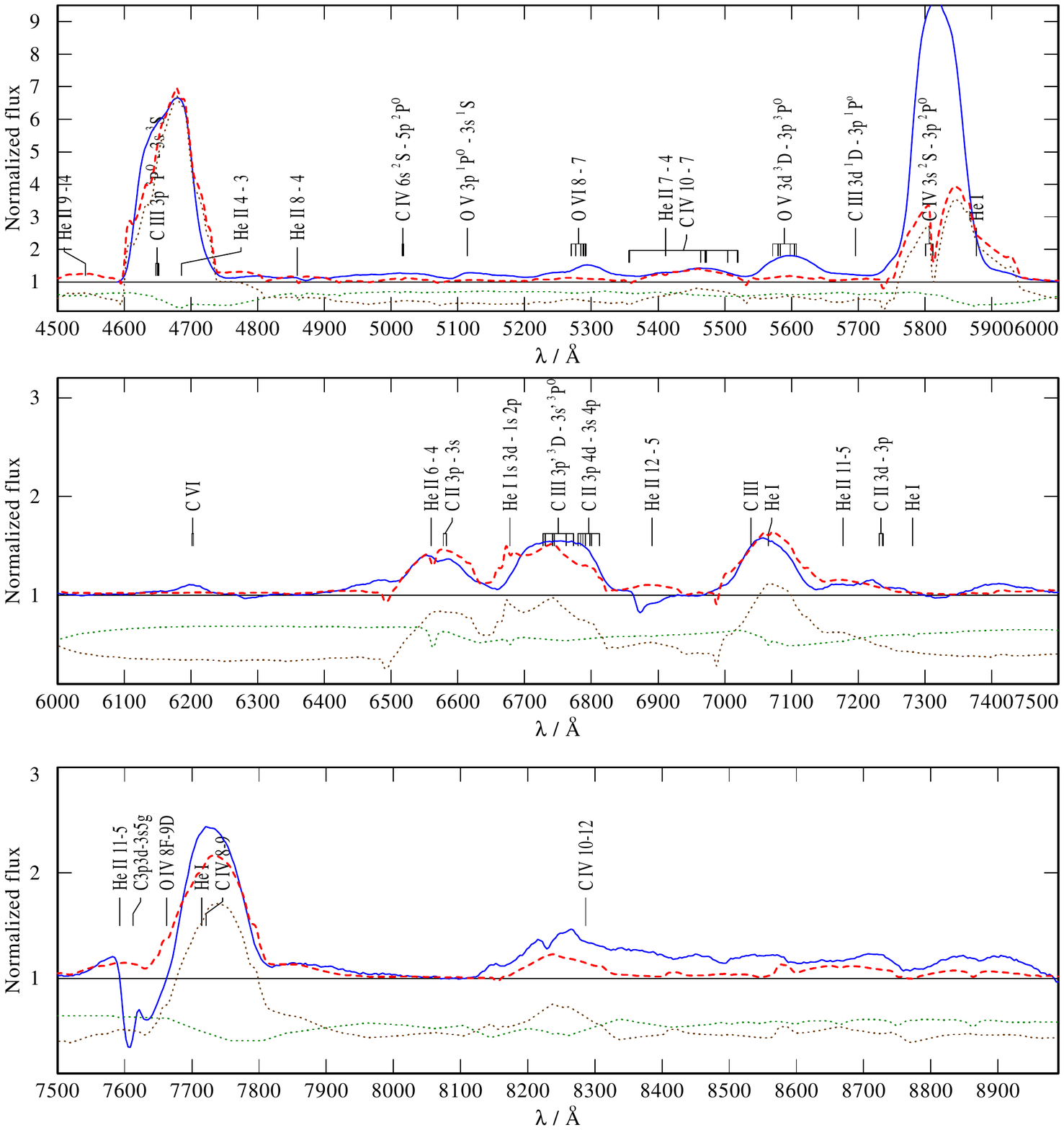}
\caption{Same as Fig.\,\ref{fig:N206FS128_1}, but shows the normalized FLAMES (panel 1-3) and FORS2 (panel 4-6) spectra. }
\label{fig:N206FS128_2}
\end{figure*}

\newpage

\section{Spectral fits of the OB stars}
\label{sect:appendixc}
We present the spectral fits of all the OB stars in the N\,206 complex analyzed
in this study. The upper panel shows the spectral energy distribution with 
photometry from UV, optical, and infra-red bands. Lower panels show the normalized VLT-FLAMES spectra depicted by blue solid lines. For a few stars the available UV spectra from HST, FUSE, IUE are also included in the lower panels. The observed spectra are over-plotted with PoWR model spectra (red dashed lines). The main parameters of the best-fit models are compiled in Table\,\ref{table:stellarparameters}, continued in Table\,\ref{table:App_stellarparameters}

\newpage
\clearpage

\pgfplotstableread{liste.dat}{\Liste}
\pgfplotstablegetrowsof{liste.dat}
\pgfmathsetmacro{\rows}{\pgfplotsretval-1}

\foreach \i in {1,...,\rows} {%
    \pgfplotstablegetelem{\i}{[index] 0}\of{\Liste} 
    \let\Name\pgfplotsretval
    \pgfplotstablegetelem{\i}{[index] 1}\of{\Liste} 
    \let\Caption\pgfplotsretval 

    \begin{figure}[!p]
    \centering
        
    \includegraphics[scale=0.9]{\Name}
    \caption{Spectral fit for N206-FS\Caption}
     \end{figure}  
     \clearpage
}

\end{document}